\documentclass[11pt,a4paper]{article}

\usepackage{ifpdf,ifthen,graphics,graphicx,lscape,indentfirst,dcolumn,amsthm,endnotes}
\usepackage{epstopdf}
\usepackage[namelimits]{amsmath,empheq}
\usepackage{epsfig,amssymb,fullpage,nicefrac,mathtools}
\usepackage[skip=0pt,singlelinecheck=false,font=small,labelfont=bf,labelsep=endash]{caption}
\usepackage[toc,page]{appendix}
\usepackage[FIGTOPCAP]{subfigure}
\usepackage[doublespacing]{setspace}
\usepackage{lipsum}
\usepackage{color}
\usepackage{bm}
\usepackage{tabularx}
\usepackage{tabu}
\usepackage{array}
\usepackage{float}
\usepackage{amsfonts}
\usepackage{bbm} 
\usepackage[bottom]{footmisc}
\usepackage{siunitx} 
\usepackage{multirow}
\usepackage{pdflscape}
\usepackage{booktabs}
\usepackage{breqn}
\usepackage{float}
\usepackage{enumerate}
\usepackage[margin=1in,showframe=false]{geometry} 
\usepackage{natbib} 
\usepackage{lscape} 
\usepackage{hyperref} 
\usepackage{url}
\usepackage{caption}
\DeclarePairedDelimiter\abs{\lvert}{\rvert}

\DeclarePairedDelimiterXPP\Aver[1]{\mathbb{E}}{[}{]}{}{

#1
}

\newcommand\iid{i.i.d.}
\newcommand\pN{\mathcal{N}}

\usepackage[scr=boondoxo,scrscaled=1.05]{mathalfa}

\usepackage{enumitem}
\SetLabelAlign{Center}{\hfil#1\hfil}
\SetLabelAlign{CenterWithParen}{\hfil(\makebox[0.9em]{#1})\hfil}

\graphicspath{{Images/}}

\usepackage{tikz}

\bibpunct[, ]{(}{)}{;}{a}{,}{,}

\title{Equity Tail Risk in the Treasury Bond Market
\footnote{The authors are grateful to Torben Andersen, Nicola Fusari and Viktor Todorov for valuable comments and helpful discussions of preliminary results. This paper has been written when Ruzzi was a Research Fellow at the Bank of Italy. The views expressed in this paper are those of the authors and do not necessarily reflect those of the Bank of Italy.}}
\author{Mirco Rubin\thanks{EDHEC Business School~(\texttt{mirco.rubin@edhec.edu})},~~~~Dario Ruzzi\thanks{Bank of Italy~(\texttt{dario.ruzzi@gmail.com})}}

\date{\today}

\begin{document}



\maketitle

\begin{abstract}
\noindent This paper quantifies the effects of equity tail risk on the US government bond market. We estimate equity tail risk with option-implied stock market volatility that stems from large negative price jumps, and we assess its value in reduced-form predictive regressions for Treasury returns and a term structure model for interest rates. We find that the left tail volatility of the stock market significantly predicts one-month excess returns on Treasuries both in- and out-of-sample. The incremental value of employing equity tail risk as a return forecasting factor can be of economic importance for a mean-variance investor trading bonds. The estimated term structure model shows that equity tail risk is priced in the US government bond market and, consistent with the theory of flight-to-safety, Treasury prices increase when the perception of tail risk is higher. 
Our results concerning the predictive power and pricing of equity tail risk extend to  major government bond markets in Europe. 
 

\end{abstract}


\thispagestyle{empty}

\textsf{\newpage}
\setcounter{page}{1}
\renewcommand{\baselinestretch}{2.0}
\small\normalsize


\section{Introduction}
\label{sec:Intro}

In times of financial distress, the disengagement from risky assets, such as stocks, and the simultaneous demand for a safe haven, such as top-tier government bonds, generate a flight-to-safety (FTS) event in the capital markets. A large body of literature examines the linkages between the stock and bond markets during crisis periods and their implications for asset pricing, see \cite{Hartmann2004}, \cite{Vayanos2004}, \cite{Chordia2005}, \cite{Connolly2005} and \cite{Adrian2015}, among others. 
We add to this literature by studying how Treasury bond prices and returns respond to changes in the perceived tail risk in the stock market. If top-tier government bonds are a major beneficiary of the FTS flows occurring when the stock market is hit by heavy losses, then we expect the downside tail risk of equity to affect bond risk premia and determine both stock and bond prices during distress periods. We investigate this conjecture by considering a Gaussian affine term structure model (ATSM) for US interest rates where the pricing factors are the principal components of the yield curve combined with the risk-neutral volatility of the US stock market that stems from large negative price jumps.
Further, we add to the existing empirical literature on bond return predictability by assessing the improvements in forecasting accuracy obtained with equity tail risk and examining whether they translate into higher risk-adjusted portfolio returns. 
Although evidence of bond return predictability based on measures of stock market uncertainty and skew has previously been found \citep{Feunou2014,Adrian2015,Crump2019}, this is, to the best of our knowledge, the first study to assess the economic gains of employing equity tail risk for predicting bond returns and examine in detail its implications for pricing Treasuries in a term structure model. 

Understanding the dynamics of bond yields is particularly useful for forecasting financial and macro variables, for making debt and monetary policy decisions and for derivative pricing. Most of these applications require the decomposition of yields into expectations of future short rates (averaged over the lifetime of the bond) and term premia, i.e.~the additional returns required by investors for bearing the risk of long-term commitment. Gaussian affine term structure models have long been used for this purpose, see, e.g., \cite{Duffee2002}, \cite{Kim2005} and \cite{Abrahams2016}. In the setup of a Gaussian ATSM, a number of pricing factors that affect bond yields are selected and assumed to evolve according to a vector autoregressive (VAR) process of order one. The yields of different maturities are all expressed as linear functions of the factors with restrictions on the coefficients that prevent arbitrage opportunities, implying that long-term yields are merely risk-adjusted expectations of future short rates.

The selection of pricing factors typically starts by extracting from the cross-section of bond yields a given number of principal components (PCs), which are linear combinations of the rates themselves. Since the seminal work of \cite{ls_jfi_1991}, the first three PCs have been prime candidates in this regard as they generally explain over 99\% of the variability in the term structure of bond yields and, due to their loadings, may be interpreted as the level, slope and curvature factor. As for the second principal component, \cite{Fama1987} and \cite{Campbell1991} showed that
variables related to the slope of the yield curve are highly informative about future bond returns. Despite the important role of the level, slope and curvature, it is well established in the literature that additional factors are needed to explain the cross-section of bond returns. For this reason, the first five principal components of the US Treasury yield curve are used as pricing factors in \cite{Adrian2013}, while \cite{Malik2016} adopt a four-factor specification for UK government bond yields. In a recent study focused on the US bond market, \cite{Feunou2018} show that a term structure model that includes the first three principal components and their own lags delivers better forecasts of excess returns than a specification using the first five principal components of yields as risk factors. 
Furthermore, several studies suggest that a great deal of information about expected excess returns -- the bond risk premium -- can be found in factors that are not principal components of the yield curve. \cite{Cochrane2005} discover a new linear combination of forward rates which is a strong predictor of future excess bond returns and, based on this evidence, \cite{Cochrane2008} use it in an ATSM along with the classical level, slope and curvature factors. More recently, \cite{Cooper2008}, \cite{Ludvigson2009}, \cite{Duffee2011}, \cite{Joslin2014}, \cite{CieslakPovala2015} and \cite{Huang2019} show that valuable information about bond risk premia is located outside of the yield curve and contained, for example, in macro variables that have little or no impact on current yields but strong predictive power for future bond returns.  

This paper explores the use of factors, other than combinations of yields, to drive the curve of US Treasury rates and explain bond returns.  In contrast to the vast majority of previous studies, however, we draw on the literature that deals with comovement in the equity and bond markets and we consider the possibility that pricing factors of Treasury bonds originate also in the stock market. The findings of \cite{Connolly2005} and \cite{Baele2010} indicate that measures linked to stock market uncertainty explain time variation in the stock-bond return relation and have important cross-market pricing effects.\footnote{\cite{Connolly2005} find that when the implied volatility from equity index options, measured by the VIX, increases to a considerable extent, bond returns tend to be higher than stock returns (flight-to-quality) and the correlation between the two assets over the next month is lower. \cite{Baele2010} show that the time-varying and sometimes negative stock-bond return correlations cannot be explained by macro variables but instead by liquidity factors and the variance risk premium, which represents the compensation demanded by investors for bearing variance risk and is defined as the difference between the risk-neutral and statistical expectations of the future return variation. Although the variance risk premium is a major contributor to the stock-bond return correlation dynamics, \cite{Baele2010} find significant exposures to it only for stock but not for bond returns.} 
Therefore, we select a risk measure which is known to predict the equity risk premium and we examine its role in the Treasury bond market. The existing literature suggests that the variance risk premium (VRP) forecasts the stock market returns at shorter horizons than do other predictors like dividend yields or price-to-earning ratios, see \cite{Bollerslev2009}, \cite{Bollerslev2014} and \cite{Bekaert2014}, among others. In view of recent studies showing that the predictive power of the VRP for the equity risk premium stems from a jump tail risk component that capture the investors' fear of a market crash (see, e.g., \cite{Andersen2015_JFE,Andersen2017_WP}, \cite{BTX2015} and \cite{Zinna2018}), we opt for the left jump volatility measure of \cite{BTX2015} to assess the impact of equity tail events on US Treasury bonds.
Building on the findings of \cite{Crump2019} that equity tail risk -- as measured by the CBOE Skew Index -- has strong in-sample predictive power for future Treasury bond returns, we are interested in understanding whether the forecast improvements afforded by equity tail risk continue to hold in a realistic out-of-sample forecasting setting and whether equity tail risk is priced in the term structure of US interest rates.
Hence, our main contribution is to provide empirical support that equity tail risk can generate economic value in bond return predictability and can be used as a bond pricing factor in a term structure model.\footnote{We stress that our pricing methodology differs from that of \cite{Farago2018}, who price Treasury bonds (and many other types of assets) using a consumption-based general equilibrium model that includes a non-risk-neutralized measure of downside risk.}


As opposed to \cite{Crump2019}, we do not rely on risk-neutral skewness to measure equity tail risk as the computation of moments higher than the second is prone to numerical errors and instability.\footnote{\cite{Liu2016} discuss the difficulties associated with the computation of risk-neutral skewness using the method by \cite{Bakshi2003}, on which the CBOE Skew Index is also based. They note how different approaches to the implementation of the \cite{Bakshi2003} method have led to mixed results in the literature of stock return predictability. With regard to this, the negative relationship between the \cite{Bakshi2003} measure of skewness and future returns found by \cite{BaliMurray2013} and \cite{Conrad2013} contrasts sharply with the positive relationship found by \cite{Rehman2012} and \cite{Stilger2016}.}
Instead, we rely on the procedures put forth by \cite{Bollerslev2011} to proxy investor fears for jump tail events. 
Specifically, we estimate equity tail risk with the model-free measure of left tail volatility developed by \cite{BTX2015} and calculated from short-dated deep out-of-the-money put options on the S\&P 500 market index. By doing so, we gauge the market's perception of jump tail risk over the following month based on the risk-neutral expectation of future return volatility associated with large negative price jumps.\footnote{As a robustness check, we also used a simple alternative measure of downside risk perceptions, the S\&P 500 implied volatility skew (or smirk), defined as the difference between the out-of-the-money put implied volatility (with delta of 0.20) and the average of the at-the-money call and put implied volatilities (with deltas of 0.50), both calculated from options with an expiration of 30 days \citep{An2014,Xing2010}. The results, which are available upon request, are very similar to those described here with the left jump volatility measure of \cite{BTX2015}.} 
The equity tail risk factor so obtained is by construction a measure of downside tail risk and in this it also differs from the CBOE VIX Index which is a symmetric risk measure that reflects compensation for both diffusive and jump risk.
With the \cite{BTX2015} measure in hand, we test whether equity tail risk is priced in the US term structure and examine whether return predictability can yield substantial gains from an investment perspective.

Our empirical analysis relies on monthly data for the US zero-coupon bond yield curve provided by \cite{GSW2007}. Given the bond yield data, we construct non-overlapping monthly excess returns on Treasuries with maturities up to ten years. Data is sampled at the end of each month between January 1996 and December 2018. For the same time period, we also compute monthly estimates of equity tail risk starting from daily observations of options on the S\&P 500 stock market index.\footnote{The option-implied left tail volatilities are computed daily and then the month-end value is recorded. To minimize the impact of outliers and help smooth out the estimation error, we also considered monthly estimates of equity tail risk obtained by averaging over the last five days of the month with the results being very similar to the ones reported below for their end-of-month counterparts.} 
The econometric framework consists of reduced-form predictive regressions that use the measure of equity tail risk to forecast monthly excess Treasury returns, and a Gaussian ATSM that uses equity tail risk to drive the curve of US interest rates. Moreover, the novel three-pass method of \cite{Giglio2019}, which delivers an estimate of a factor's risk premium that is robust to the omitted variable and measurement error problems, allows us to corroborate our conclusions regarding the pricing of equity tail risk in the bond market.

Our results can be summarized as follows.
First, there exist significant interactions between the future one-month returns of the US government bond market and the option-implied left tail volatility of the stock market.\footnote{\cite{Adrian2015} find that a nonlinear function of the VIX can predict both stock and bond returns at forecast horizons of about five months or longer. We show that the predictive power of the VIX for the future one-month returns on bonds is completely subsumed by the equity tail factor.
Our study is also related to the work of \cite{Kaminska2015}, who document the importance of global market sentiment for the term structure of UK government bonds using a VRP-based proxy of risk aversion.} 
The frequency at which we uncover the predictive power of equity tail risk for bond risk premia is considerably higher than that of the business cycle, which is normally used to interpret return predictability over forecast horizons of one quarter or longer. By contrast, the short-term predictability documented in this paper may be associated with the instantaneous reactions of market participants that, fearing a stock market crash, flock to the perceived safety of Treasuries.\footnote{The short-term predictability of the US term structure that we find is also in agreement with the fact that the investors' fear of a market crash decreases with the time horizon \citep{Zinna2018}.}  
Second, the predictability afforded by the equity tail factor continues to hold out-of-sample and can sometimes yield substantial economic value to a mean-variance investor. In fact, it is possible to achieve sizeable gains in portfolio performance when switching to a model that uses equity tail risk to predict bond returns. 
Third, turning to the results of the term structure model, the response of Treasury bond prices to a contemporaneous shock to the equity tail factor is positive and opposite to what happens in the stock market. This observation confirms the role of US government bonds as a safe haven.
Fourth, equity tail risk is strongly priced in the US term structure. We find evidence of a significant market price of equity tail risk not only with the ATSM but also with the novel framework proposed by \cite{Giglio2019} to conduct inference in the presence of omitted factors in linear asset pricing models. The equity tail factor's risk premium   that we observe in the US government bond market is consistent with the evidence in \cite{Longstaff2004} and \cite{Krishnamurthy2012}, who document the existence of a significant price for the safety and liquidity attributes of Treasuries.
Fifth, large drops in short-term bond yields and their embedded expectations of future short rates are attributable to equity tail risk. Therefore, while the Fed asset purchase programs have been a major force in lowering longer-term yields since the global financial crisis \citep{Kaminska2018}, the reduction in shorter-term yields is likely to have been caused by the investors' increased appetite for safe assets.
Finally, the strong and economically important role of equity tail risk -- estimated from both US and national index options -- extends to the Treasury bond market of the United Kingdom, Germany, Switzerland and France, while the evidence is considerably weaker in Spain and non-existent in Italy. 

The remainder of the paper is structured as follows. In Section \ref{sec:Data} we describe the bond data and the construction of the equity tail risk measure. In Section \ref{sec:Econ_Frame} we review the methodology used to assess bond return predictability and we outline the term structure modeling approach and the \cite{Giglio2019} three-pass method. 
Section \ref{sec:Empirical_Res} reports the in-sample and out-of-sample empirical results on bond return predictability and the application of equity tail risk in bond pricing. Evidence from international bond markets is also presented. Section \ref{sec:Conclusion} concludes.

\section{Data}
\label{sec:Data}

In this section we present the data sources and methods used to construct the monthly time series of excess Treasury returns and equity tail risk measure. All time series are generated over the period January 1996 to December 2018 with data recorded at the end of each month.

\subsection{Bond Returns}
\label{subsec:Data_bond}

We compute Treasury bond returns using the \cite{GSW2007} zero-coupon bond yield curve derived from observed US government bond prices.\footnote{The \cite{GSW2007} yield data are available at a daily frequency for annually spaced maturities ranging from 1 to 30 years from the Federal Reserve website \url{https://www.federalreserve.gov/pubs/feds/2006/200628/200628abs.html}~. The parameters of the Nelson-Siegel-Svensson model used by \cite{GSW2007} are also published, thus allowing to retrieve yields for any desired maturity, including the longer ones.}
We consider maturities up to ten years, for which we construct non-overlapping one-month holding period returns.\footnote{The advantages of using non-overlapping one-month returns instead of the more conventional overlapping one-year returns are explained in \cite{Gargano2019}. 
} 
Following the studies of \cite{Adrian2013}, \cite{Abrahams2016} and \cite{Gargano2019}, we define the monthly return of the bond with maturity $n$ (in months) as the return from buying an $n$-maturity bond and selling it as an  $(n-1)$-maturity bond one month later. Setting the risk-free rate equal to the $n=1$ month yield, the monthly excess log-return at date $t+1$ (i.e., from the end of month $t$ to the end of month $t+1$) for the generic bond with maturity $n$ at time $t$ gets computed as 
\begin{equation}
\label{eq:rx_b}
rx_{t+1}^{(n-1)} ~ = ~ - \frac{(n - 1)}{12} y_{t+1}^{(n - 1)} ~ + ~ \frac{n}{12} y_{t}^{(n)} -  \frac{1}{12} y_{t}^{(1)} ~,
\end{equation}
where $y_{t}^{(i)}$ is the annualized (but not in percentage) continuously compounded yield on the zero-coupon bond with maturity $i$ at time $t$, provided by \cite{GSW2007}.

Table \ref{tab:DescriptiveStats} provides descriptive statistics for one-month excess returns on US Treasury bonds with maturity $n=12,24,36,48,60,84,120$ months.\footnote{Throughout the rest of the paper, the terms ``returns'' and  ``excess returns'' are used interchangeably to indicate excess returns unless otherwise indicated by the particular context.} 
 A quick inspection of Panel A reveals that longer-term bonds are characterized by higher mean excess returns and higher volatility. However, the reward-to-volatility ratio, also known as Sharpe ratio, declines with the bond maturity. While all bond returns are leptokurtic, only returns on bonds with maturity up to 3 years display a strong positive skewness and a first-order autocorrelation coefficient above 0.1. Finally, as shown in Panel B, the cross-sectional correlation between bond returns is always above 0.5 with values well above 0.9 for maturities that are close to each other. 

\begin{center}
	[ Insert Table \ref{tab:DescriptiveStats} here ]
\end{center}



\subsection{Equity Tail Risk}
\label{subsec:Data_equity}

The equity tail risk factor of this paper corresponds to the \cite{BTX2015} measure of left jump tail volatility implied by short-dated deep out-of-the-money (OTM) put options on the US stock market index.\footnote{The interested reader is directed to \cite{BTX2015} for an in-depth description of the theoretical framework since here we limit ourselves to highlighting the distinctive features and to discussing the estimation and implementation procedures.} This measure is essentially model-free and exploits extreme value theory to characterize the density of the risk-neutral return tails. 
The intuition behind it is that short-maturity OTM options remain worthless unless the investors believe that a big jump in the underlying price will occur before the option expires. Since diffusive risk does not affect their price, these contracts are fundamentally suitable to estimate jump tail risk \citep{Bollerslev2011,BT2014}.
The calculation of the \cite{BTX2015} measure is based on two parameters that must be estimated period-per-period and represent two separate sources of independent variation in the jump intensity process. 
The first parameter is $\alpha^{-}_t$ which controls the time-varying rate of decay, or shape, of the left tail. Lower values of $\alpha^{-}_t$ are associated with a slower rate at which the put option prices decay for successively deeper OTM contracts, implying a fatter left tail of the risk neutral density. \cite{BT2014} and \cite{BTX2015} show that $\alpha^{-}_t$ can be estimated as follow,
\begin{equation}
\label{eq:alpha_m}
\hat{\alpha}^{-}_t = \underset{\alpha^{-}}{\text{arg min}} \frac{1}{N_t^{-}} \displaystyle\sum_{i=2}^{N_t^{-}} \abs*{ \log \bigg( \frac{O_{t,\tau}(k_{t,i})}{O_{t,\tau}(k_{t,i-1})} \bigg) (k_{t,i} - k_{t,i-1})^{-1} - (1 + \alpha^{-}) } ~~ ,
\end{equation}

\noindent where $O_{t,\tau}(k)$ is the time $t$ price of the OTM put option with time to expiration $\tau$ and (negative) log-forward moneyness $k$, and $N_t^{-}$ is the total number of OTM puts used in the estimation with moneyness $0 < -k_{t,1} < ... < -k_{t,N_t^{-}}$. 
The second source of variation in the jump tails comes from parameter $\phi^{-}_t$ which shifts the level of the jump intensity process through time. 
Given an estimate for $\alpha_t^{-}$, the estimate of $\phi^{-}_t$ can be calculated as follows,
\begin{equation}
\label{eq:phi_m}
\hat{\phi}^{-}_t = \underset{\phi^{-}}{\text{arg min}} \frac{1}{N_t^{-}} \displaystyle\sum_{i=1}^{N_t^{-}} \abs*{ \log \bigg( \frac{e^{r_{t,\tau}} O_{t,\tau}(k_{t,i})}{\tau F_{t,\tau}} \bigg) - (1+\hat{\alpha}_t^{-})k_{t,i} + \log (\hat{\alpha}_t^{-}+1) + \log (\hat{\alpha}_t^{-}) - \log (\phi^{-}) } ~~ ,
\end{equation}

\noindent where $r_{t,\tau}$ is the risk-free interest rate over the $[t,t+\tau]$ time interval, $F_{t,\tau}$ is the forward price of the underlying asset at time $t$ and with maturity date $t+\tau$, and the rest of the notation is as before. Following \cite{Andersen2019_JPN}, we estimate $\alpha_t^{-}$ at a weekly frequency, while we allow $\phi_t^{-}$, which is less sensitive to outliers, to vary each trading day. Furthermore, we pool data across multiple maturities for more robust estimation of both parameters.


When defining left jump tail variation, \cite{BTX2015} focus on asset price moves that are unusually large relative to the current level of risk in the economy. To this end, they use a time-varying cutoff $k_t$ for the log-jump size that identifies, for each trading day, the start of the left tail based on the market volatility level. 
In our study we let $k_t$ be the threshold for a negative tail jump at the one-month horizon and we fix it at three times the maturity-normalized  30-day at-the-money Black-Scholes implied volatility at time $t$.\footnote{The threshold that we use for the log-jump size, although smaller than that of \cite{BTX2015} and \cite{Andersen2019_JPN}, is still able to define as jumps asset price moves of greater magnitude than those corresponding to the levels of moneyness used in \cite{Bollerslev2011} and considered \emph{sufficiently ``deep'' in the tails to guarantee that the effect of the diffusive price components is minimal, and that the extreme value distribution provides a good approximation to the jump tail probabilities}. Nevertheless, we also considered larger values for the tail cutoff, resulting in similar, but less significant, interactions between the left tail volatility of the stock market and future bond returns. These results are available upon request.}
By substituting $\hat{\alpha}_t^{-}$, $\hat{\phi}_t^{-}$ and $k_t$ in the expression proposed by \cite{BTX2015} for the predictable risk-neutral left jump tail variation, we construct the equity tail risk measure of this paper as,
\begin{equation}
\label{eq:TR}
\text{TR}^{(eq)}_t = \sqrt{\hat{\phi}_t^{-} e^{-\hat{\alpha}_t^{-}|k_t|} ( \hat{\alpha}_t^{-}k_t (\hat{\alpha}_t^{-}k_t + 2) +2) / (\hat{\alpha}_t^{-})^{3} } ~~ .
\end{equation}

\begin{figure}[!ht]
	\caption{Time series of the S\&P 500 option-implied equity tail risk measure}  
	\label{fig:TR_and_CFNAI_and_REC}
	{\footnotesize \centering
	\vspace{0.2cm}
	\includegraphics[width=16.0cm, height=10cm]{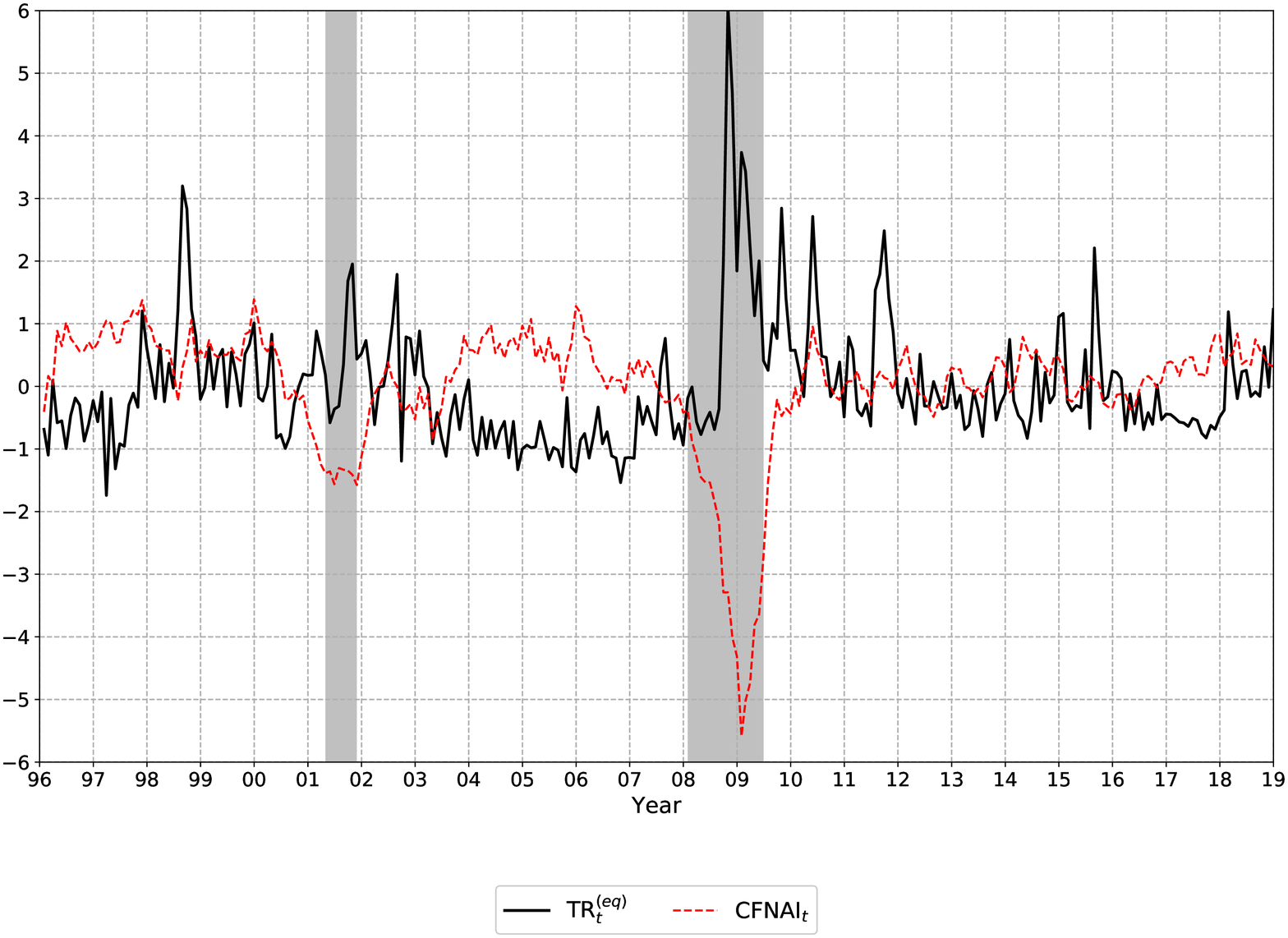} 
	}
	\begin{minipage}{15.8cm}
	\vspace{0.3cm}
	The figure displays the end-of-month values of the S\&P 500 option-implied equity tail risk measure ($\text{TR}_t^{(eq)}$) and 3-month moving average of the Chicago National Activity Index ($\text{CFNAI}_t$) from January 1996 to December 2018. For convenience, both series have been normalized to have mean zero and unit variance. Contemporaneous correlation between $\text{TR}^{(eq)}$ and CFNAI is $-0.49$. Vertical gray bars denote the National Bureau of Economic Research (NBER) based recession periods.
	\end{minipage}	
\end{figure}

To compute the equity tail risk measure in (\ref{eq:TR}), which represents the (annualized) volatility that stems from negative return jumps greater than a threshold $k_t$, we rely on daily data reported by OptionMetrics IvyDB US for the European style S\&P 500 equity-index options. We apply the following standard filters to our dataset. We discard options with a tenor of less than eight days or more than forty-five days. We discard options with missing prices, options with non-positive bid prices and options with non-positive bid-ask spread. The price of the surviving contracts is obtained as the average of bid and ask quotes. For each day in the sample, we retain only option tenors for which we have at least five pairs of call and put contracts with the same strike price. We exploit these cross sections to derive, via put-call parity, the underlying asset price adjusted for the dividend yield that apply to a given option tenor on a given day.\footnote{The risk-free rates used in the estimation of $\text{TR}^{(eq)}_t$ come from the \cite{GSW2007} dataset described in Section \ref{subsec:Data_bond}. Data for the 30-day at-the-money implied volatility used to calculate $k_t$ is from the volatility surface file of  IvyDB OptionMetrics.} We discard all in-the-money options and we retain only out-of-the-money put options with volatility-adjusted log-forward moneyness less than or equal to $-2.5$. Finally, we omit any out-of-the-money options for which the price does not decrease with the strike price. Using the data obtained from the filtering process, we compute the end-of-month values of the S\&P 500 option-implied left tail volatility $\text{TR}^{(eq)}$, which we plot in Figure \ref{fig:TR_and_CFNAI_and_REC} against the 3-month moving average of the Chicago National Activity Index (CFNAI) and the National Bureau of Economic Research (NBER) based recession periods.

\vspace{1.0em}

\begin{center}
	\noindent [ Insert Figure \ref{fig:TR_and_CFNAI_and_REC} here ]
\end{center}

\vspace{1.0em}

\noindent From Figure \ref{fig:TR_and_CFNAI_and_REC} it is clear that our equity tail risk measure is higher during periods of economic contraction. However, we note that $\text{TR}^{(eq)}$ spikes also in periods when the CFNAI is above its mean level, for instance during the Russian financial crisis in 1998 and the intensification of the European sovereign debt crisis in 2010 and 2011. Now turning to the descriptive statistics reported in Table \ref{tab:DescriptiveStats}, we find that the annualized left tail volatility of the stock market is on average 10\%. Furthermore, we observe that equity tail risk is positively correlated with the future one-month Treasury returns. The correlation coefficient is about 0.2 across all maturities. 
In the next sections, we use $\text{TR}^{(eq)}$ to gauge the market's perception of jump tail risk and examine the response of US Treasury bonds to the downside tail risk of the stock market. 

\section{Econometric Framework}
\label{sec:Econ_Frame}

In this section we describe the techniques and evaluation criteria used to investigate the predictive content of equity tail risk for future bond returns and we outline the procedures used in the assessment of equity tail risk pricing in the US government bond market.

\subsection{Reduced-form Predictive Regressions}
\label{subsec:RedForm_Regressions}

The econometric framework that we adopt to evaluate bond return predictability is based on reduced-form predictive regressions that include the equity tail risk measure in (\ref{eq:TR}) and, possibly, a certain number of PCs of bond yields that control for the forecasting information contained in the yield curve.\footnote{In Section \ref{sec:Empirical_Res} we assess the in-sample explanatory power of equity tail risk for bond risk premia by controlling for other successful return predictive factors found in the literature. Specifically, we consider the \cite{Cochrane2005} bond return predictor obtained as a linear combination of forward rates, the \cite{CieslakPovala2015} risk-premium factor obtained from a decomposition of Treasury yields into inflation expectations and maturity-specific interest-rate cycles, and the orthogonal component of the CBOE VIX with respect to $\text{TR}^{(eq)}$.}
With respect to the yield predictors, we consider both the traditional level, slope and curvature factors, which are standard in the literature on bond return predictability, and the two higher-order principal components used by \cite{Adrian2013} to explain Treasury return variation. Therefore, our bond return prediction models take the following form,
\begin{subequations}
\begin{align}
\label{eq:RX_reg_univ}
rx_{t+1}^{(n-1)} &= \beta_{0} + \beta_{1} ~ \text{TR}^{(eq)}_t + \epsilon_{t+1} ~  , \\
rx_{t+1}^{(n-1)} &= \beta_{0} + \beta_{1} ~ \text{TR}^{(eq)}_t + \beta_{2} ~ \text{PC1}_t +  \beta_{3} ~ \text{PC2}_t + \beta_{4} ~ \text{PC3}_t + \epsilon_{t+1} ~  ,
\label{eq:RX_reg_PC3} \\ 
rx_{t+1}^{(n-1)} &= \beta_{0} + \beta_{1} ~ \text{TR}^{(eq)}_t + \beta_{2} ~ \text{PC1}_t +  \beta_{3} ~ \text{PC2}_t + \beta_{4} ~ \text{PC3}_t + \beta_{5} ~ \text{PC4}_t  +  \beta_{6} ~ \text{PC5}_t + \epsilon_{t+1} ~ ,
\label{eq:RX_reg_PC5}
\end{align}
\end{subequations}

\noindent where $\text{TR}^{(eq)}$ represents the perceived tail risk in the US stock market, and $\text{PC1}$--$\text{PC5}$ are the first five principal components estimated from an eigenvalue decomposition of the variance-covariance matrix of zero-coupon bond yields. 
We include in the analysis the univariate model of equation (\ref{eq:RX_reg_univ}) not only because it is a quick and inexpensive method to gauge the strength and sign of the relation between bond returns and equity tail risk, but also because simpler models might generate more accurate out-of-sample forecasts. In the following, we will assess the forecasting performance of model (\ref{eq:RX_reg_univ}) relative to that of the Expectation Hypothesis (EH) model. The EH assumes no predictability of bond risk premia, implying that the out-of-sample model forecasts of bond returns are equal to a recursively updated constant based on the historical return mean. The performance of models (\ref{eq:RX_reg_PC3}) and (\ref{eq:RX_reg_PC5}) will be compared to that of a model that includes, respectively, the first three and five PCs of bond yields alone.

The relationship between equity tail risk and bond risk premia is firstly assessed by testing the statistical significance of the coefficient of $\text{TR}^{(eq)}$ over the full sample period. The test of $\beta_{1}=0$ is carried out not only by means of conventional inference, for which we compute the Newey-West $p$-values with a 12-lag standard error correction, but even with the more robust inference method developed by \cite{Bauer2018}. The latter addresses the small-sample distortions in bond return predictive regressions that are induced, among others, by the high persistence of the predictive variables. \cite{Bauer2018} propose a parametric bootstrap that generates yield curve data assuming that a given factor structure underlies the bond yields and that the relevant predictive information for bond returns is entirely contained in the yield curve. We compute \cite{Bauer2018} $p$-values with 5,000 artificial samples and two separate 1-month VAR processes for $\text{TR}^{(eq)}$ and the principal components of yields.\footnote{As a robustness check, we have also evaluated the strength of the relationship between equity tail risk and future Treasury bond returns using the inference method recently proposed by \cite{Crump2019}. This is a non-parametric bootstrap that accounts for the time-series and cross-sectional dependence in bond yields and generates data while remaining agnostic about the exact factor structure in the data. Based on the \cite{Crump2019} $p$-values computed with resampled data from 999 boostrap replications, we continue to observe statistically significant relationships at the 0.10 level or lower across all maturities considered. Because of space considerations, these results are not reported in the paper, but are available upon request from the authors.}

To check whether the in-sample interactions between one-month-ahead bond risk premia and equity tail risk translate into positive real-time predictive ability, we consider an out-of-sample exercise in which forecasts are recursively generated at a monthly frequency based on information available only at the forecast time.
We estimate the models in (\ref{eq:RX_reg_univ}), (\ref{eq:RX_reg_PC3}) and (\ref{eq:RX_reg_PC5}) -- and corresponding benchmarks that do not include $\text{TR}^{(eq)}$ -- recursively over expanding and rolling samples, where the first half of observations (1996:01-2007:06) constitutes the initial estimation period and the second half (2007:07-2018:12) constitutes the forecast evaluation period.
Within this out-of-sample setting, we follow the approach used by \cite{Eriksen2017} and \cite{Gargano2019}, among others, and we assess both the statistical and the economic value of bond return predictability with equity tail risk.
We evaluate statistical significance with the \cite{CT2008} $R_{OS}^2$ statistic that measures the percentage reduction in mean squared prediction error (MSPE) for the out-of-sample forecasts generated by a given model relative to a benchmark. For each one of the preferred models in (\ref{eq:RX_reg_univ}), (\ref{eq:RX_reg_PC3}) and (\ref{eq:RX_reg_PC5}), we compute the  \cite{CT2008} statistic as, 
\begin{equation}
\label{eq:R2_os}
R_{OS}^2 = 1 - \frac{\displaystyle\sum_{t=1}^{T} \Big(rx_{t+1}^{(n-1)} -  \widehat{rx}_{t+1}^{(n-1)}\Big)^2}{\displaystyle\sum_{t=1}^{T} \Big(rx_{t+1}^{(n-1)} -  \widetilde{rx}_{t+1}^{(n-1)}\Big)^2}  ~,
\end{equation}

\noindent where $\widehat{rx}_{t+1}^{(n-1)}$ and $\widetilde{rx}_{t+1}^{(n-1)}$ denote, respectively, the forecasts from one of the preferred models that include $\text{TR}^{(eq)}$ and the forecasts from its benchmark (either the PCs-only or EH model), and $T$ is the number of out-of-sample forecasts. Positive values of $R_{OS}^2$ indicate higher predictive accuracy for the bond return prediction model that includes equity tail risk.
We formally test for predictive superiority of the preferred models using the \cite{CW2007} test. This is a statistical test of the null hypothesis of $R_{OS}^2 \le 0$ against the one-sided alternative of $R_{OS}^2 > 0$. Significant predictive superiority of the model that includes equity tail risk is found in case of rejection of the null. We conduct the \cite{CW2007} test by estimating the $t$-statistic of regressing
\begin{equation}
\label{eq:CW_os}
CW_{t+1} = \Big(rx_{t+1}^{(n-1)} -  \widetilde{rx}_{t+1}^{(n-1)}\Big)^2 - \Big[ \Big(rx_{t+1}^{(n-1)} - \widehat{rx}_{t+1}^{(n-1)}\Big)^2 - \Big(\widetilde{rx}_{t+1}^{(n-1)} - \widehat{rx}_{t+1}^{(n-1)}\Big)^2  \Big] ~,
\end{equation}

\noindent on a constant term, and then computing its $p$-value according to the Newey-West and \cite{Bauer2018} inference procedures described above.\footnote{We use the \cite{Bauer2018} procedure to also bootstrap the $p$-values of the $R_{OS}^2$ statistic.} 
The statistic in (\ref{eq:CW_os}) is the difference in the preferred and benchmark model's squared prediction errors adjusted for the upward bias induced by having to estimate the additional parameter $\beta_{1}$ that is 0 under the null hypothesis. 

Finally, we examine the economic value of the predicting capability of the models in (\ref{eq:RX_reg_univ}), (\ref{eq:RX_reg_PC3}) and (\ref{eq:RX_reg_PC5}) by looking for sizeable risk-adjusted returns in asset allocation. To this end, we conduct a portfolio exercise with a mean-variance investor that every month allocates his or her wealth between a 1-month Treasury (risk-free) bond and an $n$-month Treasury (risky) bond. By solving the same expected utility maximization problem as in \cite{Eriksen2017}, at time $t$, the investor optimally allocates a proportion of:
\begin{equation}
\label{eq:w_star}
w_{t}^{(n)} = \frac{1}{\gamma} ~ \frac{\text{E}_t \Big[rx_{t+1}^{(n-1)}\Big]} {\text{Var}_t\Big[rx_{t+1}^{(n-1)}\Big]}  ~,
\end{equation}

\noindent of his or her wealth to the $n$-month bond, and $(1-w_{t}^{(n)})$ to the 1-month bond. $\text{E}_t \Big[rx_{t+1}^{(n-1)}\Big]$ denotes the conditional expectation of the $n$-month bond return, for which the investor can use the out-of-sample forecasts generated either by one of the models that include $\text{TR}^{(eq)}$ or by its benchmark that does not use the equity tail factor as predictor. $\text{Var}_t\Big[rx_{t+1}^{(n-1)}\Big]$ denotes the conditional variance of the $n$-month bond return, which we estimate with the sample variance of the returns observed over the past 10 years. $\gamma$ represents the investor's level of risk aversion. Following \cite{Thornton2012} and \cite{Gargano2019}, we assume a risk aversion coefficient of $\gamma=5$ but we also consider a less risk-averse investor characterized by $\gamma=3$. Furthermore, as in the study of \cite{Huang2019}, we prevent extreme positions by restricting the weight $w_{t}^{(n)}$ on the risky bond to lie in the interval $[-1,5]$, which amounts to a maximum short-sale of 100\% and a maximum leverage of 400\%. 
The investor's portfolio return realized at time $t+1$ is given by
\begin{equation}
\label{eq:ret_p}
r_{\mathcal{P},t+1}^{(n)} = y_t^{(1)} + w_{t}^{(n)} rx_{t+1}^{(n-1)}  ~ .
\end{equation}

\noindent where $y_t^{(1)}$ is the yield of the zero-coupon bond with 1-month maturity. The certainty equivalent return (CER) of the portfolio, which is defined as the average utility realized by the investor from using the optimal weights $w_{t}^{(n)}$, is given by
\begin{equation}
\label{eq:cer_p}
\text{CER}_{\mathcal{P}}^{(n)} = \mu_{\mathcal{P}}^{(n)}  - \frac{\gamma}{2} \sigma_{\mathcal{P}}^{2~(n)}   ~ ,
\end{equation}

\noindent where $\mu_{\mathcal{P}} = T^{-1} \sum_{t=1}^{T} r_{\mathcal{P},t+1}^{(n)}$ and $\sigma_{\mathcal{P}}^{2~(n)} = T^{-1} \sum_{t=1}^{T} \Big(r_{\mathcal{P},t+1}^{(n)} - \mu_{\mathcal{P}}\Big)^2 $. In order to establish whether an investor that relies on the investment signals generated by $\text{TR}^{(eq)}$ is able to improve upon the economic utility realized by an investor whose portfolio allocations do not rely on equity tail risk, we compute the difference between the CER for the investor that uses one of the preferred models in (\ref{eq:RX_reg_univ}), (\ref{eq:RX_reg_PC3}) and (\ref{eq:RX_reg_PC5}) and the CER for the investor that uses the corresponding benchmark. This difference, which we denote by $\Delta^{(n)}$ and we express in terms of an annualized percentage CER gain, can be interpreted as the portfolio management fee that an investor is willing to pay for the bond return forecasts produced with equity tail risk.
Following \cite{Thornton2012}, \cite{Eriksen2017} and \cite{Huang2019}, we assess portfolio performance using also the  manipulation-proof performance (MPP) measure of \cite{Goetzmann2007}. For each of the preferred models, we compute the MPP improvement relative to its benchmark as
\begin{equation}
\label{eq:mpp_improv}
\Theta^{(n)} = \frac{1}{1-\gamma} \Biggl[ \text{ln} \Bigg( T^{-1} \displaystyle\sum_{t=1}^{T} \Bigg[ \frac{1+r_{\mathcal{P},t+1,1}^{(n)}}{1+y_{t+1}^{(1)}} \Bigg]^{1-\gamma} \Bigg) - \text{ln} \Bigg( T^{-1} \displaystyle\sum_{t=1}^{T} \Bigg[ \frac{1+r_{\mathcal{P},t+1,0}^{(n)}}{1+y_{t+1}^{(1)}} \Bigg]^{1-\gamma} \Bigg) \Biggr] 
~ ,
\end{equation}

\noindent where $r_{\mathcal{P},t+1,1}^{(n)}$ and $r_{\mathcal{P},t+1,0}^{(n)}$ are the realized portfolio returns associated with the preferred and benchmark models. As with the CER gain, we report annualized percentage values for $\Theta^{(n)}$.

\subsection{Term Structure Modeling} 
\label{subsec:ATSM}

We now introduce the term structure framework adopted in this paper and we present its estimation procedure. To set up the model, we rely on the approach suggested by \cite{Adrian2013}, which has the advantage that the pricing factors of bonds are not restricted to linear combinations of yields. Factors can indeed also be of different origin, such as the international equity tail risk measure $\text{TR}^{(eq)}$ defined in Section \ref{subsec:Data_equity}. After deriving the data generating process of log excess bond returns from a dynamic asset pricing model with an exponentially affine pricing kernel, \cite{Adrian2013} propose a new regression-based estimation technique for the model parameters. The linear regressions of this simple estimator avoid the computational burden of maximum likelihood methods, which have previously been the standard approach to the pricing of interest rates.      

The formulation and estimation of the Gaussian ATSM in \cite{Adrian2013} can be summarized as follows. A $K \times 1$ vector of pricing factors, $\mathbf{X}_t$, is assumed to evolve according to a VAR process of order one:
\begin{equation}
\label{eq:VAR}
\mathbf{X}_{t+1} = \boldsymbol\mu + \boldsymbol\phi\mathbf{X}_{t} + \mathbf{v}_{t+1} ~ ,
\end{equation}
where the shocks $\mathbf{v}_{t+1} \sim \pN(\mathbf{0}, \boldsymbol\Sigma)$ are conditionally Gaussian with zero mean and variance-covariance matrix $\boldsymbol\Sigma$. Letting $P_t^{(n)}$ denote the price of a zero-coupon bond with maturity $n$ at time $t$, the assumption of no-arbitrage implies the existence of a pricing kernel $M_{t+1}$ such that,
\begin{equation}
\label{eq:NoArb}
P_t^{(n)} = \mbox{E}_t\Big[M_{t+1}P_{t+1}^{(n-1)}\Big] ~.
\end{equation} 
The pricing kernel $M_{t+1}$ is assumed to have the following exponential form:
\begin{equation}
\label{eq:Kernel}
M_{t+1} = \mbox{exp}\Big(-r_t - \frac{1}{2}\boldsymbol\lambda_t^'\boldsymbol\lambda_t - \boldsymbol\lambda_t^' \boldsymbol\Sigma^{-1/2} \mathbf{v}_{t+1} \Big) ~,
\end{equation} 
where $r_t = -\ln P_t^{(1)}$ is the continuously compounded one-period risk-free rate and $\boldsymbol\lambda_t$ is the $K \times 1$ vector of market prices of risk, which are affine in the factors as in \cite{Duffee2002}:
\begin{equation}
\label{eq:RiskPrices}
\boldsymbol\lambda_t = \boldsymbol\Sigma^{-1/2}( \boldsymbol\lambda_0 + \boldsymbol\lambda_1 \mathbf{X}_t) ~.
\end{equation}  
The log excess one-period return of a bond maturing in $n$ periods is defined as follows,
\begin{equation}
\label{eq:RX}
rx_{t+1}^{(n-1)} = \ln P_{t+1}^{(n-1)} - \ln P_{t}^{(n)} - r_t ~ .
\end{equation}
After assuming the joint normality of $\{rx_{t+1}^{(n-1)},\mathbf{v}_{t+1}\}$, \cite{Adrian2013} derive the return generating process for log excess returns, which takes the form\footnote{For the full derivation of the data generating process see Section 2.1 in \cite{Adrian2013}.},
\begin{equation}
\label{eq:RGP}
rx_{t+1}^{(n-1)} = \boldsymbol\beta^{(n-1)'}(\boldsymbol\lambda_0 + \boldsymbol\lambda_1 \mathbf{X}_t) - \frac{1}{2} (\boldsymbol\beta^{(n-1)'} \boldsymbol\Sigma  \boldsymbol\beta^{(n-1)} + \sigma^2) + \boldsymbol\beta^{(n-1)'}\mathbf{v}_{t+1} + e_{t+1}^{(n-1)}   ~ ,
\end{equation}
where the return pricing errors $e_{t+1}^{(n-1)}$ $\sim$ \iid~$ (0, \sigma^2)$ are conditionally independently and identically distributed with zero mean and variance  $\sigma^2$. Letting $N$ be the number of bond maturities available and $T$ be the number of time periods at which bond returns are observed, \cite{Adrian2013} rewrite equation (\ref{eq:RGP}) in the stacked form,
\begin{equation}
\label{eq:StackRGP}
\mathbf{rx} = \boldsymbol\beta^{'}(\boldsymbol\lambda_0\boldsymbol\iota_{T}^{'} + \boldsymbol\lambda_1 \mathbf{X}_{\rule{1ex}{.4pt}}) - \frac{1}{2} (\mathbf{B}^{*}\mbox{vec}(\boldsymbol\Sigma) + \sigma^2 \boldsymbol\iota_{N}) \boldsymbol\iota_{T}^{'} + \boldsymbol\beta^{'}\mathbf{V} + \mathbf{E}   ~ ,
\end{equation}
where $\mathbf{rx}$ is an $N \times T$ matrix of excess bond returns, $\boldsymbol\beta = \Big[\boldsymbol\beta^{(1)} ~ \boldsymbol\beta^{(2)} ~...~ \boldsymbol\beta^{(N)} \Big]$ is a $K \times N$ matrix of factor loadings, $\boldsymbol\iota_{T}$ and $\boldsymbol\iota_{N}$ are a $T \times 1$ and $N \times 1$ vector of ones, $\mathbf{X}_{\rule{1ex}{.4pt}} = [\mathbf{X}_0 ~ \mathbf{X}_1 ~...~ \mathbf{X}_{T-1}]$ is a $K \times T$ matrix of lagged pricing factors, $\mathbf{B}^{*} = \Big[\mbox{vec}(\boldsymbol\beta^{(1)}\boldsymbol\beta^{(1)'}) ~...~ \mbox{vec}(\boldsymbol\beta^{(N)}\boldsymbol\beta^{(N)'}) \Big]^{'}$ is an $N \times K^2$ matrix, $\mathbf{V}$ is a $K \times T$ matrix and $\mathbf{E}$ is an $N \times T$ matrix.     

\noindent The main novelty of the approach taken by \cite{Adrian2013} to model the term structure of interest rates is the use of ordinary least squares to estimate the parameters of equation (\ref{eq:StackRGP}). In particular, the authors propose the following three-step procedure:
\begin{enumerate}
	\item Estimate the coefficients of the VAR model in equation (\ref{eq:VAR}) by ordinary least squares.\footnote{For estimation purposes, \cite{Adrian2013} advise to set $\boldsymbol\mu=0$ in case of zero-mean pricing factors.} Stack the estimates of the innovations $\hat{\mathbf{v}}_{t+1}$ into matrix $\hat{\mathbf{V}}$ and use this to construct an estimator of the variance-covariance matrix $\hat{\boldsymbol\Sigma} = \hat{\mathbf{V}}\hat{\mathbf{V}}^{'}/T$.    
	\item From the excess return regression equation $\mathbf{rx} = \mathbf{a}\boldsymbol\iota_{T}^{'} + \boldsymbol\beta^{'}\hat{\mathbf{V}} + \mathbf{c}\mathbf{X}_{\rule{1ex}{.4pt}} + \mathbf{E}$, obtain estimates of $\hat{\mathbf{a}}$, $\hat{\boldsymbol\beta}$ and $\hat{\mathbf{c}}$. Use $\hat{\boldsymbol\beta}$ to construct $\hat{\mathbf{B}}^{*}$. Stack the residuals of the regression into matrix $\hat{\mathbf{E}}$ and use this to construct an estimator of the variance $\hat{\sigma}^2 = \mbox{tr}(\hat{\mathbf{E}}\hat{\mathbf{E}}^{'})/NT$. 
	\item Noting from equation (\ref{eq:StackRGP}) that $\mathbf{a} = \boldsymbol\beta^{'}\boldsymbol\lambda_0 - \frac{1}{2} (\mathbf{B}^{*}\mbox{vec}(\boldsymbol\Sigma) + \sigma^2 \boldsymbol\iota_{N})$ and $\mathbf{c} = \boldsymbol\beta^{'}\boldsymbol\lambda_1$, estimate the price of risk parameters $\boldsymbol\lambda_0$ and $\boldsymbol\lambda_1$ via cross-sectional regressions, 
	\begin{align}
	\hat{\boldsymbol\lambda}_0 &= (\hat{\boldsymbol\beta}\hat{\boldsymbol\beta}^{'})^{-1} \hat{\boldsymbol\beta} \Big(\hat{\mathbf{a}} + \frac{1}{2} (\hat{\mathbf{B}}^{*}\mbox{vec}(\hat{\boldsymbol\Sigma}) + \hat{\sigma}^2 \boldsymbol\iota_{N})\Big) ~ , \\
	\hat{\boldsymbol\lambda}_1 &= (\hat{\boldsymbol\beta}\hat{\boldsymbol\beta}^{'})^{-1} \hat{\boldsymbol\beta} \hat{\mathbf{c}} ~ .
	\end{align}
\end{enumerate}
The analytical expressions of the asymptotic variance and covariance of $\hat{\boldsymbol\beta}$ and $\hat{\boldsymbol\Lambda} = [\hat{\boldsymbol\lambda}_0 ~ \hat{\boldsymbol\lambda}_1]$, which we do not report here to save space, are provided in Appendix A.1 of \cite{Adrian2013}. From the estimated model parameters, \cite{Adrian2013} show how to generate a yield curve. Indeed, within the proposed framework, bond prices are exponentially affine in the pricing factors. Consequently, the yield of a zero-coupon bond with maturity $n$ at time $t$, $y_{t}^{(n)}$, can be expressed as follows,
\begin{equation} 
\label{eq:AffineY}
y_{t}^{(n)} = - \frac{1}{n} [a_n + \mathbf{b}_n^{'}\mathbf{X}_t] + u_{t}^{(n)}~ ,
\end{equation}     
where the coefficients $a_n$ and $\mathbf{b}_n$ are obtained from the following no-arbitrage recursions,
\begin{align}
\label{eq:a_n}
a_{n} &= a_{n-1} + \mathbf{b}_{n-1}^{'}(\boldsymbol\mu - \boldsymbol\lambda_0) + \frac{1}{2} (\mathbf{b}_{n-1}^{'} \boldsymbol\Sigma \mathbf{b}_{n-1} + \sigma^2) - \delta_0   ~ , \\
\label{eq:b_n}
\mathbf{b}_{n}^{'} &= \mathbf{b}_{n-1}^{'}(\boldsymbol\phi - \boldsymbol\lambda_1) - \boldsymbol\delta_1^{'}   ~ ,
\end{align} 
subject to the initial conditions $a_{0}=0$, $\mathbf{b}_{n}=\mathbf{0}$, $a_{1}= - \delta_0$ and $\mathbf{b}_{1} = - \boldsymbol\delta_1$. The parameters $\delta_0$ and $\boldsymbol\delta_1$ are estimated by regressing the short rate, $r_t = -\ln P_t^{(1)}$, on a constant and contemporaneous pricing factors according to,
\begin{equation}
\label{eq:ShortRate}
r_t = \delta_0 + \boldsymbol\delta_1 \mathbf{X}_t + \epsilon_t ~,~~ \epsilon_t \sim \iid~ (0, \sigma^2_\epsilon) ~.
\end{equation}   
By setting the price of risk parameters $\boldsymbol\lambda_0$ and $\boldsymbol\lambda_1$ to zero in equation (\ref{eq:a_n}) and (\ref{eq:b_n}), \cite{Adrian2013} obtain $a_n^{\mbox{\tiny{RN}}}$ and $\mathbf{b}_n^{\mbox{\tiny{RN}}}$, which they use to generate the risk-neutral yields, $y_{t}^{(n)~\mbox{\tiny{RN}}}$. These yields reflect the average expected short rate over the current and the subsequent $(n-1)$ periods and are computed as follows,
\begin{equation}
\label{eq:RN_Y}
y_{t}^{(n)~\mbox{\tiny{RN}}} = \frac{1}{n} \sum_{i=0}^{n-1} \mbox{E}_t [r_{t+i}] = - \frac{1}{n} [a_n^{\mbox{\tiny{RN}}} + \mathbf{b}_n^{\mbox{\tiny{RN}}'}\mathbf{X}_t] ~ .
\end{equation}   
Given equation (\ref{eq:AffineY}) and (\ref{eq:RN_Y}), the term premium $TP_t^{(n)}$, which is the additional compensation required for investing in long-term bonds relative to rolling over a series of short-term bonds, can be calculated as follows,
\begin{equation}
\label{eq:TP}
TP_t^{(n)} = y_{t}^{(n)} - y_{t}^{(n)~\mbox{\tiny{RN}}} ~ .
\end{equation} 

%

In the next sections we specify and estimate a term structure model for US interest rates following the procedure outlined above. The difference between the Gaussian ATSM in \cite{Adrian2013} and ours is that we use a different set of pricing factors. Indeed, we include in $\mathbf{X}_t$ not only the PCs of bond yields but also the equity tail factor $\text{TR}^{(eq)}$ described in Section \ref{subsec:Data_equity}.  

\subsection{Consistent Risk Premium Estimation} 
\label{subsec:GX}

In this section we briefly review the method of \cite{Giglio2019}, GX hereafter, to estimate the risk premium of an observable factor ($\text{TR}^{(eq)}$ in our case), which is valid even when the observed factor is measured with noise and the model does not fully account for all priced sources of risk in the economy. 
The new GX three-pass methodology combines principal component analysis (PCA) with two-pass regressions \citep{Fama1973} to consistently estimate the risk premium of any observed factor. 
The estimator relies on a large cross section of test assets and is valid as long as PCA can recover the entire factor space of test asset returns. In our paper we apply the GX three-pass method to the whole term structure of Treasury bond returns to estimate and test the significance of the risk premium of the equity tail factor $\text{TR}^{(eq)}$.

Unlike the term structure model described above where the pricing kernel is an exponential function of the state variables, \cite{Giglio2019} assume a linear stochastic discount factor. Working with a linear asset pricing model they can exploit the so-called ``rotation invariance'' property that allows them to estimate the risk premium $\gamma_g$ of an observable factor $g_t$ without necessarily observing or knowing all the true factors $v_t$ entering the pricing kernel. Written in matrix form, the GX model consists of the following two equations:
\begin{align}
\label{eq:GX_ret}
\bar{R} &= \beta \bar{V} + \bar{U}   ~ , \\
\label{eq:GX_g}
\bar{G} &= \eta \bar{V} + \bar{Z}  ~ ,
\end{align}
where $\bar{R}$ is the $n \times T$ matrix of demeaned excess returns of the test assets, $\bar{V}$ is the $p \times T$ matrix of demeaned true factors, $\beta$ is the $n \times p$ matrix of factor risk exposures, $\bar{U}$ is the $n \times T$ matrix of idiosyncratic errors, $\bar{G}$ is the $d \times T$ matrix of demeaned observed factors, the risk premium of which has to be estimated, $\eta$ is the $d \times p$ matrix of the loadings of the observed factors on the unobserved true factors, and $\bar{Z}$ is the $d \times T$ matrix of measurement errors. The GX estimator proceeds in three steps which can be summarized as follows: 
\begin{enumerate}
	\item \textit{PCA step}. The first pass consists of estimating the true factors and factor risk exposures by extracting the first $p$ principal components and their respective loadings from the cross section of test asset returns.
	\footnote{\cite{Giglio2019} propose a consistent estimator of $p$ in their Online Appendix I.1. They also demonstrate that as long as the number of principal components used is greater than or equal to the true number of factors, the estimator of the risk premium is consistent. In our empirical analysis we report results with respect not only to the number of principal components selected with the \cite{Giglio2019} criterion but also to higher numbers of factors to ensure robustness of the estimates.} The estimators can therefore be written as:
	\begin{equation}
	\label{eq:GX_pca}
	\widehat{V} = T^{1/2} (\xi_1 : \xi_2 : ... : \xi_p)^{\intercal} ~~~~ \text{and} ~~~~ \widehat{\beta} = T^{-1}\bar{R}\widehat{V}^{\intercal} ~,
	\end{equation}   
	where $\xi_1,...,\xi_p$ are the eigenvectors corresponding to the largest $p$ eigenvalues of  $n^{-1}T^{-1}\bar{R}^{\intercal}\bar{R}$.
	\item \textit{Cross-sectional regression step}. The second pass consists of estimating the risk premia of the latent factors by running a cross-sectional ordinary least square regression of average realized excess returns, $\bar{r}$, onto the previously estimated factor loadings, $\widehat{\beta}$:
	\begin{equation}
	\label{eq:GX_cs}
	\widehat{\gamma} = (\widehat{\beta}^{\intercal}\widehat{\beta})^{-1} \widehat{\beta}^{\intercal}\bar{r} ~.
	\end{equation}   
	\item \textit{Time-series regression step}. The third pass consists of estimating the risk premia of the factors of interest by first running a time series regression of the demeaned candidate factors onto the space of the latent factors and then combining these estimates with those of the second step. The estimator $\widehat{\eta}$ of the loadings on the latent factors and the estimator $\widehat{\gamma}_g$ of the risk premia of the observed factors of interest can therefore be written as: 
	\begin{align}
	\label{eq:GX_ts}
	&\widehat{\eta} = \bar{G}\widehat{V}^{\intercal}(\widehat{V}\widehat{V}^{\intercal})^{-1}   ~ , \\
	\label{eq:GX_gamma}
	&\widehat{\gamma}_g = \widehat{\eta} \widehat{\gamma}  ~ .
	\end{align}
\end{enumerate}

\noindent Due to space considerations, we do not provide analytical expressions for the asymptotic variance of the risk-premium estimates and we refer the reader to Section 4 in \cite{Giglio2019}. 

Another important aspect considered in the GX procedure is the noise that is contained in the observable factors and that is uncorrelated with the test asset returns. The higher the noise, the more weakly the factor is reflected in the cross section of test assets. To understand whether the factor of interest has low exposure to the fundamental factors ($\eta$ is small) or whether it is dominated by noise ($z_t$ is large), \cite{Giglio2019} define the $R^2$ of the time-series regressions in the third-pass, $R^2_g=\frac{\widehat{\eta}\widehat{V}\widehat{V}^{\intercal}\widehat{\eta}^{\intercal}}{\bar{G}\bar{G}^{\intercal}}$. Furthermore, they provide a Wald test for the null that the observed factor $g$ is weak by formulating the hypotheses $H_0: \eta = 0$ vs $H_1: \eta \neq 0$. In our empirical analysis we report the $R^2_g$ and Wald $p$-value for the strength of the observed factor $g=\text{TR}^{(eq)}$ with respect to the cross section of Treasury returns, alongside the estimate and significance of the factor's risk premium.

\section{Empirical Results}
\label{sec:Empirical_Res}

In this section we present our empirical results. We first consider in Section \ref{subsec:RX_vs_TR} the full-sample least-squares estimates for the bond return prediction models with equity tail risk. We empirically show that the equity tail factor $\text{TR}^{(eq)}$ significantly predicts monthly bond returns in- and out-of-sample and the more accurate forecasts can be of economic importance for an investor facing portfolio decisions.  In Section \ref{subsec:ATSM_with_TR} we discuss the estimates of the Gaussian ATSM which allow to explore in detail the effects of equity tail risk on bond prices and determine whether $\text{TR}^{(eq)}$ is a priced source of risk in the term structure of US interest rates. 
Section \ref{subsec:GX_with_TR} corroborates the existence of a significant market price of equity tail risk in the US government bond market using the GX three pass method. Finally, Section \ref{subsec:International} investigates to what extent equity tail risk affects the government bond market of countries other than the United States.

\subsection{Bond Return Predictability} 
\label{subsec:RX_vs_TR}

We start by examining the interactions between the one-month returns of US Treasury bonds and the S\&P 500 option-implied volatility that stems from large negative price jumps, $\text{TR}^{(eq)}$. Using the full sample (1996:01-2018:12) of monthly data, we run the predictive regressions in (\ref{eq:RX_reg_univ}), (\ref{eq:RX_reg_PC3}) and (\ref{eq:RX_reg_PC5}), for which we report in, respectively, Panels A, B, and C of Table \ref{tab:IS_results} the least-squares estimates of the slope coefficients and their corresponding $p$-values. Numbers at the bottom of each panel correspond to the adjusted $R$-squared of the predictive regressions that include and exclude $\text{TR}^{(eq)}$ as predictor, and to the $p$-value of an $F$-test of the null hypothesis that the regression that includes $\text{TR}^{(eq)}$ does not give a significantly better fit to the data than does a regression without it. In order to ease interpretation of the results, all predictors, including those discussed later, have been normalized to have a zero mean and a standard deviation of one. Here and in the rest of this section, evidence is presented for returns on the one-, two-, three-, four-, five-, seven- and ten-year Treasury bonds ($n=12,24,36,48,60,84,120$ months, respectively). The results of the analysis for other maturities are available upon request.

\vspace{1.0em}

\begin{center}
	[ Insert Table \ref{tab:IS_results} here ]
\end{center}

\vspace{1.0em}

\noindent Consider first the results of the univariate model (\ref{eq:RX_reg_univ}) presented in Panel A. The one-month-ahead returns of US Treasury bonds exhibit strong interactions with the perceived tail risk in the US stock market. The coefficient of the S\&P 500 option-implied tail risk measure $\text{TR}^{(eq)}$ is statistically significant at well below the 0.05 level across the whole yield curve. Looking at the size of the coefficient, we observe that the impact of equity tail risk on bond risk premia is monotonically increasing with the bond maturity. Our estimates suggest that a one standard deviation increase in the equity tail factor raises the expected annualized return on the 1-year and 10-year Treasury bonds by about 0.5\% and 6.2\%, respectively. 
Furthermore, we note that for all maturities considered, the sign of the coefficient is positive. 
This result is in sharp contrast with that obtained by \cite{Crump2019} with a conceptually very different measure of equity tail risk. It can however be explained in light of the opposite movements in equity and bond prices observed in times of stress and the considerations raised by previous studies that found a negative relation between future stock returns and measures of option-implied volatility, see, among others, \cite{Xing2010} and \cite{An2014}. That is, if we believe that informed traders with negative news choose the option market to trade first, then an increase in tail risk is later accompanied by lower and higher prices on, respectively, the equity and bond markets, which are slow in incorporating the information embedded in the option volatility surface. 


\noindent Since the literature on bond return predictability is more often interested in the forecasting power of a variable beyond that of the information contained in the yield curve, we now discuss the results reported in Panels B and C of Table \ref{tab:IS_results}. When controlling for yield curve factors with the first 3 and 5 PCs, the coefficient associated with $\text{TR}^{(eq)}$ remains positive and highly significant for all bond maturities.\footnote{In results available upon request, we also considered specifications of the regression equations (\ref{eq:RX_reg_PC3}) and (\ref{eq:RX_reg_PC5}) that make use of the orthogonal component of $\text{TR}^{(eq)}$ with respect to the principal components. The coefficient in front of the equity tail factor continues to be statistically significant at the 0.05 level or lower for all maturities.} We find strong significance not only with the standard Newey-West $p$-values but also with the more robust $p$-values computed with the bootstrap procedure of \cite{Bauer2018}. Furthermore, we note that the inclusion of equity tail risk in the predictive regressions determines sizeable changes in the adjusted $R^2$s, which nearly double in Panel B and increase by about 50\% in Panel C. Finally, the $F$-test results confirm the importance of $\text{TR}^{(eq)}$ for explaining the one-month-ahead variation in bond risk premia.

\noindent In addition to our baseline regressions in (\ref{eq:RX_reg_univ}), (\ref{eq:RX_reg_PC3}) and (\ref{eq:RX_reg_PC5}), we examine whether equity tail risk remains a strong predictor of future bond returns even when controlling for other successful return forecasting factors found in the literature. Specifically, we report in Panels D and E of Table \ref{tab:IS_results} the results of regressions that use the equity tail factor in combination with, respectively, the \cite{Cochrane2005} and \cite{CieslakPovala2015} factors. The \cite{Cochrane2005} bond return predictor is obtained as a linear combination of forward rates, while the \cite{CieslakPovala2015} risk-premium factor is obtained from a decomposition of Treasury yields into inflation expectations and maturity-specific interest-rate cycles.
Due to the low correlation that exists between the covariates, Treasury risk premia continue to exhibit significant interactions with both the successful predictors found in previous studies and the equity tail factor of this paper. 
Finally, we report in Panel F of Table \ref{tab:IS_results} the estimates of a regression that includes $\text{TR}^{(eq)}$ and the CBOE VIX unspanned by $\text{TR}^{(eq)}$ as predictors. The immediate point that stands out here is that the VIX components that are not related to our equity tail factor, i.e.~continuous return variation and right jump variation, are highly insignificant for almost all bond maturities. Based on this result, we can conclude that the VIX does not have predictive power over-and-above $\text{TR}^{(eq)}$ for future bond returns.

We now discuss the out-of-sample performance of the models in (\ref{eq:RX_reg_univ}), (\ref{eq:RX_reg_PC3}) and (\ref{eq:RX_reg_PC5}), which predict bond returns with the S\&P 500 option-implied tail risk measure $\text{TR}^{(eq)}$. The accuracy of the bond return forecasts of model (\ref{eq:RX_reg_univ}) is measured relative to the recursively updated forecasts from the EH model that projects returns on a constant, while the accuracy of the forecasts of models (\ref{eq:RX_reg_PC3}) and (\ref{eq:RX_reg_PC5}) is measured relative to the forecasts of the models that only include the principal components as predictors. Table \ref{tab:OOS_results} reports the \cite{CT2008} out-of-sample $R^2_{OS}$ values for each model, alongside the $p$-value of the \cite{CW2007} MSPE-adjusted statistic for testing $H_0: R^2_{OS} \le 0$ against $H_1: R^2_{OS} > 0$. We report results for both increasing and rolling windows of past data used in the estimation method. The out-of-sample period is 2007:07--2018:12.

\vspace{1.0em}

\begin{center}
	[ Insert Table \ref{tab:OOS_results} here ]
\end{center}

\vspace{1.0em}

\noindent Overall, the results in Table \ref{tab:OOS_results} suggest that the good in-sample fit provided by $\text{TR}^{(eq)}$ and discussed above translates into positive out-of-sample performance. For instance, when the benchmark is the EH model, we find that equity tail risk improves the out-of-sample bond return predictions across all maturities. The gains are in the range of 1.6\% to 4.3\% for both window estimations, with the largest improvements observed for medium-maturity bonds. We note that with the robust inference method developed by \cite{Bauer2018} the increases in the $R^2_{OS}$s are significant in a statistical sense for bond maturities greater than 2 years, while the $p$-values of the \cite{CW2007} MSPE-adjusted statistic are lower than 10\% for maturities of 5 years or longer. Similarly, we observe positive values of $R^2_{OS}$ in Panels B and C indicating higher predictive accuracy for the bond return prediction models that include $\text{TR}^{(eq)}$ compared to their PCs-only benchmark specifications. Except for the 10-year bond, the bootstrap $p$-values of both $R^2_{OS}$ and \cite{CW2007} MSPE-adjusted statistic are below 0.1, thus proving the statistical significance of the results.

Next, we examine the economic value of using equity tail risk to make one-month-ahead predictions of Treasury  bond returns. Table \ref{tab:Econ_Gains} reports values for the CER gain ($\Delta$) and \cite{Goetzmann2007} MPP improvement ($\Theta$) that an investor can achieve by switching from a benchmark to a model that uses the equity tail factor $\text{TR}^{(eq)}$ to predict bond returns. Results are based on the out-of-sample model forecasts produced for the period 2007:07--2018:12 with predictive models that are recursively estimated with a rolling window approach. 

\vspace{1.0em}

\begin{center}
	[ Insert Table \ref{tab:Econ_Gains} here ]
\end{center}

\vspace{1.0em}

\noindent From an investment perspective, the results in Table \ref{tab:Econ_Gains} indicate that predicting bond returns with equity tail risk can generate substantial risk-adjusted returns. 
This is particularly the case for an investor that can use $\text{TR}^{(eq)}$ alongside the first 5 PCs of bond yields to predict the one-month-ahead returns of Treasuries with maturities in the range of two to seven years. Specifically, we find that the investor is willing to pay from 80 up to 360 basis points per year to switch from the 5 PCs-only benchmark to the model that forecasts bond returns also with equity tail risk. Even when the benchmark is the EH model, we find that an investor trading some specific medium-term bonds is better off following the return forecasts based on equity tail risk. On the other hand, when the benchmark is the 3 PCs-only model, the investor cannot achieve any asset allocation gains by switching to the predictive model with equity tail risk.

Finally, we briefly discuss how the forecast performance of the models in (\ref{eq:RX_reg_univ}), (\ref{eq:RX_reg_PC3}) and (\ref{eq:RX_reg_PC5}) is related to the real economy. Panels A and B of Table \ref{tab:ForecastPerf_and_MacroCond} report contemporaneous correlations between the out-of-sample forecasts of one-month-ahead Treasury bond returns and the CFNAI and the macroeconomic uncertainty index ($\mathbb{U}^{\text{MACRO}}$) constructed by \cite{Jurado2015}. We note that the bond risk premia implied by any of the three models are countercyclical as they are negatively correlated with macroeconomic condition. This is a common result found in the literature on bond return predictability and is consistent with economic theories in which investors require compensation for bearing business cycle risk, see, e.g., \cite{Eriksen2017} and references therein. In order to understand whether the models that include $\text{TR}^{(eq)}$ as predictor perform well in recessions or expansion periods, Panels C and D of Table \ref{tab:ForecastPerf_and_MacroCond} report contemporaneous correlations
between the models' relative forecast and portfolio performance and the CFNAI. The relative forecast performance is defined as the difference in cumulative squared prediction error (DCSPE), while the relative portfolio performance is defined as the difference in cumulative realized utilities (DCRU). As we can see, the forecasting performance of the three models tends to be positively correlated with the CFNAI, indicating superior model performance in good times when the CFNAI is high. Looking at the relative portfolio performance gives less clear-cut results since the correlations vary substantially across maturities. In fact, asset allocation gains seem to be achievable during expansion periods for short-term bonds and during recessions for long-term bonds.

\vspace{1.0em}

\begin{center}
	[ Insert Table \ref{tab:ForecastPerf_and_MacroCond} here ]
\end{center}

\vspace{1.0em}

\subsection{Bond Pricing in ATSM} 
\label{subsec:ATSM_with_TR}   

\begin{figure}[!ht]
	\caption{Time series of US Treasury bond yields}  
	\label{fig:Yields_vs_HighTR}
	{\footnotesize \centering
		\vspace{0.2cm}
		\includegraphics[width=16.0cm, height=10cm]{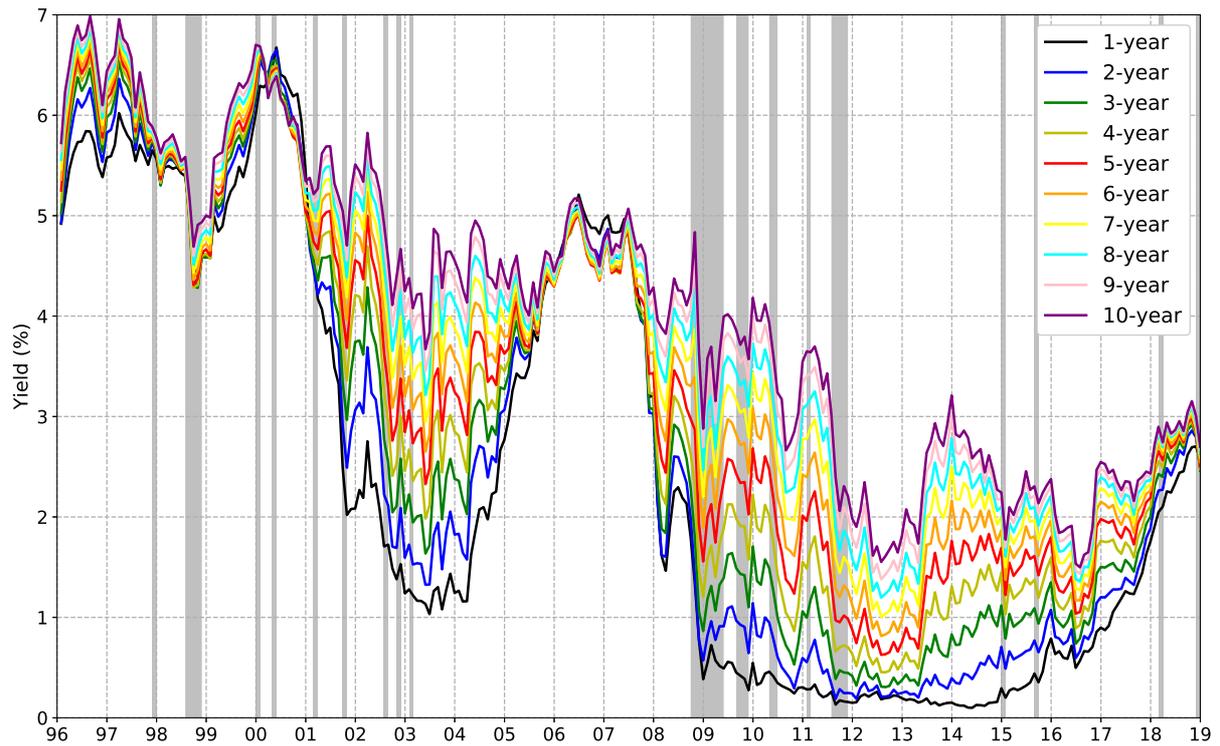} 
	}
	\begin{minipage}{15.8cm}
		\vspace{0.3cm}
		The figure displays the end-of-month values of 1- to 10-year Treasury bond yields from January 1996 to December 2018. Vertical gray bars indicate periods of elevated ($>= 85\%$-ile) equity tail risk implied by S\&P 500 index options.
	\end{minipage}	
\end{figure}

On the basis of the significant interactions observed between future Treasury returns and the equity tail factor $\text{TR}^{(eq)}$, it is of interest to examine to what extent the left tail volatility of the stock market also affects the current level of bond prices. Figure \ref{fig:Yields_vs_HighTR} shows the time trend of Treasury bond yields against periods of elevated equity tail risk, corresponding to when $\text{TR}^{(eq)}$ is above its historical 85-th percentile. 
As it can be seen in the graph, many of the most remarkable declines in Treasury rates occurred at times of elevated equity tail risk. In fact, the average contemporaneous correlation between bond yields and  $\text{TR}^{(eq)}$ is about -0.15.

\vspace{1.0em}

\begin{center}
	\noindent [ Insert Figure \ref{fig:Yields_vs_HighTR} here ]
\end{center}

\vspace{1.0em}

\begin{figure}[!ht]
	\caption{Time series of the pricing factors of US Treasuries}  
	\label{fig:X_t}
	{\footnotesize \centering
		\vspace{0.2cm}
		\includegraphics[width=16.0cm, height=19cm]{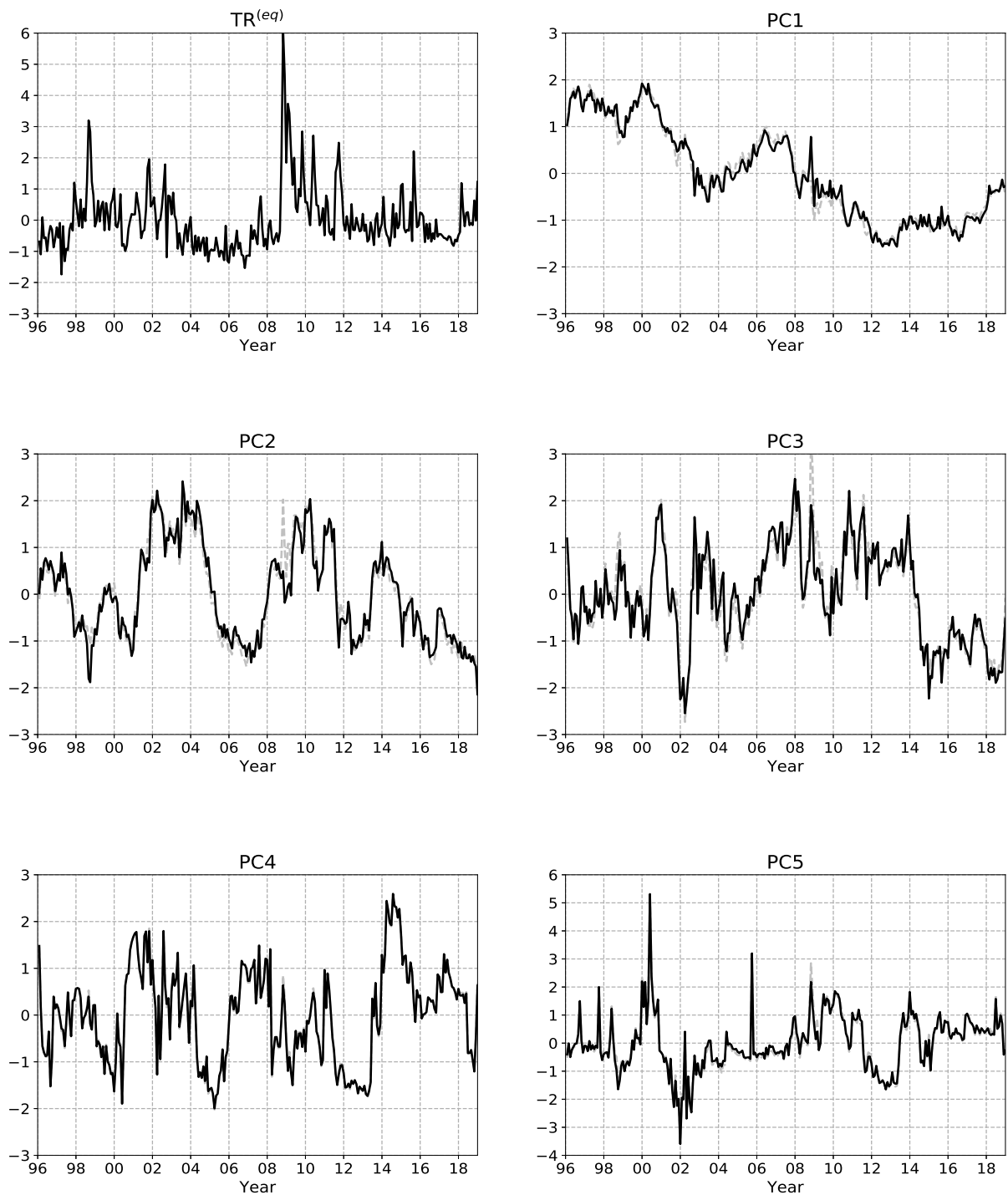} 
	}
\begin{minipage}{15.8cm}
	\vspace{0.3cm}
	The figure displays the monthly time series of the pricing factors of the proposed Gaussian ATSM with equity tail risk. The top-left panel shows the S\&P 500 option-implied equity tail risk factor $\text{TR}^{(eq)}$. The remaining panels show the first five principal components extracted from the US Treasury yields orthogonal to the $\text{TR}^{(eq)}$ factor. The light-colored dashed lines show the principal components extracted from non-orthogonalized yields, which however are not used as pricing factors in our model. All factors have been normalized to have mean zero and unit variance. 
\end{minipage}	
\end{figure}

\noindent To investigate the role of equity jump tail risk in pricing US government bonds, we now estimate the Gaussian ATSM of Section \ref{subsec:ATSM} with the inclusion of our equity tail factor in the vector of state variables.
In addition to $\text{TR}^{(eq)}$, however, we also need pricing factors that summarize the information contained in the yield curve. To this end, we extract the first five principal components of the US yield curve, which have proven to be remarkably effective in fitting the cross-section of bond yields and returns in \cite{Adrian2013}. Based on this evidence, we let these PCs drive the interest rates of our model as well, but with a slight modification of the methodology. Indeed, in order to have pricing factors that are uncorrelated with each other, we follow \cite{Cochrane2008} and extract the principal components not from the conventional yields, but instead from the yields orthogonalized to the extra factor, which in our study is $\text{TR}^{(eq)}$. By doing so, we obtain yield curve factors that are unrelated to the pricing of tail risk in the stock market, which is entirely ascribed to the $\text{TR}^{(eq)}$ factor. In view of these considerations, we employ the following set of pricing factors in our Gaussian ATSM,
\begin{equation}
\label{eq:X_t}
\mathbf{X}_t = \Big[\text{TR}^{(eq)}_t,~ \text{PC1}_t ,~ \text{PC2}_t,~ \text{PC3}_t,~ \text{PC4}_t,~ \text{PC5}_t \Big]' ~,
\end{equation}
where $\text{TR}^{(eq)}$ is the S\&P 500 option-implied measure of left tail volatility, and $\text{PC1}$--$\text{PC5}$ are the first five principal components estimated from an eigenvalue decomposition of the covariance matrix of zero-coupon bond yields of maturities $n=3,6,...,120$ months, orthogonal to $\text{TR}^{(eq)}$. All factors have mean zero and unit variance, and they are plotted in Figure \ref{fig:X_t}. The panels of $\text{PC1}$--$\text{PC5}$ also present the principal components of the conventional non-orthogonalized bond yields. We find that estimates of the factors extracted using the two yield curves track each other quite closely, with the largest differences occurring for PC2 and PC3 at the onset of the financial crisis.\footnote{In results available upon request, we have found significant relationships only between $\text{TR}^{(eq)}$ and $\text{PC2}$ and $\text{PC3}$ of the conventional non-orthogonalized bond yields. Both correlation coefficients were around $-0.24$.} Therefore, the orthogonalization of the rates with respect to $\text{TR}^{(eq)}$ does not appear to significantly alter the interpretation and role of the principal components in describing the characteristics of the US Treasury yield curve.  
\vspace{1.0em}

\begin{center}
	\noindent [ Insert Figure \ref{fig:X_t} here ]
\end{center}

\vspace{1.0em}

\begin{figure}[!ht]
	\caption{Observed and model-implied US Treasury bond yields and returns}  
	\label{fig:YRX_t}
	{\footnotesize \centering
		\vspace{0.2cm}
		\includegraphics[width=16.0cm, height=19cm]{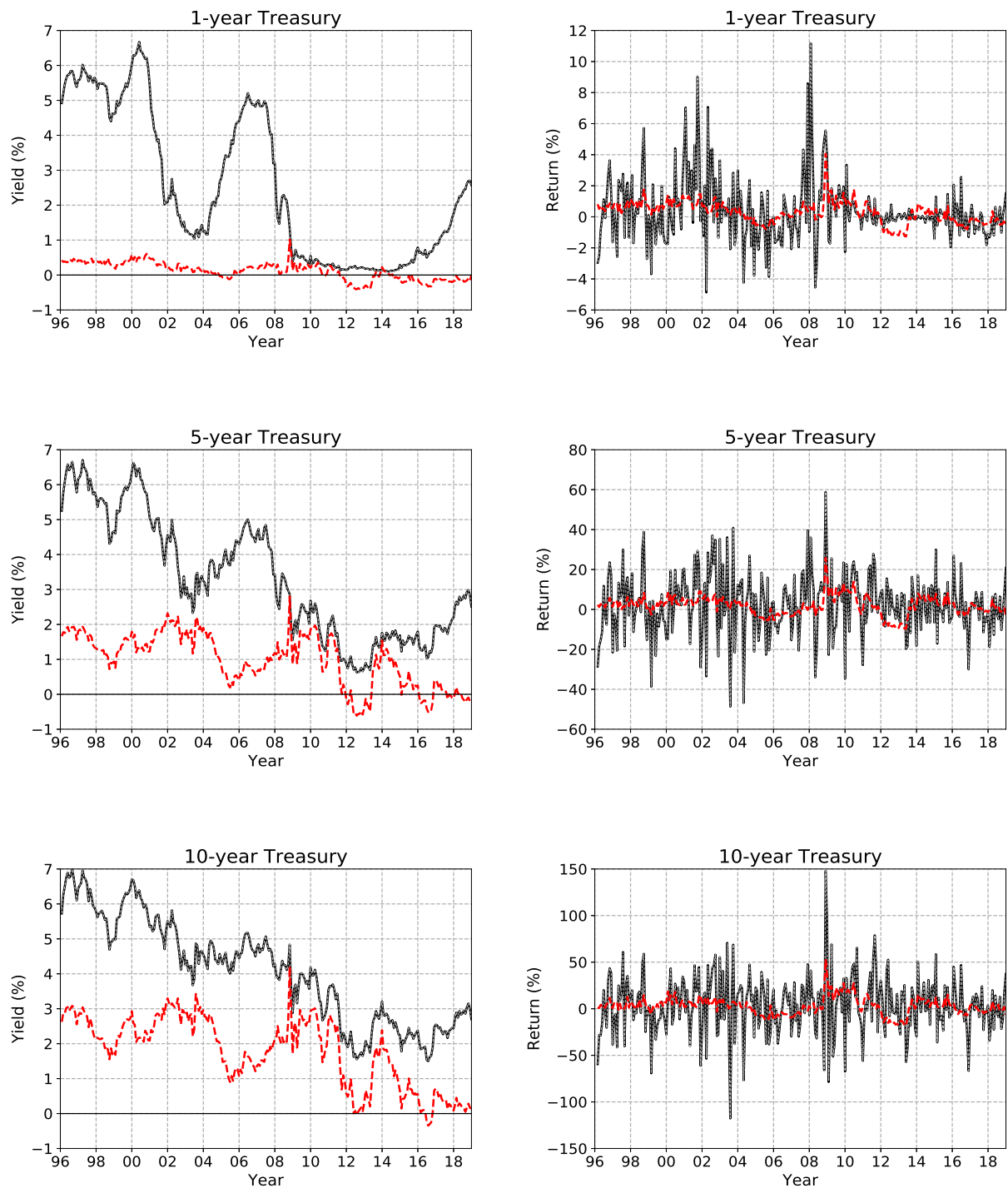} 
	}
\begin{minipage}{15.8cm}
	\vspace{0.3cm}
	The figure displays the observed and model-implied time series of yields and one-month excess returns on US Treasury bonds with 1-, 5- and 10-year maturities. In the left panels, the solid black lines show the observed yields, the dashed gray lines plot the model-implied yields, while the dashed red lines indicate the model-implied term premia. In the right panels, the solid black lines show the observed excess returns, the dashed gray lines plot the model-implied excess returns, while the dashed red lines indicate the model-implied expected excess returns.
\end{minipage}
\end{figure}

Given the vector of state variables in (\ref{eq:X_t}), we estimate our Gaussian ATSM using the method put forward by \cite{Adrian2013} and discussed in Section \ref{subsec:ATSM}. In particular, we use one-month excess returns for Treasury bonds with maturities $n=6,12,...,120$ months to fit the cross-section of yields. 
The summary statistics of the pricing errors implied by our term structure model, which accounts for equity tail risk, and a benchmark model based on only the first five PCs of the yield curve  are provided in Table \ref{tab:ATSM_Fit}.  
Overall the results indicate a good fit between the data and the proposed model with equity tail risk. Indeed, both the mean and the standard deviation of our yield pricing errors remain well below a basis point for all maturities and they never exceed, in absolute value, those of the benchmark. As for the return pricing errors, we notice that explicitly  including the equity tail risk factor $\text{TR}^{(eq)}$ in a Gaussian ATSM can improve the fit especially to the short end of the US yield curve. Moreover, consistent with the way \cite{Adrian2013} construct their framework for the term structure of interest rates, we observe a strong autocorrelation in the yield pricing errors and a negligible one in the return pricing errors, except for the 3-year bond. The success of our model in fitting the yield curve is shown graphically in the left panels of Figure \ref{fig:YRX_t}. In these plots, the solid black lines of observed yields are visually indistinguishable from the dashed gray lines of model-implied yields. Similarly, the right panels of Figure \ref{fig:YRX_t} display the tight fit between actual and fitted excess Treasury returns. The dashed red lines plot the model-implied dynamics of bond term premia in the left panels and of the expected component of excess returns in the right panels.

\vspace{1.0em}

\begin{center}
	[ Insert Table \ref{tab:ATSM_Fit} here ]
\end{center}

\begin{center}
	\noindent [ Insert Figure \ref{fig:YRX_t} here ]
\end{center}

\vspace{1.0em}

The estimation approach proposed by \cite{Adrian2013} allows for direct testing of the presence of unspanned factors, i.e.~factors that do not help explain variation in Treasury returns. The specification test is implemented as a Wald test of the null hypothesis that bond return exposures to a given factor are jointly equal to zero.
Letting $\boldsymbol\beta_i$ be the $i$-th column of $\boldsymbol\beta^{'}$, the Wald statistic, under the null $H_0: \boldsymbol\beta_i = \mathbf{0}_{N\times1}$, is defined as follows,
\begin{equation}
\label{eq:Wald_b}
W_{\beta_i} = \hat{\boldsymbol\beta}_i^{'} \hat{\mathcal{V}}^{-1}_{\beta_i} \hat{\boldsymbol\beta}_i  \stackrel{\alpha}{\sim}  \chi^2(N) ~ ,
\end{equation}
where $\hat{\mathcal{V}}_{\beta_i} $ is an  $N \times N$ diagonal matrix that contains the estimated variances of the $\hat{\boldsymbol\beta}_i$ coefficient estimates.\footnote{See Appendix A.1 in \cite{Adrian2013} for the analytical expressions of the asymptotic variance of the estimators.} The results of the Wald test on the pricing factors of both the proposed ATSM with equity tail risk and the benchmark PC-only specification are shown in Table \ref{tab:Wald_b}. As we can see, we strongly reject the hypothesis of unspanned factor for each of our state variables. This means that the data support the use of the equity tail factor $\text{TR}^{(eq)}$, together with the yield curve factors indicated by \cite{Adrian2013}, for pricing government bonds in the US market over the period 1996 -- 2018.

\vspace{1.0em}

\begin{center}
	[ Insert Table \ref{tab:Wald_b} here ]
\end{center}

\vspace{1.0em}

We now examine whether the risk factors that we use in our Gaussian ATSM are priced in the cross-section of Treasury returns. To this end, we follow \cite{Adrian2013} and perform a Wald test of the null hypothesis that the market price of risk parameters associated with a given model factor are jointly equal to zero. Letting $\boldsymbol\lambda_i^{'}$ be the $i$-th row of $\boldsymbol\Lambda = [\boldsymbol\lambda_0 ~ \boldsymbol\lambda_1]$, the Wald statistic, under the null $H_0: \boldsymbol\lambda_i^{'} = \mathbf{0}_{1\times(K+1)}$, is defined as follows,
\begin{equation}
\label{eq:Wald_Lp}
W_{\Lambda_i} = \hat{\boldsymbol\lambda}_i^{'} \hat{\mathcal{V}}^{-1}_{\lambda_i} \hat{\boldsymbol\lambda}_i  \stackrel{\alpha}{\sim}  \chi^2(K+1) ~ ,
\end{equation}
where $\hat{\mathcal{V}}_{\lambda_i}$ is a square matrix of order $(K+1)$ that contains the estimated variances of the $\hat{\boldsymbol\lambda}_i$ coefficient estimates.\footnote{See Appendix A.1 in \cite{Adrian2013} for the analytical expressions of the asymptotic variance of the estimators.} In addition, in order to test whether the market prices of risk are time-varying, \cite{Adrian2013} propose the following Wald test which focuses on $\boldsymbol\lambda_1$ and excludes the contribution of $\boldsymbol\lambda_0$. Letting $\boldsymbol\lambda_{1_i}^{'}$ be the $i$-th row of $\boldsymbol\lambda_1$, the Wald statistic of this second test, under the null $H_0: \boldsymbol\lambda_{1_i}^{'} = \mathbf{0}_{1\times(K)}$, is defined as follows,
\begin{equation}
\label{eq:Wald_Ltv}
W_{\lambda_{1_i}} = \hat{\boldsymbol\lambda}_{1_i}^{'} \hat{\mathcal{V}}^{-1}_{\lambda_{1_i}} \hat{\boldsymbol\lambda}_{1_i}  \stackrel{\alpha}{\sim}  \chi^2(K) ~ .
\end{equation}     

\noindent In Table \ref{tab:Lambda}, we report the estimates and $t$-statistics for the market price of risk parameters in the proposed Gaussian ATSM, together with the Wald statistics and $p$-values for the two tests just described. Examining the first row of the table, we note that equity tail risk, as measured by exposure to $\text{TR}^{(eq)}$, is strongly priced in our term structure model with a $p$-value of 8.5\%.
We detect statistically significant time variations in the market price of equity tail risk, which are mostly explained by the level and curvature components of bond yields. Furthermore, when looking at the
$t$-statistics in the second column of the table, we note that $\text{TR}^{(eq)}$ is an important driver of the market price of level risk. Finally, we observe that $\text{PC2}$ carries a significant price of risk in our term structure model. This result, together with the fact that \cite{Adrian2013} find a significant market price of slope risk only after adding an unspanned real activity factor to their framework, corroborates the hypothesis that valuable information about bond premia is located outside of the yield curve.

\vspace{1.0em}

\begin{center}
	[ Insert Table \ref{tab:Lambda} here ]
\end{center}

\vspace{1.0em}

\begin{figure}[!ht]
	\caption{Model-implied yield loadings on the pricing factors of US Treasuries}  
	\label{fig:Y_load}
	{\footnotesize \centering
		\vspace{0.4cm}
		\includegraphics[width=16.0cm, height=19cm]{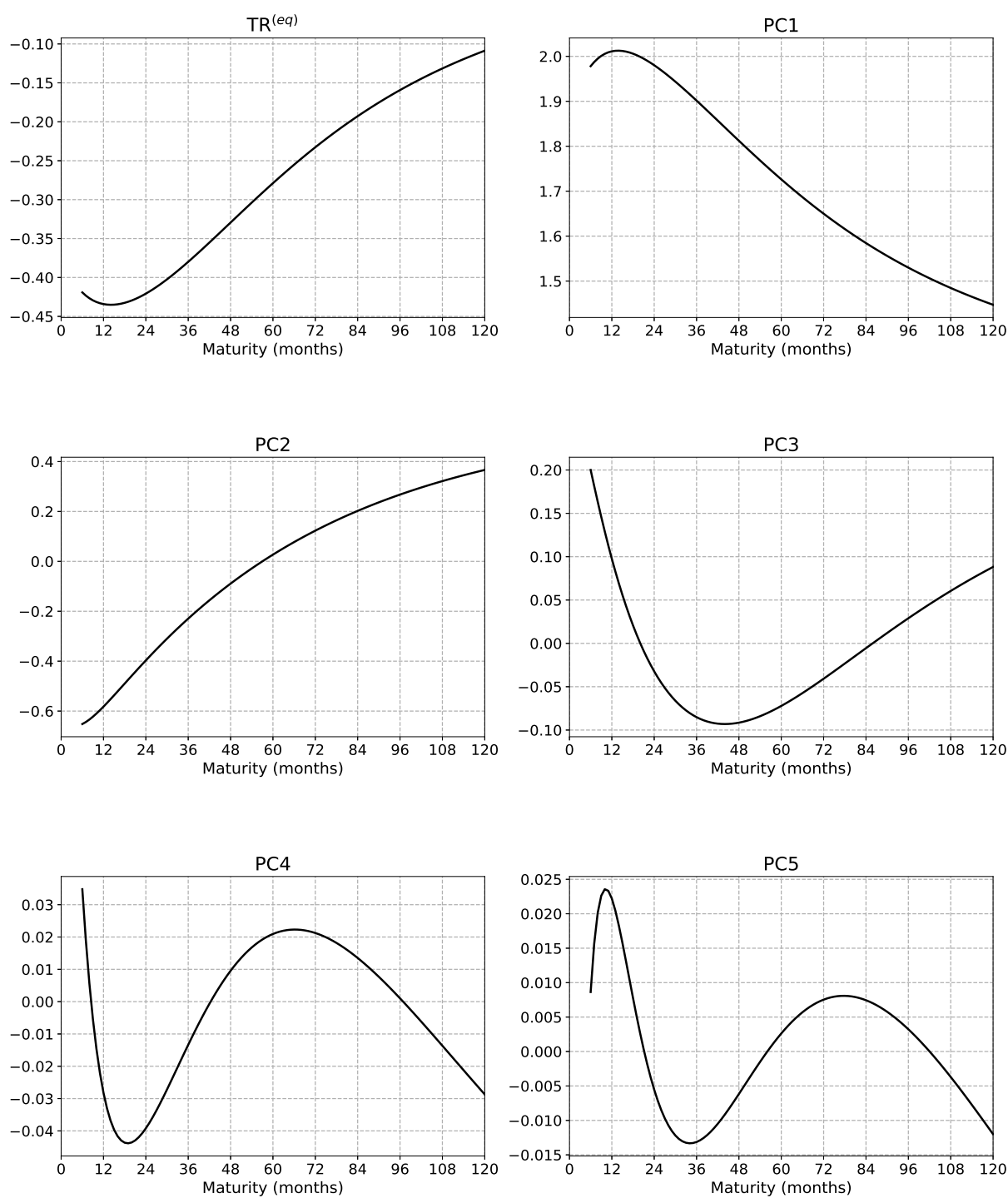} 
	}
\begin{minipage}{15.8cm}
	\vspace{0.3cm}
	The figure displays the model-implied yield loadings on the pricing factors of the proposed ATSM with equity tail risk. These coefficients are calculated as $-(1/n)\mathbf{b}_{n}$ and can be interpreted as the response of the $n$-month yield (expressed in annualized percentage terms) to a contemporaneous shock to the respective factor. $\text{TR}^{(eq)}$ represents the S\&P 500 option-implied equity tail risk factor, normalized to have mean zero and unit variance. $\text{PC1}$ -- $\text{PC5}$ denote the first five standardized principal components extracted from the US Treasury yields orthogonal with respect to the $\text{TR}^{(eq)}$ factor.
\end{minipage}
\end{figure}
\begin{figure}[!ht]
	\caption{Model-implied return loadings on the pricing factors of US Treasuries}  
	\label{fig:RX_load}
	{\footnotesize \centering
		\vspace{0.4cm}
		\includegraphics[width=16.0cm, height=19cm]{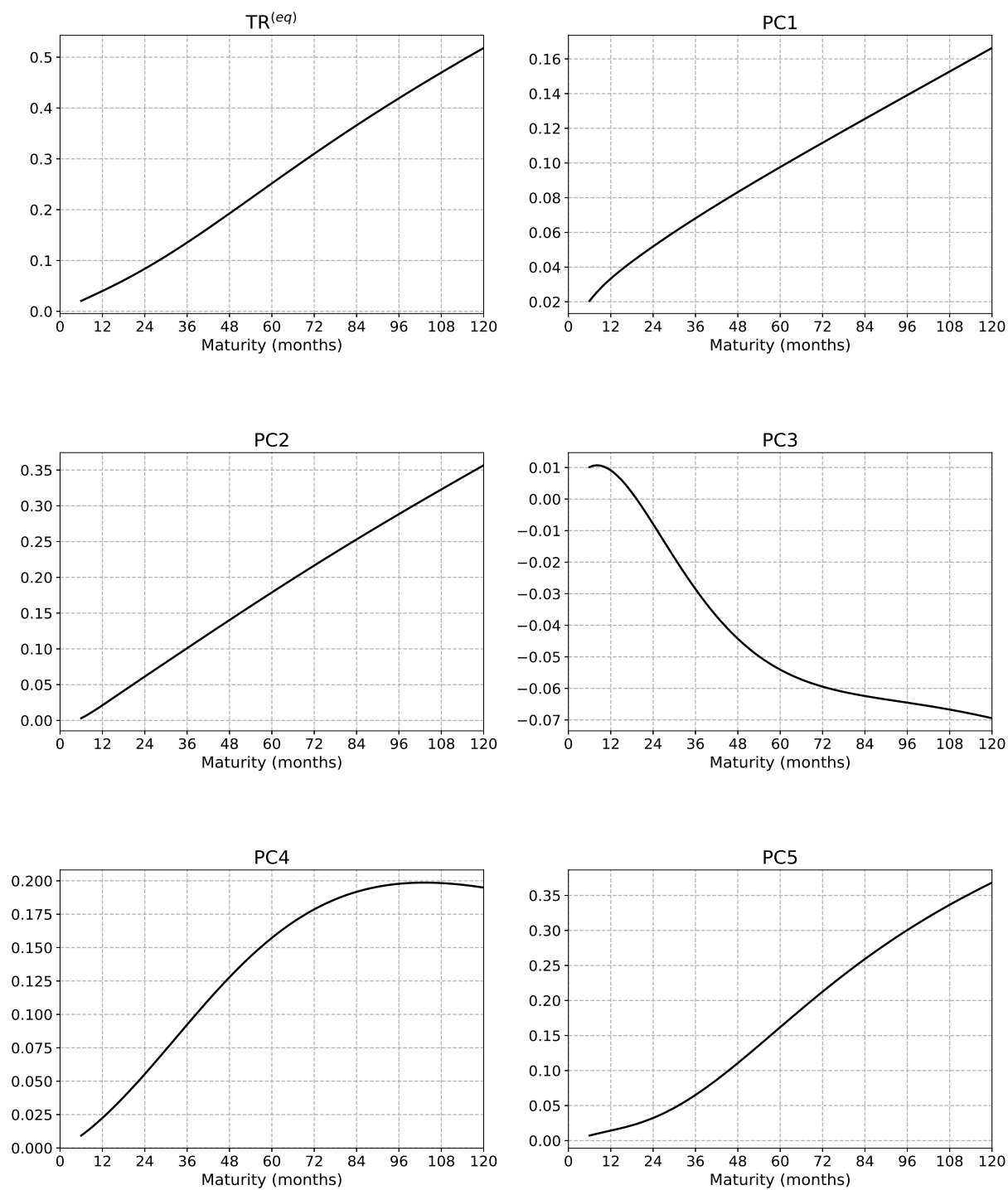} 
	}
\begin{minipage}{15.8cm}
	\vspace{0.3cm}
	The figure displays the model-implied excess return loadings on the pricing factors of the proposed ATSM with equity tail risk. These coefficients are calculated as $\mathbf{b}_{n}^{'}\boldsymbol\lambda_1$ and can be interpreted as the response of the expected one-month excess return (expressed in percentage not annualized terms) on the $n$-month bond to a contemporaneous shock to the respective factor. $\text{TR}^{(eq)}$ represents the S\&P 500 option-implied equity tail risk factor, normalized to have mean zero and unit variance. $\text{PC1}$ -- $\text{PC5}$ denote the first five standardized principal components extracted from the US Treasury yields orthogonal with respect to the $\text{TR}^{(eq)}$ factor.
\end{minipage}
\end{figure}

We now discuss the impact of the state variables of our Gaussian ATSM on the pricing of Treasury bonds. The loadings of the yields on all model factors are reported in Figure \ref{fig:Y_load}, whereas the loadings of the expected one-month excess returns are displayed in Figure \ref{fig:RX_load}. From an examination of the state variables that are in common with the work of \cite{Adrian2013}, we can see that our results are broadly consistent with the well-established role of these factors. Indeed, given the sign of the yield loadings on PC1, PC2 and PC3, we can argue that the first three principal components of yields preserve in our study the interpretation of, respectively, level, slope and curvature of the term structure. Moreover, the yield loadings on PC4 and PC5 are both quite small, reflecting the modest variability of bond rates explained by these factors. As can be seen from  Figure \ref{fig:RX_load}, however, all the principal components, including the higher order ones, are important to explain variation in Treasury returns. Specifically, in line with previous findings concerning the predictability of bond returns with yield spreads, our evidence suggests that an increase in the slope factor forecasts higher expected excess returns on bonds of all maturities. Now turning to the new pricing factor that we propose in this paper, we observe from the top left panel of Figure \ref{fig:Y_load} that the yield loadings on $\text{TR}^{(eq)}$ are negative across all maturities. These results imply that bond prices, which move inversely to yields, rise in response to a contemporaneous shock to the equity left tail factor. And since, by construction, $\text{TR}^{(eq)}$ is associated with a downturn in the stock market, we confirm the hypothesis that US Treasury bonds benefit from flight-to-safety flows during periods of turmoil.\footnote{In results available upon request, we found that the contemporaneous correlation between $\text{TR}^{(eq)}$ and the \cite{ff_jfe_1993} market factor is -0.35. Also, there is a negative but insignificant relation between $\text{TR}^{(eq)}$ and the one-month-ahead stock market returns, as measured by the \cite{ff_jfe_1993} market factor.}
Judging by the magnitude of the coefficients, the immediate flight-to-safety effect is stronger on shorter-term bonds. We find that a one standard deviation increase in the $\text{TR}^{(eq)}$ factor is associated with a reduction of about 40 basis point in the yields of Treasuries with maturities ranging from six months to three years. Further, it is worth noting that, according to the size of the loadings, the contemporaneous effect of the equity left tail factor on the yield curve is not negligible compared to that of the first three principal components.
The expected return loadings on $\text{TR}^{(eq)}$ displayed in the top left panel of Figure \ref{fig:RX_load} confirm the previously established positive relation between the left tail volatility of the stock market and the one-month-ahead risk premia of the US government bond market.
Due to the convenient orthogonalization of pricing factors described at the start of this section, we are able to quantify the effects of a shock to the equity tail factor on the bond risk premia. In particular, we find that a one standard deviation increase in the $\text{TR}^{(eq)}$ factor raises the annualized expected excess return by approximately 1\% for the 2-year bond and 6\% for the 10-year bond. The effect is linearly related to the bond maturity.

\vspace{1.0em}

\begin{center}
	\noindent [ Insert Figure \ref{fig:Y_load} here ]
\end{center}

\begin{center}
	\noindent [ Insert Figure \ref{fig:RX_load} here ]
\end{center}

\vspace{1.0em}

\begin{figure}[!ht]
	\caption{Impact over time of equity tail risk on US Treasury bond yields and components}
	\label{fig:Impact_on_Y_RNY_TP}
	{\footnotesize \centering
		\vspace{0.2cm}
		\includegraphics[width=16.0cm, height=19cm]{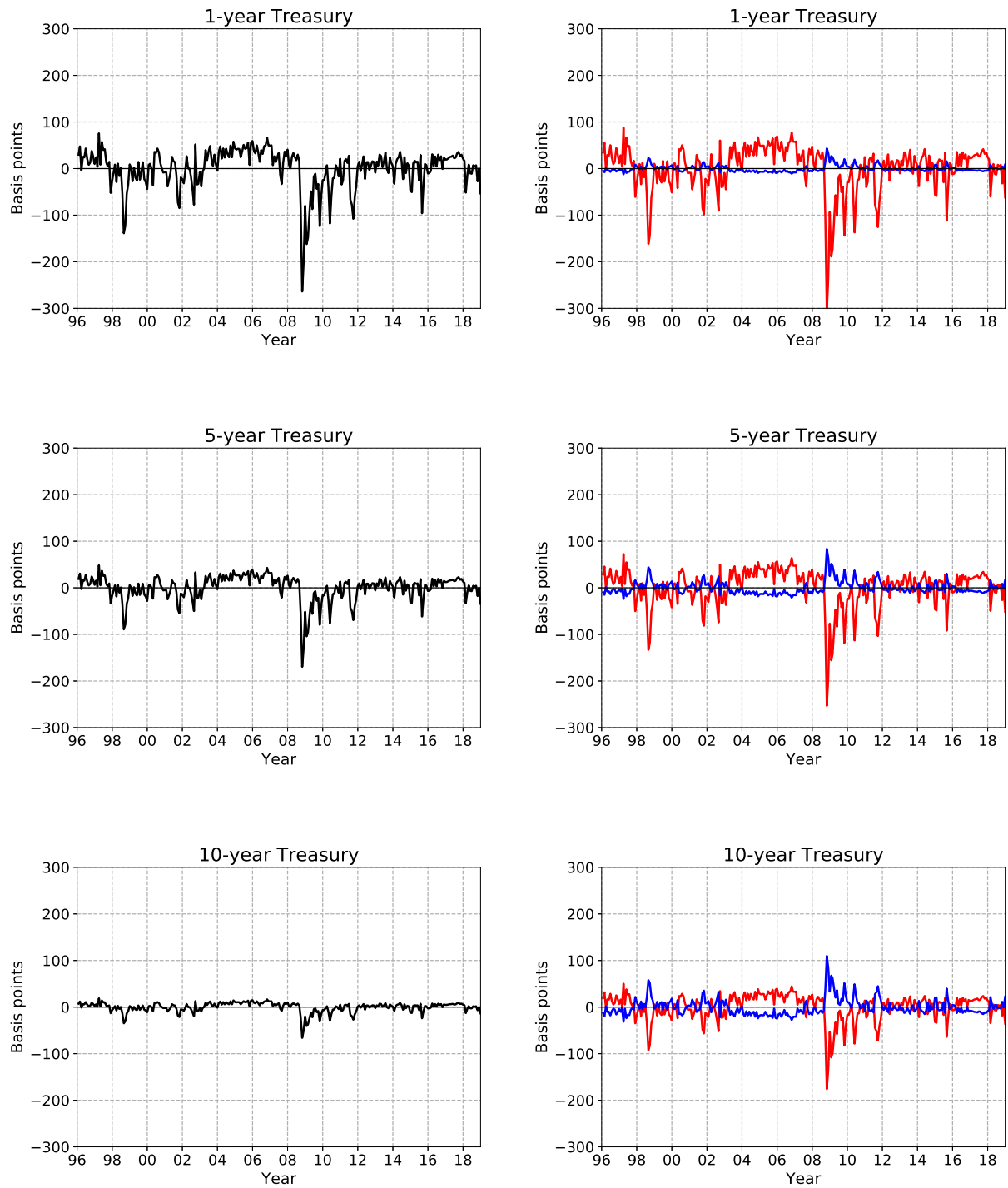} 
	}
	\begin{minipage}{15.8cm}
		\vspace{0.3cm}
		The figure displays the impact over time of the S\&P 500 option-implied equity tail risk factor $\text{TR}_t^{(eq)}$ on the 1-, 5- and 10-year US Treasury bond yields (black lines) and on their two components, i.e~average expected future short rate (red lines) and term premium (blue lines). 
	\end{minipage}	
\end{figure}

We conclude this section by discussing how equity tail risk has affected the trend of yields, risk-neutral rates and term premia over the course of time.
To this end, we calculate the component of fitted yields in equation (\ref{eq:AffineY}) and the component of their risk-neutral counterparts in equation (\ref{eq:RN_Y}) that the model attributes to the equity left tail factor $\text{TR}^{(eq)}$. Similarly, we determine the contribution of equity tail risk to the bond term premia in equation (\ref{eq:TP}) as the difference between the component of fitted yields and the component of their risk-neutral counterparts that the model ascribes to $\text{TR}^{(eq)}$. 
The left panels of Figure \ref{fig:Impact_on_Y_RNY_TP} illustrate the effect of the equity left tail factor $\text{TR}^{(eq)}$ on the dynamics of the 1-, 5- and 10-year Treasury yields, whereas the right panels display the effects on the expected future short rate and term premium embedded in those rates. 
The following remarks can be made by observing Figure \ref{fig:Impact_on_Y_RNY_TP}. 
The effect of equity tail risk is much smaller (in absolute value) for the bond term premium than for the expectation of future short rates. Therefore, when the equity left tail factor $\text{TR}^{(eq)}$ increases, the reduction in the expected future short rate more than offsets the increase in the term premium. As a result, bond yields fall in periods of elevated equity tail risk. 
However, it is interesting to see that, although the same pattern is observed for all yields in Figure \ref{fig:Yields_vs_HighTR}, the equity left tail factor $\text{TR}^{(eq)}$ has influenced the downward trend of rates differently depending on the bond maturity. 
Indeed, from the left panels of Figure \ref{fig:Impact_on_Y_RNY_TP}, it appears that the dynamics of short-maturity bond yields was strongly affected by equity tail risk, whereas the response of longer-maturity rates was consistently negligible.
This further corroborates our previous conclusion that short-term bonds provide a more effective shelter against equity market losses than long-term bonds do. 

\vspace{1.0em}

\begin{center}
	\noindent [ Insert Figure \ref{fig:Impact_on_Y_RNY_TP} here ]
\end{center}

\vspace{1.0em}

\begin{figure}[!ht]
	\caption{Impact of equity tail risk on US Treasury bond yields}
	\label{fig:Impact_on_Y_dates}
	{\footnotesize \centering
		\vspace{0.2cm}
		\includegraphics[width=16.0cm, height=10cm]{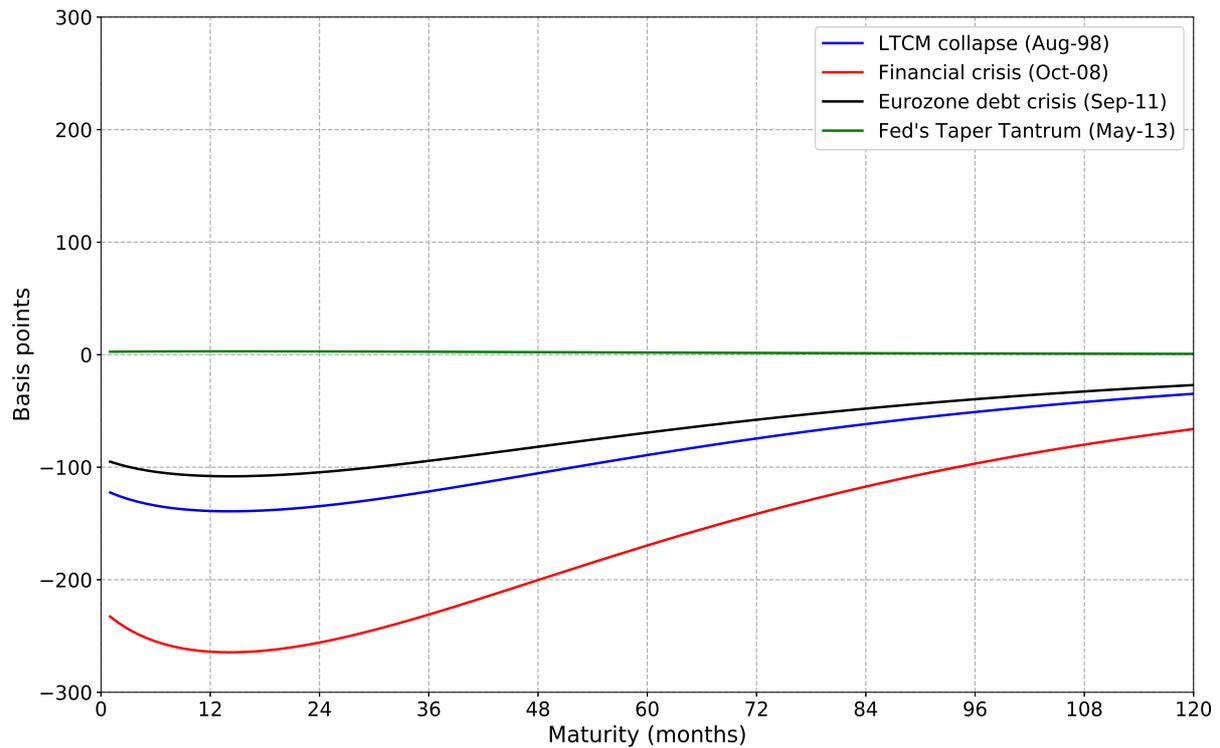} 
	}
	\begin{minipage}{15.8cm}
		\vspace{0.3cm}
		The figure displays the impact (in basis points) of the S\&P 500 option-implied equity tail risk factor $\text{TR}_t^{(eq)}$ on the term structure of US interest rates for selected dates: Russian financial crisis and collapse of Long Term Capital Management fund (Aug-98), onset of 2008-09 financial crisis with bankruptcy of Lehman Brothers (Oct-08), intensification of European sovereign debt crisis (Sep-11), announcement of the Federal Reserve's ``taper tantrum'' (May-13). Interest rates fell on all dates except for May-13, when yields markedly rose.  
	\end{minipage}	
\end{figure}

To better visualize how the impact of equity tail risk varies across maturities and in time, Figure \ref{fig:Impact_on_Y_dates} shows the effect of the $\text{TR}^{(eq)}$ factor for the whole term structure calculated on selected dates: August 1998, October 2008, September 2011, and May 2013. Interest rates fell on all dates except for May 2013, when yields markedly rose with the announcement of the Federal Reserve's ``taper tantrum''. On that occasion, as it can be seen from the figure,  $\text{TR}^{(eq)}$ did not play any role in the yield changes. On the other hand, at the peak of the 2008-09 financial crisis, we measure the impact of equity tail risk on bond yields to be larger than -200 basis points for Treasuries with maturities up to four years, while it is reduced to only -66 basis points for the 10-year Treasury. 
The rates showed strong downward oscillations also in the summer of 1998 and the second half of 2011, when the equity left tail factor increased in response to, respectively, the collapse of Long Term Capital Management fund and the intensification of the European sovereign debt crisis. In both these instances, the extent of the reduction in short-term bond rates that can be credited to equity tail risk is approximately 100 basis points.

\vspace{1.0em}

\begin{center}
	\noindent [ Insert Figure \ref{fig:Impact_on_Y_dates} here ]
\end{center}

\vspace{1.0em}

In conclusion, we can state that equity jump tail risk has been a dominant factor for the evolution of the short end of the US Treasury yield curve. In particular, while the unconventional monetary policies introduced by central banks to mitigate the severity of the financial crisis have been a major force in lowering longer-term yields \citep{Kaminska2018}, the reduction in shorter-term yields can be associated with the investors' increased fear of a stock market crash. 


\subsection{Three-Pass Method Estimates}
\label{subsec:GX_with_TR}   

To address the concern that the rows in the price of risk parameters $\boldsymbol\lambda_0$ and $\boldsymbol\lambda_1$ in (\ref{eq:RiskPrices}) corresponding to $\text{TR}^{(eq)}$ can only be weakly identified because our equity tail factor is weakly spanned by bond yields, we now provide further evidence for a significant price of equity tail risk in the US government bond market. 
This is done by estimating the risk premium of $\text{TR}^{(eq)}$ with the novel three-pass procedure of \cite{Giglio2019}. The results of the GX three-pass method applied to the whole term structure of Treasury bond returns are reported in Table \ref{tab:GX_RiskPremia_US}.

\vspace{1.0em}

\begin{center}
	[ Insert Table \ref{tab:GX_RiskPremia_US} here ]
\end{center}

\vspace{1.0em}

\noindent We start by examining the results reported in column $p=5$, which corresponds to the number of principal components of bond returns selected with the criterion of \cite{Giglio2019}.\footnote{See Online Appendix I.1 in \cite{Giglio2019} for a consistent estimator of $p$.} For this number of latent factors, we find that the estimated risk premium of $\text{TR}^{(eq)}$ in the US Treasury bond market is statistically significant at the 10\% level. Although obtained with a different asset pricing model, this results is well in line with the estimates of the ATSM presented in the previous section. 
Furthermore, we provide evidence against the hypothesis that $\text{TR}^{(eq)}$ is measured with noise or weakly reflected in the cross section of government bond returns. In fact, the $R^2$ of the time-series regression in the third-pass of the GX procedure amounts to 0.09 and we reject, at the 5\% significance level, the null of $\text{TR}^{(eq)}$ being a weak factor.   
If we now look at the estimates obtained with a higher number of latent factors, we observe robustness of our empirical results with respect to the choice of $p$. Even when using eight principal components, the market price of equity tail risk is still significant at the 0.1 level. However, including the principal components beyond the fifth one does not result in further noticeable improvement in the regression $R^2$. On the other hand, we find that much of the information about equity tail risk is contained in the slope factor, with the $R^2$ that jumps from 0.04 to 0.08 when the second principal component is included in the model.

\subsection{International Evidence}
\label{subsec:International}   

\begin{figure}[!ht]
	\caption{Time series of international equity tail risk measures}  
	\label{fig:WORLD_TR}
	{\footnotesize \centering
		\vspace{0.2cm}
		\includegraphics[width=16.0cm, height=19cm]{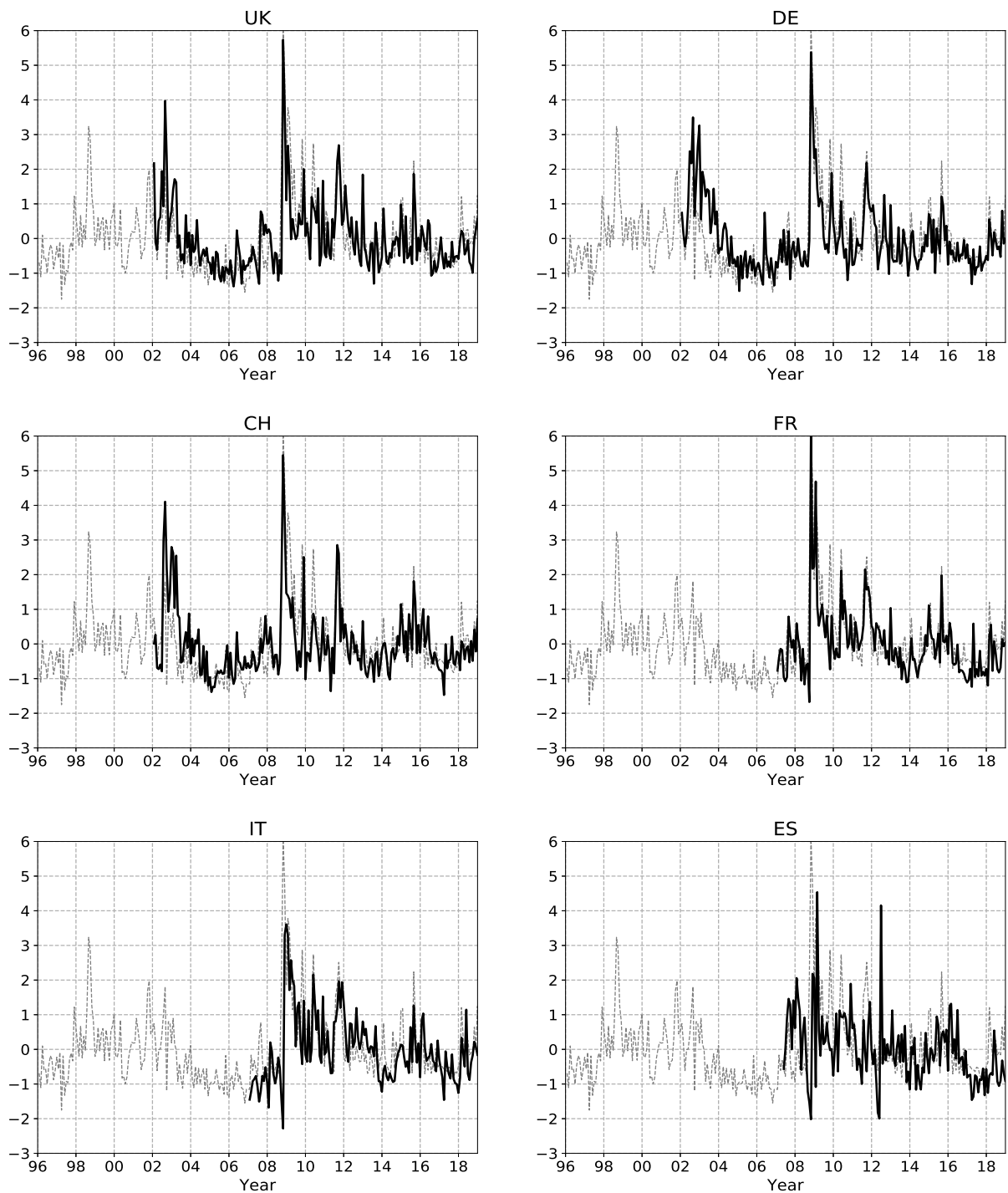} 
	}
	\begin{minipage}{15.8cm}
		\vspace{0.3cm}
		The figure displays the international equity tail risk measures calculated using options on the FTSE 100 (UK), DAX (DE), SMI (CH), CAC 40 (FR), FTSE MIB (IT) and IBEX 35 (ES) equity index. All series have been normalized to have mean zero and unit variance. The solid black lines show the equity tail risk measure of the country of interest, while the dashed gray lines show, for comparison, the S\&P 500 option-implied equity tail risk measure $\text{TR}^{(eq)}$. 
	\end{minipage}	
\end{figure}

In this subsection, we extend our empirical analysis of bond pricing and return predictability to the Treasury market of United Kingdom, Germany, Switzerland, France, Italy and Spain. First, we explore to what extent the S\&P 500 option-implied tail risk measure $\text{TR}^{(eq)}$ affects the  Treasury market of countries other than the United States. Then, we estimate country-specific measures of equity tail risk and investigate the relation between these measures and the government bond market in the corresponding European country.
To compute the one-month holding period returns on Treasuries in Europe, we construct a data set of end-of-month zero-coupon interest rates that extends from January 1996 to December 2018. 
We collect data for the United Kingdom (UK) from the Bank of England, for Germany (DE) from the Bundesbank and BIS database, for Switzerland (CH) from the Swiss National Bank and BIS database, for Italy (IT) and Spain (ES) from the BIS database, while for France (FR) we fit a Nelson-Siegel-Svensson model to the constant maturity yields from Datastream. As for the country-specific measures of equity tail risk, we follow the methodology outlined in Section \ref{subsec:Data_equity} and estimate option-implied volatility that stems from large negative price jumps using daily data reported by OptionMetrics IvyDB Europe for the European style FTSE 100 (UK), DAX 30 (DE), SMI (CH), CAC 40 (FR), FTSE MIB (IT), and IBEX 35 (ES) equity-index options. Data is available from January 2002 to December 2018 for UK, DE and CH, from January 2007 to December 2018 for IT and FR, and from May 2007 to December 2018 for ES. We use option-implied left tail volatilities recorded at the end of the month for UK, DE and CH, while we use the average value over the last five days of the month for FR, IT, and ES since their less liquid option markets yield a much noisier measure of equity tail risk. Figure \ref{fig:WORLD_TR} displays the time series of these international equity tail risk measures along with the S\&P 500 option-implied measure $\text{TR}^{(eq)}$. 
Comparing the left tail volatility of the US stock market to that of the UK, German, Swiss and French stock markets, we note a strong coherence between the series with all the correlation coefficients above 0.70. At the same time, however, there are also some important differences. In particular, we note that in 2002-03 the UK, DE and CH tail risk measures attained higher values and remained elevated for a much longer period of time than $\text{TR}^{(eq)}$, which however exhibits more pronounced peaks in the aftermath of the recent financial crisis.  
With regard to the equity tail risk measures of Italy and Spain, their series diverge quite substantially from that of the US measure with correlation coefficients of only 0.50 and 0.20, respectively.

\vspace{1.0em}

\begin{center}
	\noindent [ Insert Figure \ref{fig:WORLD_TR} here ]
\end{center}

\vspace{1.0em}

We begin by assessing the predictive power of the left tail volatility of the US stock market for future one-month returns on the government bond market of the European countries. 
To this end, we estimate the predictive regressions in (\ref{eq:RX_reg_univ}), (\ref{eq:RX_reg_PC3}) and (\ref{eq:RX_reg_PC5}) using international bond returns on the left hand side of the equations and $\text{TR}^{(eq)}$, combined with the principal components of the country-specific yield curves, on the right hand side. For each Treasury market, Table \ref{tab:IS_inter_vs_US} reports the full-sample estimates of the coefficient of $\text{TR}^{(eq)}$ and the corresponding $p$-values computed with both Newey-West and \cite{Bauer2018} inference procedures.

\vspace{1.0em}

\begin{center}
	[ Insert Table \ref{tab:IS_inter_vs_US} here ]
\end{center}

\vspace{1.0em}

\noindent Overall, the results in Table \ref{tab:IS_inter_vs_US} indicate  that the perceived tail risk in the US stock market has significant explanatory power for future returns on Treasury bonds in the UK, Germany, Switzerland and France. 
When we do not control for yield curve factors in the return predictive regressions, the coefficient of $\text{TR}^{(eq)}$, assessed with the robust inference method developed by \cite{Bauer2018}, is statistically significant at the 0.05 level or lower for all maturities of UK, DE and CH  bonds, and at the 0.10 level or lower for all maturities of FR bonds. Controlling with the first three or five principal components of bond yields does not change the results for the UK and DE Treasuries, while it reduces the significance for the longer maturities of CH and FR bonds.
Consistent with the results in Table \ref{tab:IS_results}, the sign of the coefficient is positive, implying that higher equity tail risk is associated with an increase in the one-month-ahead bond risk premia.
In contrast to the results obtained with the UK, DE, CH and FR bonds, the equity tail risk factor $\text{TR}^{(eq)}$ does not seem to help explain time variations in the bond risk premia of Italy and Spain. In fact, the explanatory power of $\text{TR}^{(eq)}$ is never statistically significant at the 10\% level for IT bonds with maturity greater than one year, and is at most significant at that level for the short-term ES bonds. These results point to the possible role that country risk may play in the identification of a safe asset when the equity market tumbles. It is indeed possible that, in periods of stress, international investors shift their holdings into instruments like the ``safe'' German Bund rather than  debt issued by fiscally weak sovereigns, such as Italy and Spain. 
Due to the mostly insignificant interactions observed in-sample between $\text{TR}^{(eq)}$ and Treasury bonds of Italy and Spain, we do not consider the out-of-sample forecast improvements afforded by equity tail risk for bond returns in these two countries. For all other countries, Table \ref{tab:OOS_inter_vs_US} reports the out-of-sample relative forecast and portfolio performance of the models in (\ref{eq:RX_reg_univ}), (\ref{eq:RX_reg_PC3}) and (\ref{eq:RX_reg_PC5}), which predict international bond returns with the S\&P 500 option-implied tail risk measure $\text{TR}^{(eq)}$. Results are based on the out-of-sample setting described in Section \ref{subsec:RedForm_Regressions}, with predictive regressions that are recursively estimated with a rolling window approach and the assumption that the investor's level of risk aversion is $\gamma=5$.

\vspace{1.0em}

\begin{center}
	[ Insert Table \ref{tab:OOS_inter_vs_US} here ]
\end{center}

\vspace{1.0em}

\noindent From an examination of the \cite{CT2008} out-of-sample $R^2_{OS}$s in Table \ref{tab:OOS_inter_vs_US}, we note that the models that include equity tail risk systematically outperform the benchmarks in predicting returns of the UK and Germany Treasury markets. The same holds true for short- and medium-maturity bonds in Switzerland and France. The reductions in the MSPE for the forecasts generated by the model that includes $\text{TR}^{(eq)}$ are in the range of 4\% to 24\% for UK bond returns and in the range of 0.5\% to 11\% for DE bond returns. On the basis of the \cite{CW2007} test results, however, the gains of predictability in international bond returns are only marginally statistically significant. When assessing the portfolio performance afforded by equity tail risk, we observe that $\text{TR}^{(eq)}$ can generate substantial risk-adjusted returns for investors trading bonds in all four countries, but especially in the UK and Germany. For instance, when the benchmark is the 3 PCs-only model, we find that an investor trading the 5-year UK (DE) Treasury bond is willing to pay approximately 165 (213) basis points per year to switch from the benchmark to the model that predicts bond returns with equity tail risk.
     
Having identified significant associations between the left tail volatility of the US stock market and the future returns on some of the major international government bond markets, the natural question that arises is whether equity tail risk is also a key determinant of the current level of prices of those bonds. To answer this question, we estimate the risk premium of $\text{TR}^{(eq)}$ by applying the GX three-pass method to the term structures of Treasury bonds in the UK, Germany, Switzerland, France, Italy and Spain. The results are reported in Table \ref{tab:GX_RiskPremia_inter_vs_US}.

\vspace{1.0em}

\begin{center}
	[ Insert Table \ref{tab:GX_RiskPremia_inter_vs_US} here ]
\end{center}

\vspace{1.0em}

\noindent As we did for the US term structure, we assess robustness of the estimates by reporting results also for a higher number of latent factors than those selected with the \cite{Giglio2019} criterion, which points to 5 principal components for all Treasury markets except for the UK where 4 factors are selected. Examining the significance of the risk premium estimates $\gamma_g$, we can see that $\text{TR}^{(eq)}$ carries a significant price of risk in the Treasury bond market of not only Germany, Switzerland and France, for which we found strong return predictability, but also Spain, where the evidence on predictability was much weaker. However, we do not reject the null of $\text{TR}^{(eq)}$  being a weak factor for the ES term structure. Surprisingly, equity tail risk is not priced in the UK Treasury market, where $\text{TR}^{(eq)}$ has strong predictive power for future returns. As for the Italian government bond market, we confirm the lack of a connection with equity tail risk. 
Furthermore, it can be seen from the time-series regression $R^2$s that the equity left tail factor is mostly spanned by the second and third principal components of the Treasury returns.

We end this section by relating the returns of the international government bond markets to the perceived tail risk in the stock market of the home country. We do this by running the predictive regressions in (\ref{eq:RX_reg_univ}), (\ref{eq:RX_reg_PC3}) and (\ref{eq:RX_reg_PC5}) with the country-specific equity tail risk measures displayed in Figure \ref{fig:WORLD_TR} and estimating their risk premium with the GX three-pass procedure. Due to the limited availability of option data on the European stock market indices, we only consider the in-sample performance of the predictive models in (\ref{eq:RX_reg_univ}), (\ref{eq:RX_reg_PC3}) and (\ref{eq:RX_reg_PC5}). The full-sample estimates of the coefficients of the country-specific equity tail risk measures are reported in Table \ref{tab:IS_inter_vs_cs} while the results of the GX three-pass regression procedure are shown in Table \ref{tab:GX_RiskPremia_inter_vs_cs}. 

\vspace{1.0em}

\begin{center}
	\noindent [ Insert Table \ref{tab:IS_inter_vs_cs} here ]
\end{center}

\begin{center}
	\noindent [ Insert Table \ref{tab:GX_RiskPremia_inter_vs_cs} here ]
\end{center}

\vspace{1.0em}

\noindent A quick inspection of Table  \ref{tab:IS_inter_vs_cs}  reveals that the future one-month returns of UK, DE and CH Treasury bonds are strongly associated not only with the S\&P 500 option-implied left tail factor $\text{TR}^{(eq)}$ but also with the corresponding country-specific measure of equity tail risk. On the other hand, we do not find any statistically significant relationship between the FR, IT and ES bond returns and the perceived tail risk in the stock market of the home country. Finally, the results in Table \ref{tab:GX_RiskPremia_inter_vs_cs} support our previous observations on the existence of a significant market price of equity tail risk in the Treasury bond market of Germany, Switzerland and France.

In conclusion, our findings concerning the predictive power and pricing of equity tail risk are robust to alternative data sets. In fact, there is clear evidence that equity tail risk carries significant information about the dynamics of Treasury bond yields and returns not only in the US but also in major government bond markets in Europe.



\section{Conclusion}
\label{sec:Conclusion}

In this paper, we study how US Treasury bonds respond to changes in the perceived tail risk in the stock market.
We estimate equity tail risk with the risk-neutral expectation of future volatility that stems from large negative price jumps and we examine how it relates to the future one-month returns on bonds in reduced-form predictive regressions. 
Also, we propose an affine term structure model in which the main drivers of interest rates are the principal components of the zero-coupon yield curve and the equity tail risk factor. While earlier approaches to pricing bonds with factors other than combinations of yields have proven useful when macro variables are considered, we focus here on the observed comovement in stock and bond markets during crisis periods and use a state variable that originates in the equity option market.

The results of our main application to the US government bond and S\&P 500 index option markets are summarized as follows. First, there exist significant interactions between the one-month-ahead risk premia in Treasury bonds and the left tail volatility of the stock market. 
Second, the strong predictive power of equity tail risk for future bond returns is confirmed in a real-time out-of-sample exercise, where this predictability can be exploited to improve the economic utility of a mean-variance investor.
Third, the left tail volatility of the stock market is a priced state variable in the US term structure. We find evidence of a significant market price of equity tail risk not only with the ATSM but also with the novel three-pass method proposed by \cite{Giglio2019}.  
Fourth, consistent with the theory of flight-to-safety, bond prices rise in
response to a contemporaneous shock to the equity left tail factor. 
Fifth, large drops in short-term bond yields and expected future short rates are attributable to equity tail risk. 
Finally, our results concerning the predictive power and pricing of equity tail risk are robust to alternative data sets. When extending the analysis to major government bond markets in Europe, we find that equity tail risk carries significant information about the dynamics of Treasury bond yields and returns in United Kingdom, Germany, Switzerland and France, while the evidence is considerably weaker in Spain and non-existent in Italy.

Given our findings with a measure of downside tail risk of the stock market, a natural direction for future research would be to assess the impact on the yield curve of a tail factor implied by Treasury options. For instance, it would be interesting to see whether the downside, or even the upside, tail risk of the bond market receives compensation in a term structure model and how its pricing differs from that of equity tail risk. This would contribute to the recent literature on the auxiliary role of Treasury variance and jump risk in explaining bond risk premia, see \citep{Wright2009, Mueller2016}. We leave investigation of such possibilities to future research.  

\clearpage \newpage


\bibliographystyle{ecta}
\bibliography{Bib_skew_paper}




\clearpage \newpage

\begin{table}[!ht]
	\caption{Descriptive statistics: bond risk premia and equity tail risk} 
	\label{tab:DescriptiveStats}
	\fontsize{9}{11} \selectfont
	\renewcommand{\tabcolsep}{6.2pt}
	\sisetup{ input-symbols = {()},
		table-format = -0.2,
		table-space-text-post = ***,
		table-align-text-post = false,
		group-digits = false,
		explicit-sign}
	\centering 
	\begin{tabular}{lSSSSSSSS} \hline \hline
		\vspace{-0.15cm}\\
		& {$\text{RX}_{t+1}^{(12)}$} & {$\text{RX}_{t+1}^{(24)}$} & {$\text{RX}_{t+1}^{(36)}$} & {$\text{RX}_{t+1}^{(48)}$} & {$\text{RX}_{t+1}^{(60)}$} & {$\text{RX}_{t+1}^{(84)}$} & {$\text{RX}_{t+1}^{(120)}$} & {$\text{TR}_{t}^{(eq)}$}  \vspace{+0.12cm}\\
		\hline
		\multicolumn{9}{l}{\rule[-2.2mm]{0mm}{6.5mm}{Panel A: Descriptive Statistics } }\\
		\hline
		\vspace{-0.15cm}\\
		{Mean} & 0.385 & 0.953 & 1.511 & 2.030 & 2.498 & 3.274 & 4.097 & 0.100 \\
		\\[-0.3cm]
		{Std. dev.} & 0.597 & 1.535 & 2.556 & 3.561 & 4.530 & 6.382 & 8.998 & 0.041 \\
		\\[-0.3cm]
		{Skewness} & 1.386 & 0.489 & 0.140 & 0.004 & -0.029 & 0.020 & 0.071 & 2.210 \\
		\\[-0.3cm]
		{Kurtosis} & 7.542 & 4.604 & 3.894 & 3.693 & 3.752 & 4.187 & 4.981 & 10.575 \\
		\\[-0.3cm]
		{$\rho(1)$} & 0.214 & 0.156 & 0.113 & 0.085 & 0.067 & 0.048 & 0.028 & 0.657 \\
		\\[-0.3cm]
		{$\rho(6)$} & 0.093 & -0.024 & -0.072 & -0.088 & -0.093 & -0.093 & -0.083 & 0.260 \\
		\\[-0.3cm]
		{$\rho(12)$} & 0.082 & 0.133 & 0.137 & 0.127 & 0.110 & 0.065 & 0.010 & 0.172 \\
		\\[-0.3cm]
		{SR} & 0.645 & 0.621 & 0.591 & 0.570 & 0.551 & 0.513 & 0.455 &   \\   
		\\[-0.2cm]
		\hline
		\multicolumn{9}{l}{\rule[-2.2mm]{0mm}{6.5mm}{Panel B: Correlation Matrix} }\\
		\hline
		\vspace{-0.15cm}\\
		{$\text{RX}_{t+1}^{(12)}$}   & 1.000 &  &  &  &  &  &  & \\
		\\[-0.3cm]
		{$\text{RX}_{t+1}^{(24)}$}   & 0.926 & 1.000 &  &  &  &  &  & \\
		\\[-0.3cm]
		{$\text{RX}_{t+1}^{(36)}$}   & 0.849 & 0.981 & 1.000 &  &  &  &  & \\
		\\[-0.3cm]
		{$\text{RX}_{t+1}^{(48)}$}   & 0.790 & 0.946 & 0.990 & 1.000 &  &  &  & \\
		\\[-0.3cm]
		{$\text{RX}_{t+1}^{(60)}$}   & 0.739 & 0.905 & 0.966 & 0.993 & 1.000 &  &  & \\
		\\[-0.3cm]
		{$\text{RX}_{t+1}^{(84)}$}   & 0.652 & 0.821 & 0.899 & 0.949 & 0.980 & 1.000 &  & \\
		\\[-0.3cm]
		{$\text{RX}_{t+1}^{(120)}$}  & 0.549 & 0.711 & 0.799 & 0.865 & 0.915 & 0.975 & 1.000 & \\
		\\[-0.3cm]
		{$\text{TR}_{t}^{(eq)}$}          & 0.223 & 0.190 & 0.185 & 0.189 & 0.194 & 0.201 & 0.199 & 1.000 \\
		\vspace{-0.15cm}\\
		\hline \hline
		\vspace{-2.5mm}
		\end{tabular}
	\begin{minipage}{15.8cm}
		Notes: This table contains descriptive statistics for the one-month excess US Treasury bond returns $\text{RX}_{t+1}^{(n)}$, with maturity $n=12,24,36,48,60,84,120$ months, and for the S\&P 500 option-implied equity tail risk measure $\text{TR}_t^{(eq)}$ used as predictor in the empirical analyses. Panel A reports the sample mean, standard deviation, skewness, kurtosis and autocorrelation coefficients of order one, six and twelve for each of the variables. Return means and standard deviations are expressed in annualized percentage terms. The annualized Sharpe ratio (SR) is also reported for the Treasury bonds. Panel B reports the correlation coefficients calculated with the future bond returns and contemporaneous $\text{TR}^{(eq)}$ factor. The sample uses end-of-month data for 1996:01--2018:12.
	\end{minipage}
\end{table}

\clearpage \newpage

\begin{table}[!ht]
	\caption{In-sample forecasts of Treasury returns with equity tail risk} 
	\label{tab:IS_results}
	\fontsize{9}{11} \selectfont
	\renewcommand{\tabcolsep}{5.7pt}  
	\sisetup{ input-symbols = {()},
		table-format = -0.2,
		table-space-text-post = ***,
		table-align-text-post = false,
		group-digits = false,
		explicit-sign}
	\centering 
	\begin{tabular}{llSSSSSSS} \hline \hline
		\vspace{-0.15cm}\\
		& & {$n=12$} & {$n=24$} & {$n=36$}  & {$n=48$} & {$n=60$} & {$n=84$} & {$n=120$} \vspace{+0.12cm}\\
		\hline
		\multicolumn{9}{l}{\rule[-2.2mm]{0mm}{6.5mm}{\textbf{Panel A: No control for bond return forecasting factors}}}\\  
		\hline
		\vspace{-0.15cm}\\
		{$\text{TR}_{t}^{(eq)}$} & {$\beta$}  & 0.460 & 1.009 & 1.638 & 2.322 & 3.032 & 4.429 & 6.206 \\
		& {$p$-value}  & 0.000 & 0.000 & 0.001 & 0.001 & 0.001 & 0.002 & 0.005 \\
		& {$p$-value \tiny{(b)}} & 0.008 & 0.037 & 0.012 & 0.010 & 0.006 & 0.001 & 0.001 \\
		\vspace{-0.2cm}\\
		\hline
		\vspace{-0.2cm}\\
		{Adj. $R^2$\tiny{(\%)}} &   & 4.622 & 3.258 & 3.081 & 3.204 & 3.394 & 3.676 & 3.627 \\  
		{Adj. $R^2$\tiny{(\%)}} & {\tiny{no $\text{TR}^{(eq)}$}}  & 0.000 & 0.000 & 0.000 & 0.000 & 0.000 & 0.000 & 0.000 \\
		{$F$-test} &   & 0.000 & 0.002 & 0.002 & 0.002 & 0.001 & 0.001 & 0.001 \\
		\\
		\vspace{-0.15cm}\\
		\hline
		\multicolumn{9}{l}{\rule[-2.2mm]{0mm}{6.5mm}{\textbf{Panel B: Control for yield curve factors with 3 PCs}}}\\
		\hline
		\vspace{-0.15cm}\\
		{$\text{TR}_{t}^{(eq)}$} & {$\beta$}  & 0.419 & 0.910 & 1.490 & 2.115 & 2.751 & 3.970 & 5.499 \\
		& {$p$-value}  & 0.001 & 0.008 & 0.011 & 0.011 & 0.010 & 0.010 & 0.014 \\
		& {$p$-value \tiny{(b)}} & 0.001 & 0.010 & 0.007 & 0.008 & 0.007 & 0.007 & 0.008 \\
		\\[-0.2cm]
		{$\text{PC1}_{t}$} & {$\beta$}  & 0.355 & 0.542 & 0.692 & 0.834 & 0.968 & 1.208 & 1.532 \\
		& {$p$-value}  & 0.015 & 0.126 & 0.218 & 0.277 & 0.316 & 0.370 & 0.414 \\
		\\[-0.2cm]
		{$\text{PC2}_{t}$} & {$\beta$}  & 0.240 & 0.746 & 1.246 & 1.743 & 2.234 & 3.182 & 4.478 \\
		& {$p$-value}  & 0.032 & 0.007 & 0.004 & 0.002 & 0.001 & 0.000 & 0.000 \\
		\\[-0.2cm]
		{$\text{PC3}_{t}$} & {$\beta$}  & 0.158 & 0.043 & -0.126 & -0.255 & -0.322 & -0.325 & -0.308 \\
		& {$p$-value}  & 0.471 & 0.929 & 0.864 & 0.791 & 0.784 & 0.835 & 0.881 \\
		\vspace{-0.2cm}\\
		\hline
		\vspace{-0.2cm}\\
		{Adj. $R^2$\tiny{(\%)}} &   & 8.477 & 5.211 & 4.629 & 4.638 & 4.779 & 4.994 & 4.859 \\
		{Adj. $R^2$\tiny{(\%)}} & {\tiny{no $\text{TR}^{(eq)}$}}  & 5.198 & 2.986 & 2.491 & 2.404 & 2.429 & 2.510 & 2.475 \\
		{$F$-test} &   & 0.001 & 0.007 & 0.008 & 0.007 & 0.006 & 0.005 & 0.006 \\
		\\
		\vspace{-0.15cm}\\
		\hline
		\multicolumn{9}{l}{\rule[-2.2mm]{0mm}{6.5mm}{\textbf{Panel C: Control for yield curve factors with 5 PCs}}}\\
		\hline
		\vspace{-0.15cm}\\
		{$\text{TR}_{t}^{(eq)}$} & {$\beta$}  & 0.411 & 0.895 & 1.453 & 2.042 & 2.635 & 3.770 & 5.202 \\
		& {$p$-value}  & 0.001 & 0.006 & 0.010 & 0.010 & 0.010 & 0.010 & 0.013 \\
		& {$p$-value \tiny{(b)}} & 0.002 & 0.009 & 0.007 & 0.009 & 0.010 & 0.009 & 0.011 \\
		\\[-0.2cm]
		{$\text{PC1}_{t}$} & {$\beta$}  & 0.353 & 0.540 & 0.686 & 0.822 & 0.950 & 1.177 & 1.487 \\
		& {$p$-value}  & 0.014 & 0.122 & 0.232 & 0.310 & 0.363 & 0.426 & 0.462 \\
		\\[-0.2cm]
		{$\text{PC2}_{t}$} & {$\beta$}  & 0.241 & 0.747 & 1.250 & 1.753 & 2.253 & 3.219 & 4.539 \\
		& {$p$-value}  & 0.045 & 0.009 & 0.006 & 0.004 & 0.003 & 0.002 & 0.000 \\
		\\[-0.2cm]
		{$\text{PC3}_{t}$} & {$\beta$}  & 0.160 & 0.046 & -0.117 & -0.237 & -0.294 & -0.276 & -0.235 \\
		& {$p$-value}  & 0.450 & 0.921 & 0.870 & 0.802 & 0.802 & 0.861 & 0.911 \\
		\\[-0.2cm]
		{$\text{PC4}_{t}$} & {$\beta$}  & 0.194 & 0.569 & 0.958 & 1.282 & 1.515 & 1.698 & 1.468 \\
		& {$p$-value}  & 0.239 & 0.092 & 0.075 & 0.095 & 0.134 & 0.257 & 0.497 \\
		\\[-0.2cm]
		{$\text{PC5}_{t}$} & {$\beta$}  & -0.171 & -0.387 & -0.814 & -1.406 & -2.057 & -3.258 & -4.514 \\
		& {$p$-value}  & 0.178 & 0.213 & 0.121 & 0.059 & 0.032 & 0.015 & 0.015 \\
		\vspace{-0.2cm}\\
		\hline
		\vspace{-0.2cm}\\
		{Adj. $R^2$\tiny{(\%)}} &   & 9.399 & 6.221 & 5.984 & 6.366 & 6.783 & 7.112 & 6.521 \\
		{Adj. $R^2$\tiny{(\%)}} & {\tiny{no $\text{TR}^{(eq)}$}}  & 6.250 & 4.075 & 3.962 & 4.300 & 4.645 & 4.895 & 4.415 \\
		{$F$-test} &   & 0.001 & 0.008 & 0.010 & 0.009 & 0.008 & 0.007 & 0.008 \\
		\\
		\hline
		\hline
		\vspace{-2.5mm}
	\end{tabular}
\end{table}

\clearpage \newpage

\begin{table}[!ht]
	\ContinuedFloat
	\caption{In-sample forecasts of Treasury returns with equity tail risk (continued)}
	\fontsize{9}{11} \selectfont
	\renewcommand{\tabcolsep}{6.2pt}  
	\sisetup{ input-symbols = {()},
		table-format = -0.2,
		table-space-text-post = ***,
		table-align-text-post = false,
		group-digits = false,
		explicit-sign}
	\centering 
	\begin{tabular}{llSSSSSSS} \hline \hline
		\vspace{-0.15cm}\\
		& & {$n=12$} & {$n=24$} & {$n=36$}  & {$n=48$} & {$n=60$} & {$n=84$} & {$n=120$} \vspace{+0.12cm}\\
		\hline
		\multicolumn{9}{l}{\rule[-2.2mm]{0mm}{6.5mm}{\textbf{Panel D: Control for Cochrane-Piazzesi (CP) factor}}}\\
		\hline
		\vspace{-0.15cm}\\
		{$\text{TR}_{t}^{(eq)}$} & {$\beta$}  & 0.365 & 0.798 & 1.303 & 1.853 & 2.425 & 3.563 & 5.033 \\
		& {$p$-value}  & 0.002 & 0.008 & 0.009 & 0.008 & 0.007 & 0.007 & 0.011 \\
		\\[-0.2cm]
		{$\text{CP}_{t}$} & {$\beta$}  & 0.473 & 1.053 & 1.668 & 2.339 & 3.026 & 4.312 & 5.847 \\
		& {$p$-value}  & 0.000 & 0.000 & 0.000 & 0.000 & 0.000 & 0.000 & 0.000 \\
		\vspace{-0.2cm}\\
		\hline
		\vspace{-0.2cm}\\
		{Adj. $R^2$\tiny{(\%)}} &   & 9.357 & 6.704 & 6.170 & 6.337 & 6.647 & 7.013 & 6.688 \\
		{Adj. $R^2$\tiny{(\%)}} & {\tiny{no $\text{TR}^{(eq)}$}}  & 6.673 & 4.871 & 4.419 & 4.499 & 4.680 & 4.842 & 4.509 \\
		{$F$-test} &   & 0.003 & 0.012 & 0.014 & 0.012 & 0.010 & 0.007 & 0.007 \\
		\\
		\vspace{-0.15cm}\\
		\hline
		\multicolumn{9}{l}{\rule[-2.2mm]{0mm}{6.5mm}{\textbf{Panel E: Control for Cieslak-Povala (CiP) factor}}}\\
		\hline
		\vspace{-0.15cm}\\
		{$\text{TR}_{t}^{(eq)}$} & {$\beta$}  & 0.438 & 0.934 & 1.507 & 2.136 & 2.792 & 4.086 & 5.721 \\
		& {$p$-value}  & 0.001 & 0.006 & 0.007 & 0.007 & 0.006 & 0.005 & 0.008 \\
		\\[-0.2cm]
		{$\text{CiP}_{t}$} & {$\beta$}  & 0.307 & 1.029 & 1.794 & 2.554 & 3.292 & 4.695 & 6.652 \\
		& {$p$-value}  & 0.005 & 0.000 & 0.000 & 0.000 & 0.000 & 0.000 & 0.000 \\
		\vspace{-0.2cm}\\
		\hline
		\vspace{-0.2cm}\\
		{Adj. $R^2$\tiny{(\%)}} &   & 6.488 & 6.669 & 6.852 & 7.158 & 7.465 & 7.857 & 7.853 \\
		{Adj. $R^2$\tiny{(\%)}} & {\tiny{no $\text{TR}^{(eq)}$}}  & 2.340 & 3.923 & 4.292 & 4.495 & 4.633 & 4.772 & 4.815 \\
		{$F$-test} &   & 0.000 & 0.003 & 0.004 & 0.003 & 0.002 & 0.002 & 0.002 \\
		\\
		\vspace{-0.15cm}\\
		\hline
		\multicolumn{9}{l}{\rule[-2.2mm]{0mm}{6.5mm}{\textbf{Panel F: Control for $\text{VIX}^{\perp}$}}}\\
		\hline
		\vspace{-0.15cm}\\
		{$\text{TR}_{t}^{(eq)}$} & {$\beta$}  & 0.460 & 1.009 & 1.638 & 2.322 & 3.032 & 4.429 & 6.206 \\
		& {$p$-value}  & 0.000 & 0.001 & 0.001 & 0.001 & 0.001 & 0.002 & 0.006 \\
		\\[-0.2cm]
		{$\text{VIX}_{t}^{\perp}$} & {$\beta$}  & 0.427 & 0.701 & 0.815 & 0.812 & 0.711 & 0.281 & -0.693 \\
		& {$p$-value}  & 0.018 & 0.241 & 0.380 & 0.392 & 0.374 & 0.371 & 0.482 \\
		\vspace{-0.2cm}\\
		\hline
		\vspace{-0.2cm}\\
		{Adj. $R^2$\tiny{(\%)}} &   & 8.578 & 4.658 & 3.580 & 3.286 & 3.247 & 3.338 & 3.323 \\
		{Adj. $R^2$\tiny{(\%)}} & {\tiny{only $\text{TR}^{(eq)}$}}  & 4.622 & 3.258 & 3.081 & 3.204 & 3.394 & 3.676 & 3.627 \\
		{$F$-test} &   & 0.000 & 0.026 & 0.121 & 0.268 & 0.445 & 0.830 & 0.708 \\
		\\
		\hline
		\hline
		\vspace{-2.5mm}
	\end{tabular}
	\begin{minipage}{15.8cm}
		Notes: This table reports the slope estimates and $p$-values from predictive regressions of one-month US Treasury bond returns on the S\&P 500 option-implied equity tail risk measure $\text{TR}^{(eq)}$. $n$ denotes the bond maturity in months. Panel A reports the results of a regression that only uses $\text{TR}^{(eq)}$ as predictor. Panels B to E report the results of regressions that control for bond return predictors identified in the literature: $\text{PC1}$ -- $\text{PC5}$ are the first five principal components extracted from the Treasury bond yields, $\text{CP}$ is the \cite{Cochrane2005} bond return predictor obtained as a linear combination of forward rates, $\text{CiP}$ is the \cite{CieslakPovala2015} risk-premium factor obtained from a decomposition of Treasury yields into inflation expectations and maturity-specific interest-rate cycles. Panel F reports the results of a regression that uses $\text{TR}^{(eq)}$ and the orthogonal component of the CBOE VIX with respect to $\text{TR}^{(eq)}$. All predictors have been normalized to have mean zero and unit variance. For all predictors we report the Newey-West $p$-values computed with a 12-lag standard error correction. In addition, for the $\text{TR}^{(eq)}$ factor used alone or alongside the principal components in the predictive regressions, we report the $p$-value \tiny{(b)} \footnotesize computed with the bootstrap procedure of \cite{Bauer2018}. For each regression we report the adjusted $R$-squared in percentage. This measure is also reported for a regression that excludes the $\text{TR}^{(eq)}$ factor as predictor in Panels A to E, and for a regression that only uses $\text{TR}^{(eq)}$ as predictor in Panel F. We also report the $p$-value of an $F$-test of the null hypothesis that the regression that includes the $\text{TR}^{(eq)}$ as predictor does not give a significantly better fit to the data than does a regression without it in Panels A to E, and the $p$-value of an $F$-test of the null hypothesis that the regression that includes $\text{VIX}^{\perp}$ does not give a significantly better fit to the data than does a regression that only uses $\text{TR}^{(eq)}$ in Panel F. The in-sample period is 1996:01--2018:12.
	\end{minipage}
\end{table}

\clearpage\newpage


\begin{table}[!ht]
	\caption{Out-of-sample forecasts of Treasury returns with equity tail risk}  
	\label{tab:OOS_results}
	\fontsize{9}{11} \selectfont
	\renewcommand{\tabcolsep}{6.8pt}
	\sisetup{ input-symbols = {()},
		table-format = -0.2,
		table-space-text-post = ***,
		table-align-text-post = false,
		group-digits = false,
		explicit-sign}
	\centering 
	\begin{tabular}{llSSSSSSS} \hline \hline
		\vspace{-0.15cm}\\
		& & {$n=12$} & {$n=24$} & {$n=36$}  & {$n=48$} & {$n=60$} & {$n=84$} & {$n=120$} \vspace{+0.12cm}\\
		\hline
		\vspace{-0.15cm}\\
		\multicolumn{9}{c}{\textbf{Panel A: Benchmark predictor is EH model (no predictability)}} \\
		\vspace{-0.30cm}\\
		\hline
		\multicolumn{9}{l}{\rule[-2.2mm]{0mm}{6.5mm}{Panel A1: Increasing windows}}\\
		\hline
		\vspace{-0.15cm}\\
		{$R^2_{OS}$ (\%)} & & 1.642 & 2.112 & 3.074 & 3.876 & 4.295 & 4.081 & 2.745 \\
		{$p$-value} & \scriptsize($CW$) & 0.118 & 0.107 & 0.093 & 0.079 & 0.066 & 0.043 & 0.018 \\
		{$p$-value \tiny{(b)}} & \scriptsize($CW$) & 0.161 & 0.213 & 0.134 & 0.111 & 0.090 & 0.073 & 0.079 \\
		{$p$-value \tiny{(b)}} & \scriptsize($R^2_{OS}$) & 0.102 & 0.097 & 0.024 & 0.014 & 0.006 & 0.003 & 0.008 \\
		\\[-0.2cm]
		\hline
		\multicolumn{9}{l}{\rule[-2.2mm]{0mm}{6.5mm}{Panel A2: Rolling windows}}\\
		\hline
		\vspace{-0.15cm}\\
		{$R^2_{OS}$ (\%)} & & 0.867 & 1.571 & 2.671 & 3.581 & 4.096 & 4.039 & 2.834 \\
		{$p$-value} & \scriptsize($CW$) & 0.129 & 0.114 & 0.094 & 0.074 & 0.056 & 0.028 & 0.006 \\
		{$p$-value \tiny{(b)}} & \scriptsize($CW$) & 0.175 & 0.224 & 0.139 & 0.107 & 0.085 & 0.063 & 0.062 \\
		{$p$-value \tiny{(b)}} & \scriptsize($R^2_{OS}$) & 0.166 & 0.135 & 0.036 & 0.018 & 0.009 & 0.004 & 0.008 \\
		\\[-0.2cm]
		\hline
		\vspace{-0.10cm}\\
		\multicolumn{9}{c}{\textbf{Panel B: Benchmark predictor is 3 PCs-only model}} \\
		\vspace{-0.30cm}\\
		\hline
		\multicolumn{9}{l}{\rule[-2.2mm]{0mm}{6.5mm}{Panel B1: Increasing windows}}\\
		\hline
		\vspace{-0.15cm}\\
		{$R^2_{OS}$ (\%)} & & 1.837 & 1.272 & 1.984 & 2.681 & 3.062 & 2.924 & 1.868 \\
		{$p$-value} & \scriptsize($CW$) & 0.039 & 0.061 & 0.060 & 0.051 & 0.041 & 0.022 & 0.011 \\
		{$p$-value \tiny{(b)}} & \scriptsize($CW$) & 0.040 & 0.067 & 0.059 & 0.067 & 0.055 & 0.058 & 0.091 \\
		{$p$-value \tiny{(b)}} & \scriptsize($R^2_{OS}$) & 0.022 & 0.045 & 0.017 & 0.013 & 0.006 & 0.008 & 0.027 \\
		\\[-0.2cm]
		\hline
		\multicolumn{9}{l}{\rule[-2.2mm]{0mm}{6.5mm}{Panel B2: Rolling windows}}\\
		\hline
		\vspace{-0.15cm}\\
		{$R^2_{OS}$ (\%)} & & 1.967 & 1.370 & 1.767 & 2.233 & 2.472 & 2.191 & 0.953 \\
		{$p$-value} & \scriptsize($CW$) & 0.056 & 0.069 & 0.053 & 0.036 & 0.024 & 0.016 & 0.078 \\
		{$p$-value \tiny{(b)}} & \scriptsize($CW$) & 0.049 & 0.066 & 0.052 & 0.052 & 0.046 & 0.060 & 0.127 \\
		{$p$-value \tiny{(b)}} & \scriptsize($R^2_{OS}$) & 0.024 & 0.045 & 0.027 & 0.021 & 0.017 & 0.022 & 0.075 \\
		\\[-0.2cm]
		\hline
		\vspace{-0.10cm}\\
		\multicolumn{9}{c}{\textbf{Panel C:  Benchmark predictor is 5 PCs-only model}} \\
		\vspace{-0.30cm}\\
		\hline
		\multicolumn{9}{l}{\rule[-2.2mm]{0mm}{6.5mm}{Panel C1: Increasing windows}}\\
		\hline
		\vspace{-0.15cm}\\
		{$R^2_{OS}$ (\%)} & & 1.067 & 1.460 & 1.975 & 2.338 & 2.481 & 2.155 & 1.139 \\
		{$p$-value} & \scriptsize($CW$) & 0.081 & 0.123 & 0.130 & 0.124 & 0.112 & 0.080 & 0.038 \\
		{$p$-value \tiny{(b)}} & \scriptsize($CW$) & 0.051 & 0.080 & 0.078 & 0.087 & 0.084 & 0.104 & 0.142 \\
		{$p$-value \tiny{(b)}} & \scriptsize($R^2_{OS}$) & 0.060 & 0.042 & 0.019 & 0.017 & 0.013 & 0.020 & 0.061 \\
		\\[-0.2cm]
		\hline
		\multicolumn{9}{l}{\rule[-2.2mm]{0mm}{6.5mm}{Panel C2: Rolling windows}}\\
		\hline
		\vspace{-0.15cm}\\
		{$R^2_{OS}$ (\%)} & & 2.759 & 3.155 & 3.448 & 3.663 & 3.654 & 3.002 & 1.602 \\
		{$p$-value} & \scriptsize($CW$) & 0.055 & 0.058 & 0.053 & 0.043 & 0.030 & 0.011 & 0.049 \\
		{$p$-value \tiny{(b)}} & \scriptsize($CW$) & 0.039 & 0.043 & 0.038 & 0.047 & 0.045 & 0.057 & 0.100 \\
		{$p$-value \tiny{(b)}} & \scriptsize($R^2_{OS}$) & 0.016 & 0.013 & 0.006 & 0.006 & 0.005 & 0.012 & 0.041 \\
		\\
		\hline
		\hline
		\vspace{-2.5mm}
	\end{tabular}
	\begin{minipage}{15.8cm}
		Notes: This table reports the \cite{CT2008} out-of-sample $R^2_{OS}$s of predicting one-month returns on the $n$-month US Treasury bond with the S\&P 500 option-implied equity tail risk measure $\text{TR}^{(eq)}$. These $R^2_{OS}$ statistics represent the percentage reduction in the MSPE for the forecasts generated by a preferred model that includes $\text{TR}^{(eq)}$ relative to a benchmark that does not use it as predictor. Panel A: the preferred model uses the $\text{TR}^{(eq)}$ factor alone, while the benchmark model complies with the expectation hypothesis that assumes no predictability of bond returns. Panel B: the preferred model includes $\text{TR}^{(eq)}$ and the first 3 principal components of bond yields, while the benchmark model only includes the 3 principal components. Panel C: the preferred model includes $\text{TR}^{(eq)}$ and the first 5 principal components of bond yields, while the benchmark model only includes the 5 principal components. Predictive regressions are recursively estimated with both expanding and rolling window approach. The out-of-sample period is 2007:07--2018:12. Statistical significance for $R^2_{OS}$ is based on the $p$-value of the \cite{CW2007} MSPE-adjusted statistic ($CW$) for testing $H_0: R^2_{OS} \le 0$ against $H_1: R^2_{OS} > 0$. For the $CW$ statistics we report both the Newey-West $p$-value computed with a 12-lag standard error correction and the $p$-value \tiny{(b)} \footnotesize computed with the bootstrap procedure of \cite{Bauer2018}. For the out-of-sample $R^2_{OS}$ we only report the bootstrap $p$-value \tiny{(b)} \footnotesize.
	\end{minipage}
\end{table}

\clearpage\newpage

\begin{table}[!ht]
	\caption{Asset allocation gains of equity tail risk}
	\label{tab:Econ_Gains}
	\fontsize{9}{11} \selectfont
	\renewcommand{\tabcolsep}{9.0pt}
	\sisetup{ input-symbols = {()},
		table-format = -0.2,
		table-space-text-post = ***,
		table-align-text-post = false,
		group-digits = false,
		explicit-sign}
	\centering 
	\begin{tabular}{lcSSSSSSS} \hline \hline
		\vspace{-0.15cm}\\
		& & {$n=12$} & {$n=24$} & {$n=36$}  & {$n=48$} & {$n=60$} & {$n=84$} & {$n=120$} \vspace{+0.12cm}\\
		\hline
		\vspace{-0.15cm}\\
		\multicolumn{9}{c}{\textbf{Panel A: Benchmark predictor is EH model (no predictability)}} \\
		\vspace{-0.30cm}\\
		\hline
		\multicolumn{9}{l}{\rule[-2.2mm]{0mm}{6.5mm}{Panel A1: Risk aversion $\gamma=3$}}\\
		\hline
		\vspace{-0.15cm}\\
		{$\Delta$ \scriptsize(\%)} &  & 0.028 & -0.438 & -0.993 & -0.890 & -0.054 & 0.571 & -3.350 \\
		{$\Theta$ \scriptsize(\%)} & & 0.028 & -0.448 & -1.025 & -0.908 & -0.002 & 1.099 & -2.711 \\
		\\[-0.2cm]
		\hline
		\multicolumn{9}{l}{\rule[-2.2mm]{0mm}{6.5mm}{Panel A2: Risk aversion $\gamma=5$}}\\
		\hline
		\vspace{-0.15cm}\\
		{$\Delta$ \scriptsize(\%)} &  & 0.023 & -0.675 & -0.410 & 0.787 & 0.481 & -2.082 & -4.887 \\
		{$\Theta$ \scriptsize(\%)} & & 0.022 & -0.713 & -0.421 & 0.877 & 0.763 & -0.513 & -3.925 \\
		\\[-0.2cm]
		\hline  
		\vspace{-0.10cm}\\
		\multicolumn{9}{c}{\textbf{Panel B: Benchmark predictor is 3 PCs-only model}} \\
		\vspace{-0.30cm}\\
		\hline
		\multicolumn{9}{l}{\rule[-2.2mm]{0mm}{6.5mm}{Panel B1: Risk aversion $\gamma=3$}}\\
		\hline
		\vspace{-0.15cm}\\
		{$\Delta$ \scriptsize(\%)} &  & -0.411 & -1.087 & -1.263 & -1.732 & -1.897 & -0.597 & -1.298 \\
		{$\Theta$ \scriptsize(\%)} & & -0.412 & -1.093 & -1.253 & -1.731 & -1.911 & -0.541 & -1.357 \\
		\\[-0.2cm]
		\hline
		\multicolumn{9}{l}{\rule[-2.2mm]{0mm}{6.5mm}{Panel B2: Risk aversion $\gamma=5$}}\\
		\hline
		\vspace{-0.15cm}\\
		{$\Delta$ \scriptsize(\%)} &  & -0.325 & -0.869 & -0.964 & -0.853 & -0.408 & -0.244 & -1.054 \\
		{$\Theta$ \scriptsize(\%)} & & -0.330 & -0.880 & -0.962 & -0.834 & -0.359 & 0.147 & -0.213 \\
		\\[-0.2cm]
		\hline
		\vspace{-0.10cm}\\
		\multicolumn{9}{c}{\textbf{Panel C:  Benchmark predictor is 5 PCs-only model}} \\
		\vspace{-0.30cm}\\
		\hline
		\multicolumn{9}{l}{\rule[-2.2mm]{0mm}{6.5mm}{Panel C1: Risk aversion $\gamma=3$}}\\
		\hline
		\vspace{-0.15cm}\\
		{$\Delta$ \scriptsize(\%)} &  & -0.687 & -0.079 & 0.627 & 1.973 & 3.559 & 3.401 & -4.345 \\
		{$\Theta$ \scriptsize(\%)} & & -0.691 & -0.080 & 0.630 & 2.029 & 3.719 & 3.525 & -4.906 \\
		\\[-0.2cm]
		\hline
		\multicolumn{9}{l}{\rule[-2.2mm]{0mm}{6.5mm}{Panel C2: Risk aversion $\gamma=5$}}\\
		\hline
		\vspace{-0.15cm}\\
		{$\Delta$ \scriptsize(\%)} &  & -0.708 & 0.048 & 0.883 & 1.862 & 2.492 & 1.005 & -2.951 \\
		{$\Theta$ \scriptsize(\%)} & & -0.719 & 0.048 & 0.895 & 2.027 & 2.909 & 0.793 & -2.262 \\
		\\
		\hline
		\hline
		\vspace{-2.5mm}
	\end{tabular}
	\begin{minipage}{15.8cm}
		Notes: This table reports the asset allocation gains of predicting one-month US Treasury bond returns with the S\&P 500 option-implied equity tail risk measure $\text{TR}^{(eq)}$. $n$ denotes the maturity of the bond in months. We assume a mean-variance investor with risk aversion $\gamma=3$ or $\gamma=5$ that every month allocates his or her wealth between a 1-month Treasury (risk-free) bond and an $n$-month Treasury bond. Investment decisions are based on the expected return forecasts of the $n$-month bond which are generated by a preferred model that includes $\text{TR}^{(eq)}$ or by a benchmark model that does not use $\text{TR}^{(eq)}$ as predictor. Panel A: the preferred model uses the $\text{TR}^{(eq)}$ factor alone, while the benchmark model complies with the expectation hypothesis that assumes no predictability of bond returns, implying that model forecasts are based on historical return means. Panel B: the preferred model includes $\text{TR}^{(eq)}$ and the first three principal components of bond yields, while the benchmark model only includes the three principal components. Panel C: the preferred model includes $\text{TR}^{(eq)}$ and the first five principal components of bond yields, while the benchmark model only includes the five principal components. Predictive models are recursively estimated with a rolling window approach. The (out-of-sample) investment period is 2007:07--2018:12. We report two measures for the performance of the preferred model relative to that of the benchmark model: certainty equivalent return gain ($\Delta$) and \cite{Goetzmann2007} manipulation-proof performance improvement ($\Theta$). Both measures are expressed in annualized percentage terms. 
	\end{minipage}
\end{table}

\clearpage\newpage


\begin{table}[!ht]
	\caption{Expected returns, forecasting performance and macroeconomic condition}
	\label{tab:ForecastPerf_and_MacroCond}
	\fontsize{9}{11} \selectfont
	\renewcommand{\tabcolsep}{7.7pt}
	\sisetup{ input-symbols = {()},
		table-format = -0.2,
		table-space-text-post = ***,
		table-align-text-post = false,
		group-digits = false,
		explicit-sign}
	\centering 
	\begin{tabular}{lSSSSSSS} \hline \hline
		\vspace{-0.15cm}\\
		& {$n=12$} & {$n=24$} & {$n=36$}  & {$n=48$} & {$n=60$} & {$n=84$} & {$n=120$} \vspace{+0.12cm}\\
		\hline
		\vspace{-0.2cm}\\
		\multicolumn{8}{l}{\textbf{Panel A: \normalfont{$\rho\big(\mathbb{E}_t[\text{RX}_{t+1}^{(n)}], \text{CFNAI}_t\big)$}}} \\
		\vspace{-0.25cm}\\
		\hline
		\vspace{-0.15cm}\\
		{$\text{TR}^{(eq)}$} & -0.664 & -0.645 & -0.634 & -0.625 & -0.614 & -0.583 & -0.534 \\
		{$\text{TR}^{(eq)}$ + 3PCs} & -0.504 & -0.457 & -0.442 & -0.442 & -0.447 & -0.450 & -0.415 \\
		{$\text{TR}^{(eq)}$ + 5PCs} & -0.373 & -0.216 & -0.151 & -0.140 & -0.153 & -0.202 & -0.277 \\
		\\
		\hline
		\vspace{-0.2cm}\\
		\multicolumn{8}{l}{\textbf{Panel B: \normalfont{$\rho\big(\mathbb{E}_t[\text{RX}_{t+1}^{(n)}], \mathbb{U}^{\text{MACRO}}_t\big)$}}} \\
		\vspace{-0.25cm}\\
		\hline
		\vspace{-0.15cm}\\
		{$\text{TR}^{(eq)}$} & 0.697 & 0.657 & 0.632 & 0.615 & 0.599 & 0.566 & 0.519 \\
		{$\text{TR}^{(eq)}$ + 3PCs} & 0.590 & 0.547 & 0.527 & 0.522 & 0.525 & 0.526 & 0.495 \\
		{$\text{TR}^{(eq)}$ + 5PCs} & 0.522 & 0.364 & 0.291 & 0.277 & 0.291 & 0.347 & 0.432 \\
		\\
		\hline
		\vspace{-0.2cm}\\
		\multicolumn{8}{l}{\textbf{Panel C: \normalfont{$\rho\big(\text{DCSPE}_t, \text{CFNAI}_t\big)$}}} \\
		\vspace{-0.25cm}\\
		\hline
		\vspace{-0.15cm}\\
		{$\text{TR}^{(eq)}$} & -0.448 & -0.277 & -0.079 & 0.056 & 0.140 & 0.235 & 0.302 \\
		{$\text{TR}^{(eq)}$ + 3PCs} & 0.444 & 0.390 & 0.357 & 0.343 & 0.335 & 0.303 & 0.184 \\
		{$\text{TR}^{(eq)}$ + 5PCs} & 0.412 & 0.407 & 0.372 & 0.351 & 0.341 & 0.313 & 0.200 \\
		\\
		\hline
		\vspace{-0.2cm}\\
		\multicolumn{8}{l}{\textbf{Panel D: \normalfont{$\rho\big(\text{DCRU}_t, \text{CFNAI}_t\big)$}}}  \\
		\vspace{-0.25cm}\\
		\hline
		\vspace{-0.15cm}\\
		{$\text{TR}^{(eq)}$} & 0.182 & 0.372 & 0.122 & -0.183 & -0.296 & -0.344 & -0.364 \\
		{$\text{TR}^{(eq)}$ + 3PCs} & 0.481 & -0.200 & -0.216 & -0.032 & -0.063 & -0.135 & -0.082 \\
		{$\text{TR}^{(eq)}$ + 5PCs} & 0.078 & 0.159 & 0.167 & -0.004 & 0.098 & -0.040 & -0.224 \\
		\\
		\hline
		\hline
		\vspace{-2.5mm}
	\end{tabular}
	\begin{minipage}{15.8cm}
		Notes: This table reports contemporaneous correlations between economic variables and the expected bond risk premia and forecasting performance obtained with the S\&P 500 option-implied equity tail risk measure $\text{TR}^{(eq)}$. $n$ denotes the bond maturity in months. Panels A and B report contemporaneous correlations between the out-of-sample forecasts of the one-month-ahead Treasury bond returns obtained by one of the three models that use $\text{TR}^{(eq)}$ as predictor and the Chicago Fed National Activity Index ($\text{CFNAI}$) and the macroeconomic uncertainty index ($\mathbb{U}^{\text{MACRO}}$) constructed by \cite{Jurado2015}. Panels C and D report contemporaneous correlations between relative forecast and portfolio performance obtained by one of the three models that use $\text{TR}^{(eq)}$ as predictor (relative to its benchmark that does not use $\text{TR}^{(eq)}$ to predict bond returns) and the CFNAI. Relative forecast performance is defined as the difference in cumulative squared prediction error (DCSPE) and portfolio performance is defined as the difference in cumulative realized utilities (DCRU). The out-of-sample evaluation period is 2007:07--2018:12. The predictive models are recursively estimated with a rolling window approach. The investor's risk aversion coefficient is $\gamma=5$. 
	\end{minipage}
\end{table}

\clearpage \newpage


\begin{table}[!ht]
	\caption{Fit diagnostics of the ATSM with equity tail risk}
	\label{tab:ATSM_Fit}
	\fontsize{9}{11} \selectfont
	\renewcommand{\tabcolsep}{8.2pt}
	\sisetup{ input-symbols = {()},
		table-format = -0.2,
		table-space-text-post = ***,
		table-align-text-post = false,
		group-digits = false,
		explicit-sign}
	\centering 
	\begin{tabular}{lccSSSSSS} \hline \hline
		\vspace{+0.05cm}\\
		\multicolumn{9}{c}{\textbf{Panel A: Equity Tail Risk ATSM}} \vspace{+0.22cm}\\ \hline
		\vspace{-0.18cm}\\
		& & & {$n=12$} & {$n=24$} & {$n=36$} & {$n=60$} & {$n=84$} & {$n=120$}  \vspace{+0.12cm}\\
		\hline
		\multicolumn{9}{l}{\rule[-2.2mm]{0mm}{6.5mm}{Panel A1: Yield Pricing Errors } }\\
		\hline
		\vspace{-0.15cm}\\
		{Mean}                & & & -0.001 &  0.000 &  0.001 &  -0.001 &  -0.001 &  -0.001  \\
		{Standard Deviation}  & & & 0.004 &  0.005 &  0.003 &  0.004 &  0.003 &  0.006 \\
		{Skewness}            & & & -0.390 &  0.843 &  0.228 &  -0.236 &  0.595 &  -0.394 \\
		{Kurtosis}            & & & 4.399 &  4.182 &  1.994 &  3.292 &  3.166 &  3.343  \\
		{$\rho(1)$}           & & & 0.867 &  0.807 &  0.909 &  0.897 &  0.839 &  0.860  \\
		{$\rho(6)$}           & & & 0.530 &  0.370 &  0.767 &  0.587 &  0.451 &  0.497 \\
		\\[-0.2cm]
		\hline
		\multicolumn{9}{l}{\rule[-2.2mm]{0mm}{6.5mm}{Panel A2: Return Pricing Errors} }\\
		\hline
		\vspace{-0.15cm}\\
		{Mean}                & & & 0.000 &  0.002 &  -0.001 &  -0.004 &  0.004 &  -0.024  \\
		{Standard Deviation}  & & & 0.047 &  0.074 &  0.069 &  0.113 &  0.117 &  0.389  \\
		{Skewness}            & & & -0.278 &  -0.501 &  -0.407 &  -0.021 &  -0.292 &  -0.234  \\
		{Kurtosis}            & & & 5.650 &  6.903 &  13.366 &  5.479 &  5.999 &  4.885 \\
		{$\rho(1)$}           & & & 0.020 &  0.050 &  0.245 &  -0.005 &  -0.055 &  -0.021   \\
		{$\rho(6)$}           & & & 0.153 &  0.214 &  0.274 &  0.031 &  0.132 &  0.052 \\
		\vspace{-0.15cm}\\
		\hline
		\vspace{+0.2cm}\\
		\multicolumn{9}{c}{\textbf{Panel B: PC-only ATSM}} \vspace{+0.22cm}\\ \hline
		\vspace{-0.18cm}\\
		& & & {$n=12$} & {$n=24$} & {$n=36$} & {$n=60$} & {$n=84$} & {$n=120$}  \vspace{+0.12cm}\\
		\hline
		\multicolumn{9}{l}{\rule[-2.2mm]{0mm}{6.5mm}{Panel B1: Yield Pricing Errors } }\\
		\hline
		\vspace{-0.15cm}\\
		{Mean}                & & & -0.004 &  -0.001 &  -0.001 &  -0.003 &  -0.003 &  -0.002  \\
		{Standard Deviation}  & & &  0.006 &  0.006 &  0.003 &  0.005 &  0.003 &  0.006  \\
		{Skewness}            & & & -0.080 &  0.772 &  -0.050 &  -0.089 &  0.121 &  -0.371 \\
		{Kurtosis}            & & & 3.875 &  3.994 &  1.838 &  3.181 &  2.301 &  3.312 \\
		{$\rho(1)$}           & & & 0.902 &  0.812 &  0.952 &  0.920 &  0.896 &  0.862  \\
		{$\rho(6)$}           & & & 0.606 &  0.398 &  0.875 &  0.649 &  0.690 &  0.551  \\
		\\[-0.2cm]
		\hline
		\multicolumn{9}{l}{\rule[-2.2mm]{0mm}{6.5mm}{Panel B2: Return Pricing Errors} }\\
		\hline
		\vspace{-0.15cm}\\
		{Mean}                & & & -0.001 &  0.002 &  -0.004 &  -0.009 &  -0.001 &  -0.013   \\
		{Standard Deviation}  & & & 0.052 &  0.076 &  0.067 &  0.128 &  0.114 &  0.383  \\
		{Skewness}            & & & -0.376 &  -0.431 &  -0.699 &  -0.078 &  0.174 &  -0.233 \\
		{Kurtosis}            & & & 5.474 &  7.669 &  13.211 &  5.848 &  6.281 &  5.275 \\
		{$\rho(1)$}           & & & 0.118 &  0.009 &  0.348 &  0.076 &  -0.192 &  -0.088 \\
		{$\rho(6)$}           & & & 0.104 &  0.218 &  0.334 &  0.003 &  0.139 &  0.036 \\
		\vspace{-0.15cm}\\
		\hline
		\hline
		\vspace{-2.5mm}
	\end{tabular}
	\begin{minipage}{15.8cm}
		Notes: This table contains the summary statistics of the pricing errors implied by the Gaussian ATSM that includes the S\&P 500 option-implied equity tail risk measure $\text{TR}^{(eq)}$ (Panel A) and by the benchmark model that only uses the first five PCs of the yield curve (Panel B). Models are estimated over the period 1996 to 2018. Reported are the sample mean, standard deviation, skewness, kurtosis and the autocorrelation coefficients of order one and six. Panels A1 and B1: properties of the yield pricing errors $\hat{u}$. Panels A2 and B2: properties of the return pricing errors $\hat{e}$. $n$ denotes the maturity of the bonds in months.
	\end{minipage}
\end{table}

\clearpage \newpage


\begin{table}[!ht]
\caption{Factor risk exposures in the ATSM with equity tail risk}
\label{tab:Wald_b}
\fontsize{9}{11} \selectfont
\renewcommand{\tabcolsep}{19.7pt} 
\sisetup{ input-symbols = (),
  				table-format = -0.2,
  				table-space-text-post = ***,
  				table-align-text-post = false,
  				group-digits = false}
\centering \begin{tabular}{lcSccSc}\hline \hline
\vspace{-0.25cm}\\
& & \multicolumn{2}{c}{\hspace{-0.5em}\rule[-2mm]{0mm}{6mm} {\textbf{Equity Tail Risk ATSM}}} & &
\multicolumn{2}{c}{\hspace{-0.5em}\rule[-2mm]{0mm}{6mm} {\textbf{PC-only ATSM}}} \\
\vspace{-0.25cm}\\
\hline
\vspace{-0.18cm}\\
{Factor} & &  {$W_{\beta_i}$~~~~~~~~~} & {$p$-value} & & {$W_{\beta_i}$~~~~~~~~~} & {$p$-value} \\
\vspace{-0.2cm}\\
\hline
\vspace{-0.1cm}\\
{$\text{TR}^{(eq)}$}       & &  9471518.154 & 0.000 & & {-~~~~~~~~~~} &   {-} \\[0.15cm]
{PC1}         		       & & 29773988.504 & 0.000 & &  31625802.379 & 0.000 \\[0.15cm]
{PC2}                      & &  5640992.750 & 0.000 & &   6114464.179 & 0.000 \\[0.15cm]
{PC3}                      & &   933067.335 & 0.000 & &    942985.226 & 0.000 \\[0.15cm]
{PC4}                      & &   174656.368 & 0.000 & &    176667.454 & 0.000 \\[0.15cm]
{PC5}                      & &    33311.223 & 0.000 & &     33261.513 & 0.000 \\
\vspace{-0.15cm}\\
\hline \hline
\vspace{-2.5mm}
\end{tabular}
\begin{minipage}{15.8cm}
Notes: This table provides the Wald statistics and corresponding $p$-values for the Wald test of whether the exposures of bond returns to a given model factor are jointly zero. Under the null $H_0: \boldsymbol\beta_i = \mathbf{0}_{N\times1}$ the $i$-th pricing factor is unspanned, i.e.~Treasury returns are not exposed to it. The test is conducted on the pricing factors of both the proposed ATSM specified with the S\&P 500 option-implied equity tail risk measure $\text{TR}^{(eq)}$, and a benchmark PC-only model specification.
\end{minipage}
\end{table}

\clearpage \newpage


\begin{table}[!ht]
\caption{Market prices of risk in the ATSM with equity tail risk}
\label{tab:Lambda}
\fontsize{9}{11} \selectfont
\renewcommand{\tabcolsep}{4.3pt}
\sisetup{ input-symbols = {()},
  				table-format = -0.2,
  				table-space-text-post = ***,
  				table-align-text-post = false,
  				group-digits = false,
  				explicit-sign}
\centering
\begin{tabular}{lSSSSSSS|SS}\hline \hline
\vspace{-0.15cm}\\
{Factor} & {$\lambda_0$} & {$\lambda_{1,1}$} & {$\lambda_{1,2}$} & {$\lambda_{1,3}$} & {$\lambda_{1,4}$} & {$\lambda_{1,5}$} & {$\lambda_{1,6}$} & {$W_{\Lambda_i}$} & {$W_{\lambda_{1_i}}$}  \\
\vspace{-0.15cm}\\
\hline
\vspace{-0.15cm}\\
{$\text{TR}^{(eq)}$}& 0.138 &  -0.164 &  0.413 &  0.152 &  -0.403 &  -0.103 &  0.227 &  12.493 &  12.190   \\
					& (0.935) &  {$($}-1.070{$)$} &  (2.454) &  (1.017) &  {$($}-2.463{$)$} &  {$($}-0.701{$)$} &  (1.482) &  (0.085) &  (0.058) \\[0.15cm]
{PC1}           	& 0.003 &  -0.054 &  0.051 &  0.004 &  -0.059 &  -0.032 &  0.018 &  12.201 &  12.193  \\
					& (0.144) &  {$($}-2.173{$)$} &  (1.901) &  (0.182) &  {$($}-2.231{$)$} &  {$($}-1.322{$)$} &  (0.713) &  (0.094) &  (0.058) \\[0.15cm]
{PC2}          		& -0.047 &  0.012 &  -0.096 &  -0.069 &  0.122 &  0.019 &  -0.097 &  15.532 &  14.973 \\
					& {$($}-1.137{$)$} &  (0.279) &  {$($}-2.073{$)$} &  {$($}-1.650{$)$} &  (2.696) &  (0.467) &  {$($}-2.262{$)$} &  (0.030) &  (0.020) \\[0.15cm]
{PC3}           	& 0.004 &  0.031 &  -0.150 &  0.009 &  0.038 &  0.074 &  -0.068 &  12.150 &  12.043\\
					& (0.081) &  (0.646) &  {$($}-2.889{$)$} &  (0.183) &  (0.745) &  (1.573) &  {$($}-1.410{$)$} &  (0.096) &  (0.061) \\[0.15cm]
{PC4}          		& 0.020 &  -0.045 &  -0.064 &  0.086 &  -0.028 &  -0.037 &  -0.074 &  18.343 &  17.951 \\
					& (0.591) &  {$($}-1.317{$)$} &  {$($}-1.873{$)$} &  (2.499) &  {$($}-0.813{$)$} &  {$($}-1.073{$)$} &  {$($}-2.156{$)$} &  (0.011) &  (0.006)\\[0.15cm]
{PC5}          		& -0.071 &  -0.003 &  -0.040 &  -0.127 &  0.050 &  -0.030 &  -0.141 &  16.187 &  14.546 \\
					& {$($}-1.403{$)$} &  {$($}-0.049{$)$} &  {$($}-0.748{$)$} &  {$($}-2.484{$)$} &  (0.939) &  {$($}-0.591{$)$} &  {$($}-2.733{$)$} &  (0.023) &  (0.024) \\
\vspace{-0.15cm}\\
\hline \hline
\vspace{-2.5mm}
\end{tabular}
\begin{minipage}{15.8cm}
Notes: This table provides the estimates of the market price of risk parameters $\boldsymbol\lambda_0$ and $\boldsymbol\lambda_1$ in equation (\ref{eq:RiskPrices}) for the Gaussian ATSM specified with the S\&P 500 option-implied equity tail risk measure $\text{TR}^{(eq)}$. Estimated $t$-statistics are reported in parentheses. Wald statistics for tests of the rows of $\boldsymbol\Lambda$ and of $\boldsymbol\lambda_1$ being different from zero are reported along each row, with the corresponding $p$-values in parentheses below. The null hypothesis underlying $W_{\Lambda_i}$ is that the risk related to a given factor is not priced in the term structure model. The null hypothesis underlying $W_{\lambda_{1_i}}$ is that the price of risk associated with a given factor does not vary over time.      
\end{minipage}
\end{table}

\clearpage\newpage


\begin{table}[!ht]
	\caption{Market price of equity tail risk with GX procedure}  
	\label{tab:GX_RiskPremia_US}
	\fontsize{9}{11} \selectfont
	\renewcommand{\tabcolsep}{6.7pt} 
	\sisetup{ input-symbols = {()},
		table-format = -0.2,
		table-space-text-post = ***,
		table-align-text-post = false,
		group-digits = false,
		explicit-sign}
	\centering 
	\begin{tabular}{lSSSSSSSS} \hline \hline
		\vspace{-0.15cm}\\
		& {$p=1$} & {$p=2$} & {$p=3$}  & {$p=4$} & {$p=5$} & {$p=6$} & {$p=7$} & {$p=8$}\vspace{+0.12cm}\\ 
		\hline
		\vspace{-0.15cm}\\
		{$\gamma_g$} & 0.028{*} & 0.045{*} & 0.050{*} & 0.053{*} & 0.053{*} & 0.053{*} & 0.052{*} & 0.052{*} \\
		& (0.015) & (0.027) & (0.029) & (0.030) & (0.030) & (0.030) & (0.029) & (0.029) \\
		\vspace{-0.3cm}\\
		{$R^{2}_g$} & 0.038 & 0.079 & 0.085 & 0.087 & 0.093 & 0.094 & 0.097 & 0.097 \\
		\vspace{-0.3cm}\\
		{$p$-value} & 0.000 & 0.006 & 0.011 & 0.023 & 0.019 & 0.025 & 0.041 & 0.024 \\
		{$g$ weak} &  &  &  &  &  &  &  &  \\
		\\
		\hline
		\hline
		\vspace{-2.5mm}
	\end{tabular}
	\begin{minipage}{15.8cm}
		Notes: This table reports the results of the three-pass regression procedure of \cite{Giglio2019} to estimate the risk premium of the S\&P 500 option-implied equity tail risk measure $\text{TR}^{(eq)}$ in the US Treasury bond market. $p$ denotes the number of latent factors used in the three-pass estimator. For each number of latent factors, we report the estimate of the market price of risk $\gamma_g$ of the observable factor $g = \text{TR}^{(eq)}$ with standard errors in parentheses, the $R$-squared of the time series regression of the observable factor $g$ onto the $p$ latent factors, and the $p$-value of the Wald test of testing the null hypothesis that the observable factor is weak. * (resp. **, and ***) denote statistical significance at the 10\% (resp. 5\%, and 1\%) level. 
	\end{minipage}
\end{table}

\clearpage \newpage

\begin{table}[!ht]
	\caption{In-sample forecasts of international bond returns with equity tail risk}
	\label{tab:IS_inter_vs_US}
	\fontsize{9}{11} \selectfont
	\renewcommand{\tabcolsep}{7.4pt}  
	\sisetup{ input-symbols = {()},
		table-format = -0.2,
		table-space-text-post = ***,
		table-align-text-post = false,
		group-digits = false,
		explicit-sign}
	\centering 
	\begin{tabular}{llSSSSSSS} \hline \hline
		\vspace{-0.15cm}\\
		& & {$n=12$} & {$n=24$} & {$n=36$}  & {$n=48$} & {$n=60$} & {$n=84$} & {$n=120$} \vspace{+0.12cm}\\
		\hline
		\multicolumn{9}{l}{\rule[-2.2mm]{0mm}{6.5mm}{\textbf{Panel A: No control for bond return forecasting factors}}}\\  
		\hline
		\vspace{-0.15cm}\\
		{UK} & {$\beta$}  & 0.975 & 1.806 & 2.450 & 3.002 & 3.549 & 4.520 & 5.754 \\
		& {$p$-value}  & 0.001 & 0.000 & 0.000 & 0.000 & 0.000 & 0.000 & 0.000 \\
		& {$p$-value \tiny{(b)}} & 0.000 & 0.000 & 0.000 & 0.000 & 0.000 & 0.000 & 0.000 \\
		\\[-0.2cm]
		{DE} & {$\beta$}  & 0.542 & 1.030 & 1.448 & 1.827 & 2.172 & 2.758 & 3.378 \\
		& {$p$-value}  & 0.002 & 0.001 & 0.001 & 0.001 & 0.002 & 0.004 & 0.009 \\
		& {$p$-value \tiny{(b)}} & 0.000 & 0.001 & 0.001 & 0.003 & 0.002 & 0.010 & 0.026 \\
		\\[-0.2cm]
		{CH} & {$\beta$}  & 0.547 & 0.654 & 0.799 & 1.033 & 1.295 & 1.776 & 2.397 \\
		& {$p$-value}  & 0.036 & 0.002 & 0.000 & 0.001 & 0.001 & 0.002 & 0.009 \\
		& {$p$-value \tiny{(b)}} & 0.000 & 0.011 & 0.035 & 0.037 & 0.032 & 0.034 & 0.047 \\
		\\[-0.2cm]
		{FR} & {$\beta$}  & 0.468 & 0.893 & 1.231 & 1.511 & 1.754 & 2.170 & 2.667 \\
		& {$p$-value}  & 0.003 & 0.003 & 0.004 & 0.007 & 0.012 & 0.035 & 0.097 \\
		& {$p$-value \tiny{(b)}} & 0.003 & 0.008 & 0.010 & 0.014 & 0.024 & 0.042 & 0.083 \\
		\\[-0.2cm]
		{IT} & {$\beta$}  & 0.402 & 0.454 & 0.679 & 1.008 & 1.336 & 1.811 & 2.084 \\
		& {$p$-value}  & 0.103 & 0.343 & 0.350 & 0.285 & 0.237 & 0.205 & 0.240 \\
		& {$p$-value \tiny{(b)}} & 0.116 & 0.417 & 0.402 & 0.317 & 0.269 & 0.245 & 0.285 \\
		\\[-0.2cm]
		{ES} & {$\beta$}  & 0.619 & 1.179 & 1.662 & 2.079 & 2.445 & 3.067 & 3.866 \\
		& {$p$-value}  & 0.000 & 0.000 & 0.000 & 0.001 & 0.002 & 0.005 & 0.011 \\
		& {$p$-value \tiny{(b)}} & 0.017 & 0.045 & 0.054 & 0.072 & 0.085 & 0.112 & 0.156 \\
		\\[-0.15cm]
		\vspace{-0.15cm}\\
		\hline
		\multicolumn{9}{l}{\rule[-2.2mm]{0mm}{6.5mm}{\textbf{Panel B: Control for yield curve factors with 3 PCs (country-specific)}}}\\
		\hline
		\vspace{-0.15cm}\\
		{UK} & {$\beta$}  & 0.966 & 1.758 & 2.351 & 2.862 & 3.376 & 4.289 & 5.460 \\
		& {$p$-value}  & 0.001 & 0.000 & 0.000 & 0.000 & 0.000 & 0.000 & 0.000 \\
		& {$p$-value \tiny{(b)}} & 0.000 & 0.000 & 0.000 & 0.000 & 0.000 & 0.000 & 0.001 \\
		\\[-0.2cm]
		{DE} & {$\beta$}  & 0.548 & 1.009 & 1.393 & 1.732 & 2.034 & 2.527 & 3.001 \\
		& {$p$-value}  & 0.002 & 0.001 & 0.001 & 0.002 & 0.003 & 0.007 & 0.022 \\
		& {$p$-value \tiny{(b)}} & 0.000 & 0.000 & 0.001 & 0.003 & 0.004 & 0.010 & 0.035 \\
		\\[-0.2cm]
		{CH} & {$\beta$}  & 0.559 & 0.693 & 0.828 & 1.008 & 1.194 & 1.523 & 1.992 \\
		& {$p$-value}  & 0.034 & 0.003 & 0.001 & 0.002 & 0.004 & 0.011 & 0.036 \\
		& {$p$-value \tiny{(b)}} & 0.000 & 0.007 & 0.030 & 0.041 & 0.041 & 0.055 & 0.074 \\
		\\[-0.2cm]
		{FR} & {$\beta$}  & 0.452 & 0.830 & 1.099 & 1.295 & 1.447 & 1.678 & 1.923 \\
		& {$p$-value}  & 0.007 & 0.010 & 0.015 & 0.028 & 0.050 & 0.123 & 0.265 \\
		& {$p$-value \tiny{(b)}} & 0.000 & 0.002 & 0.007 & 0.019 & 0.035 & 0.094 & 0.222 \\
		\\[-0.2cm]
		{IT} & {$\beta$}  & 0.446 & 0.597 & 0.841 & 1.117 & 1.344 & 1.549 & 1.374 \\
		& {$p$-value}  & 0.111 & 0.283 & 0.312 & 0.288 & 0.272 & 0.290 & 0.427 \\
		& {$p$-value \tiny{(b)}} & 0.068 & 0.269 & 0.299 & 0.279 & 0.260 & 0.318 & 0.469 \\
		\\[-0.2cm]
		{ES} & {$\beta$}  & 0.500 & 1.015 & 1.416 & 1.746 & 2.031 & 2.512 & 3.127 \\
		& {$p$-value}  & 0.003 & 0.003 & 0.008 & 0.014 & 0.020 & 0.031 & 0.050 \\
		& {$p$-value \tiny{(b)}} & 0.051 & 0.082 & 0.111 & 0.143 & 0.170 & 0.199 & 0.253 \\
		\\
		\hline
		\vspace{-2.5mm}
	\end{tabular}
\end{table}

\clearpage \newpage

\begin{table}[!ht]
	\ContinuedFloat
	\caption{In-sample forecasts of international bond returns with equity tail risk (continued)}
	\fontsize{9}{11} \selectfont
	\renewcommand{\tabcolsep}{7.4pt}  
	\sisetup{ input-symbols = {()},
		table-format = -0.2,
		table-space-text-post = ***,
		table-align-text-post = false,
		group-digits = false,
		explicit-sign}
	\centering 
	\begin{tabular}{llSSSSSSS} \hline \hline
		\vspace{-0.15cm}\\
		& & {$n=12$} & {$n=24$} & {$n=36$}  & {$n=48$} & {$n=60$} & {$n=84$} & {$n=120$} \vspace{+0.12cm}\\
		\hline		
		\multicolumn{9}{l}{\rule[-2.2mm]{0mm}{6.5mm}{\textbf{Panel C: Control for yield curve factors with 5 PCs (country-specific)}}}\\
		\hline
		\vspace{-0.15cm}\\
		{UK} & {$\beta$}  & 0.941 & 1.752 & 2.343 & 2.877 & 3.431 & 4.481 & 5.876 \\
		& {$p$-value}  & 0.002 & 0.000 & 0.000 & 0.000 & 0.000 & 0.000 & 0.001 \\
		& {$p$-value \tiny{(b)}} & 0.000 & 0.000 & 0.000 & 0.000 & 0.000 & 0.000 & 0.000 \\
		\\[-0.2cm]
		{DE} & {$\beta$}  & 0.479 & 0.908 & 1.279 & 1.615 & 1.915 & 2.399 & 2.846 \\
		& {$p$-value}  & 0.000 & 0.000 & 0.000 & 0.001 & 0.002 & 0.006 & 0.023 \\
		& {$p$-value \tiny{(b)}} & 0.000 & 0.001 & 0.003 & 0.006 & 0.008 & 0.017 & 0.044 \\
		\\[-0.2cm]
		{CH} & {$\beta$}  & 0.464 & 0.605 & 0.767 & 0.939 & 1.081 & 1.281 & 1.585 \\
		& {$p$-value}  & 0.014 & 0.001 & 0.001 & 0.005 & 0.010 & 0.027 & 0.080 \\
		& {$p$-value \tiny{(b)}} & 0.000 & 0.018 & 0.047 & 0.057 & 0.068 & 0.110 & 0.155 \\
		\\[-0.2cm]
		{FR} & {$\beta$}  & 0.441 & 0.817 & 1.093 & 1.298 & 1.454 & 1.673 & 1.856 \\
		& {$p$-value}  & 0.016 & 0.015 & 0.017 & 0.026 & 0.047 & 0.127 & 0.297 \\
		& {$p$-value \tiny{(b)}} & 0.000 & 0.003 & 0.007 & 0.017 & 0.034 & 0.096 & 0.234 \\
		\\[-0.2cm]
		{IT} & {$\beta$}  & 0.586 & 0.837 & 1.134 & 1.426 & 1.657 & 1.900 & 1.882 \\
		& {$p$-value}  & 0.027 & 0.099 & 0.123 & 0.119 & 0.118 & 0.140 & 0.237 \\
		& {$p$-value \tiny{(b)}} & 0.017 & 0.115 & 0.155 & 0.170 & 0.170 & 0.214 & 0.331 \\
		\\[-0.2cm]
		{ES} & {$\beta$}  & 0.678 & 1.189 & 1.605 & 2.009 & 2.395 & 3.085 & 3.957 \\
		& {$p$-value}  & 0.008 & 0.009 & 0.012 & 0.015 & 0.017 & 0.027 & 0.050 \\
		& {$p$-value \tiny{(b)}} & 0.006 & 0.039 & 0.071 & 0.094 & 0.100 & 0.118 & 0.155 \\
		\\
		\hline
		\hline
		\vspace{-2.5mm}
	\end{tabular}
	\begin{minipage}{15.8cm}
		\footnotesize
		Notes: This table reports the slope estimates and $p$-values associated with the S\&P 500 option-implied equity tail risk measure $\text{TR}^{(eq)}$ used in return predictive regressions of Treasury bonds of United Kingdom (UK), Germany (DE), Switzerland (CH), France (FR), Italy (IT), and Spain (ES). $n$ denotes the bond maturity in months. Panel A reports the results of a regression that only uses $\text{TR}_t^{(eq)}$ as predictor. Panel B (resp.~C) reports the results of a predictive regression that controls for country-specific yield curve factors represented by the first three (resp.~five) principal components of Treasury bond yields. Predictors have been normalized to have mean zero and unit variance. We report the Newey-West $p$-values computed with a 12-lag standard error correction, and the $p$-value \tiny{(b)} \footnotesize computed with the bootstrap procedure of \cite{Bauer2018}. The in-sample period is 1996:01--2018:12.
	\end{minipage}
\end{table}

\clearpage \newpage

\begin{table}[!ht]
	\caption{Out-of-sample forecasts of international bond returns with equity tail risk}
	\label{tab:OOS_inter_vs_US}
	\fontsize{9}{11} \selectfont
	\renewcommand{\tabcolsep}{7.6pt}  
	\sisetup{ input-symbols = {()},
		table-format = -0.2,
		table-space-text-post = ***,
		table-align-text-post = false,
		group-digits = false,
		explicit-sign}
	\centering 
	\begin{tabular}{llSSSSSSS} \hline \hline
		\vspace{-0.15cm}\\
		& & {$n=12$} & {$n=24$} & {$n=36$}  & {$n=48$} & {$n=60$} & {$n=84$} & {$n=120$} \vspace{+0.12cm}\\
		\hline
		\multicolumn{9}{l}{\rule[-2.2mm]{0mm}{6.5mm}{\textbf{Panel A: No control for bond return forecasting factors}}}\\  
		\hline
		\vspace{-0.15cm}\\
		{UK} & {$R^2_{OS}$ (\%)}  & 24.245 & 18.207 & 14.287 & 12.136 & 10.904 & 8.411 & 5.991 \\
		& {$p$-value}  & 0.103 & 0.078 & 0.064 & 0.056 & 0.051 & 0.044 & 0.034 \\
		\\[-0.3cm]
		& {$\Delta$ \scriptsize(\%)} & 0.009 & 0.243 & 1.012 & 2.093 & 3.433 & 1.786 & -4.959 \\
		& {$\Theta$ \scriptsize(\%)} & 0.007 & 0.264 & 1.110 & 2.284 & 3.774 & 2.419 & -5.765 \\
		\\[-0.1cm]
		{DE} & {$R^2_{OS}$ (\%)}  & 9.637 & 7.644 & 6.923 & 6.498 & 6.038 & 4.941 & 3.278 \\
		& {$p$-value}  & 0.046 & 0.046 & 0.041 & 0.034 & 0.028 & 0.020 & 0.014 \\
		\\[-0.3cm]
		& {$\Delta$ \scriptsize(\%)} & -0.011 & 0.070 & 0.188 & 0.420 & 0.785 & 1.359 & 0.121 \\
		& {$\Theta$ \scriptsize(\%)} & -0.011 & 0.072 & 0.220 & 0.505 & 0.986 & 1.858 & 0.800 \\
		\\[-0.1cm]
		{CH} & {$R^2_{OS}$ (\%)}  & 9.422 & 5.735 & 2.465 & 2.594 & 3.182 & 3.362 & 2.476 \\
		& {$p$-value}  & 0.100 & 0.056 & 0.033 & 0.032 & 0.030 & 0.016 & 0.001 \\
		\\[-0.3cm]
		& {$\Delta$ \scriptsize(\%)} & 0.105 & 0.332 & 0.061 & -0.192 & 0.070 & 0.112 & 1.241 \\
		& {$\Theta$ \scriptsize(\%)} & 0.104 & 0.330 & 0.059 & -0.251 & 0.028 & 0.183 & 1.972 \\
		\\[-0.1cm]
		{FR} & {$R^2_{OS}$ (\%)}  & 7.269 & 5.474 & 4.397 & 3.480 & 2.757 & 1.763 & 0.754 \\
		& {$p$-value}  & 0.102 & 0.116 & 0.118 & 0.117 & 0.120 & 0.139 & 0.196 \\
		\\[-0.3cm]
		& {$\Delta$ \scriptsize(\%)} & -0.001 & 0.005 & 0.080 & 0.140 & 0.140 & 0.321 & -3.606 \\
		& {$\Theta$ \scriptsize(\%)} & -0.001 & 0.004 & 0.093 & 0.178 & 0.250 & 0.640 & -4.526 \\
		\\[-0.15cm]
		\vspace{-0.15cm}\\
		\hline
		\multicolumn{9}{l}{\rule[-2.2mm]{0mm}{6.5mm}{\textbf{Panel B: Control for yield curve factors with 3 PCs (country-specific)}}}\\
		\hline
		\vspace{-0.15cm}\\
		{UK} & {$R^2_{OS}$ (\%)}  & 20.834 & 14.966 & 11.143 & 9.289 & 8.254 & 6.173 & 4.187 \\
		& {$p$-value}  & 0.096 & 0.097 & 0.097 & 0.090 & 0.081 & 0.063 & 0.049 \\
		\\[-0.3cm]
		& {$\Delta$ \scriptsize(\%)} & 0.422 & 0.449 & 0.526 & 1.553 & 1.649 & 0.868 & -1.859 \\
		& {$\Theta$ \scriptsize(\%)} & 0.435 & 0.503 & 0.621 & 1.680 & 1.794 & 1.076 & -0.406 \\
		\\[-0.1cm]
		{DE} & {$R^2_{OS}$ (\%)}  & 7.717 & 5.934 & 5.218 & 4.500 & 3.714 & 2.205 & 0.545 \\
		& {$p$-value}  & 0.084 & 0.103 & 0.105 & 0.101 & 0.096 & 0.089 & 0.112 \\
		\\[-0.3cm]
		& {$\Delta$ \scriptsize(\%)} & 0.285 & 1.671 & 2.281 & 2.433 & 2.134 & 1.769 & 0.151 \\
		& {$\Theta$ \scriptsize(\%)} & 0.284 & 1.677 & 2.301 & 2.497 & 2.243 & 2.053 & 0.123 \\
		\\[-0.1cm]
		{CH} & {$R^2_{OS}$ (\%)}  & 8.870 & 5.372 & 2.211 & 1.918 & 1.832 & 0.927 & -0.577 \\
		& {$p$-value}  & 0.102 & 0.071 & 0.081 & 0.102 & 0.111 & 0.106 & 0.223 \\
		\\[-0.3cm]
		& {$\Delta$ \scriptsize(\%)} & 0.839 & 0.566 & 1.104 & 1.175 & 1.184 & 0.941 & 0.235 \\
		& {$\Theta$ \scriptsize(\%)} & 0.839 & 0.572 & 1.120 & 1.203 & 1.232 & 1.056 & 0.086 \\
		\\[-0.1cm]
		{FR} & {$R^2_{OS}$ (\%)}  & 6.493 & 4.489 & 3.136 & 1.950 & 1.009 & -0.249 & -1.334 \\
		& {$p$-value}  & 0.115 & 0.151 & 0.173 & 0.197 & 0.231 & 0.339 & 0.621 \\
		\\[-0.3cm]
		& {$\Delta$ \scriptsize(\%)} & 0.017 & -0.239 & -0.194 & -0.196 & -0.594 & -2.188 & -4.100 \\
		& {$\Theta$ \scriptsize(\%)} & 0.019 & -0.237 & -0.153 & -0.066 & -0.359 & -1.748 & -3.270 \\
		\\
		\hline
	\end{tabular}
\end{table}

\clearpage \newpage

\begin{table}[!ht]
	\ContinuedFloat
	\caption{Out-of-sample forecasts of international bond returns with equity tail risk (continued)}
	\fontsize{9}{11} \selectfont
	\renewcommand{\tabcolsep}{7.6pt}  
	\sisetup{ input-symbols = {()},
		table-format = -0.2,
		table-space-text-post = ***,
		table-align-text-post = false,
		group-digits = false,
		explicit-sign}
	\centering 
	\begin{tabular}{llSSSSSSS} \hline \hline
		\vspace{-0.15cm}\\
		& & {$n=12$} & {$n=24$} & {$n=36$}  & {$n=48$} & {$n=60$} & {$n=84$} & {$n=120$} \vspace{+0.12cm}\\
		\hline
		\multicolumn{9}{l}{\rule[-2.2mm]{0mm}{6.5mm}{\textbf{Panel C: Control for yield curve factors with 5 PCs (country-specific)}}}\\
		\hline
		\vspace{-0.15cm}\\
		{UK} & {$R^2_{OS}$ (\%)}  & 20.461 & 15.758 & 12.521 & 11.005 & 10.215 & 8.208 & 5.754 \\
		& {$p$-value}  & 0.096 & 0.086 & 0.080 & 0.072 & 0.063 & 0.049 & 0.043 \\
		\\[-0.3cm]
		& {$\Delta$ \scriptsize(\%)} & 0.481 & 0.701 & 0.497 & 1.682 & 2.288 & 0.996 & 1.044 \\
		& {$\Theta$ \scriptsize(\%)} & 0.494 & 0.761 & 0.586 & 1.773 & 2.266 & 0.793 & 0.954 \\
		\\[-0.1cm]
		{DE} & {$R^2_{OS}$ (\%)}  & 10.968 & 8.402 & 6.647 & 5.373 & 4.320 & 2.580 & 0.694 \\
		& {$p$-value}  & 0.030 & 0.048 & 0.051 & 0.051 & 0.051 & 0.054 & 0.107 \\
		\\[-0.3cm]
		& {$\Delta$ \scriptsize(\%)} & 0.036 & -0.207 & -0.311 & 0.138 & 0.428 & 0.053 & -0.889 \\
		& {$\Theta$ \scriptsize(\%)} & 0.036 & -0.209 & -0.294 & 0.190 & 0.473 & 0.197 & -1.056 \\
		\\[-0.1cm]
		{CH} & {$R^2_{OS}$ (\%)}  & 8.631 & 3.860 & 1.082 & 0.770 & 0.775 & 0.395 & -0.737 \\
		& {$p$-value}  & 0.059 & 0.027 & 0.060 & 0.097 & 0.116 & 0.136 & 0.484 \\
		\\[-0.3cm]
		& {$\Delta$ \scriptsize(\%)} & -0.114 & 0.398 & 0.593 & 0.543 & 0.611 & 0.639 & 0.112 \\
		& {$\Theta$ \scriptsize(\%)} & -0.114 & 0.397 & 0.605 & 0.562 & 0.617 & 0.623 & -0.176 \\
		\\[-0.1cm]
		{FR} & {$R^2_{OS}$ (\%)}  & 7.846 & 5.277 & 3.554 & 2.268 & 1.283 & -0.110 & -1.416 \\
		& {$p$-value}  & 0.103 & 0.149 & 0.179 & 0.204 & 0.233 & 0.334 & 0.666 \\
		\\[-0.3cm]
		& {$\Delta$ \scriptsize(\%)} & -0.058 & -0.101 & 0.766 & 1.339 & 0.783 & -0.557 & -3.398 \\
		& {$\Theta$ \scriptsize(\%)} & -0.057 & -0.091 & 0.801 & 1.483 & 1.050 & -0.090 & -3.877 \\
		\\[-0.1cm]
		\hline
		\hline
		\vspace{-2.5mm}
	\end{tabular}
	\begin{minipage}{15.8cm}
		Notes: This table reports the \cite{CT2008} out-of-sample $R^2_{OS}$s of predicting one-month Treasury bond returns in United Kingdom (UK), Germany (DE), Switzerland (CH), and France (FR) with the S\&P 500 option-implied equity tail risk measure $\text{TR}^{(eq)}$. These $R^2_{OS}$ statistics represent the percentage reduction in the MSPE for the forecasts generated by a preferred model that includes $\text{TR}^{(eq)}$ relative to a benchmark that does not use it as predictor. The preferred model uses the $\text{TR}^{(eq)}$ factor alone in Panel A, and alongside the country-specific first three (resp.~five) principal components of bond yields in Panel B (resp.~C). Statistical significance for $R^2_{OS}$ is based on the \cite{CW2007} MSPE-adjusted statistic, for which we report Newey-West $p$-values computed with a 12-lag standard error correction. To assess the portfolio performance afforded by $\text{TR}^{(eq)}$ relative to the benchmark models, we report the certainty equivalent return gain ($\Delta$) and \cite{Goetzmann2007} manipulation-proof performance improvement ($\Theta$) in annualized percentage terms. The out-of-sample period is 2007:07--2018:12.  Predictive regressions are recursively estimated with a rolling window approach. The investor's risk aversion coefficient $\gamma$ is set equal to 5.
	\end{minipage}
\end{table}

\clearpage \newpage

\begin{table}[!ht]
	\caption{Market price of equity tail risk in international bond markets}
	\label{tab:GX_RiskPremia_inter_vs_US}
	\fontsize{9}{11} \selectfont
	\renewcommand{\tabcolsep}{5.4pt}  
	\sisetup{ input-symbols = {()},
		table-format = -0.2,
		table-space-text-post = ***,
		table-align-text-post = false,
		group-digits = false,
		explicit-sign}
	\centering 
	\begin{tabular}{llSSSSSSSS} \hline \hline
		\vspace{-0.15cm}\\
		& & {$p=1$} & {$p=2$} & {$p=3$}  & {$p=4$} & {$p=5$} & {$p=6$} & {$p=7$} & {$p=8$} \vspace{+0.12cm}\\
		\hline
		\vspace{-0.15cm}\\
		{UK} & {$\gamma_g$} & 0.045{**} & 0.068 & 0.082{*} & 0.080 & 0.082 & 0.085 & 0.085 & 0.086 \\
		& & (0.023) & (0.042) & (0.049) & (0.050) & (0.051) & (0.052) & (0.052) & (0.053) \\
		\vspace{-0.3cm}\\
		& {$R^{2}_g$} & 0.066 & 0.152 & 0.218 & 0.237 & 0.254 & 0.259 & 0.260 & 0.260 \\
		\vspace{-0.3cm}\\
		& {$p$-value} & 0.000 & 0.001 & 0.000 & 0.000 & 0.000 & 0.000 & 0.000 & 0.000 \\
		& {$g$ weak} &  &  &  &  &  &  &  &  \\
		\\[-0.2cm]
		{DE} & {$\gamma_g$} & 0.066{**} & 0.080{*} & 0.086{*} & 0.105{*} & 0.104{*} & 0.110{*} & 0.107{*} & 0.108{*} \\
		& & (0.033) & (0.045) & (0.051) & (0.059) & (0.059) & (0.063) & (0.062) & (0.063) \\
		\vspace{-0.3cm}\\
		& {$R^{2}_g$} & 0.074 & 0.156 & 0.212 & 0.238 & 0.241 & 0.255 & 0.257 & 0.260 \\
		\vspace{-0.3cm}\\
		& {$p$-value} & 0.002 & 0.004 & 0.001 & 0.000 & 0.000 & 0.000 & 0.000 & 0.000 \\
		& {$g$ weak} &  &  &  &  &  &  &  &  \\
		\\[-0.2cm]
		{CH} & {$\gamma_g$} & 0.053{**} & 0.059{**} & 0.058{**} & 0.075{**} & 0.111{**} & 0.115{**} & 0.125{*} & 0.136{*} \\
		& & (0.023) & (0.025) & (0.026) & (0.037) & (0.047) & (0.054) & (0.068) & (0.075) \\
		\vspace{-0.3cm}\\
		& {$R^{2}_g$} & 0.047 & 0.065 & 0.080 & 0.129 & 0.158 & 0.159 & 0.164 & 0.173 \\
		\vspace{-0.3cm}\\
		& {$p$-value} & 0.001 & 0.000 & 0.000 & 0.001 & 0.001 & 0.001 & 0.000 & 0.002 \\
		& {$g$ weak} &  &  &  &  &  &  &  &  \\
		\\[-0.2cm]
		{FR} & {$\gamma_g$} & 0.044{**} & 0.096{**} & 0.096{**} & 0.110{**} & 0.143{**} & 0.168{**} & 0.171{**} & 0.176{**} \\
		& & (0.022) & (0.047) & (0.048) & (0.055) & (0.064) & (0.071) & (0.073) & (0.073) \\
		\vspace{-0.3cm}\\
		& {$R^{2}_g$} & 0.035 & 0.138 & 0.142 & 0.163 & 0.194 & 0.215 & 0.218 & 0.225 \\
		\vspace{-0.3cm}\\
		& {$p$-value} & 0.006 & 0.004 & 0.005 & 0.014 & 0.007 & 0.007 & 0.013 & 0.008 \\
		& {$g$ weak} &  &  &  &  &  &  &  &  \\
		\\[-0.2cm]
		{IT} & {$\gamma_g$} & -0.003 & -0.001 & 0.001 & 0.004 & 0.007 & 0.007 & 0.007 & 0.007 \\
		& & (0.011) & (0.013) & (0.012) & (0.014) & (0.016) & (0.015) & (0.016) & (0.016) \\
		\vspace{-0.3cm}\\
		& {$R^{2}_g$} & 0.000 & 0.006 & 0.008 & 0.018 & 0.022 & 0.023 & 0.023 & 0.024 \\
		\vspace{-0.3cm}\\
		& {$p$-value} & 0.803 & 0.487 & 0.592 & 0.717 & 0.748 & 0.293 & 0.245 & 0.027 \\
		& {$g$ weak} &  &  &  &  &  &  &  &  \\
		\\[-0.2cm]
		{ES} & {$\gamma_g$} & 0.008 & 0.024 & 0.026 & 0.040 & 0.049 & 0.106{**} & 0.108{*} & 0.119{**} \\
		& & (0.008) & (0.018) & (0.021) & (0.030) & (0.038) & (0.054) & (0.056) & (0.057) \\
		\vspace{-0.3cm}\\
		& {$R^{2}_g$} & 0.002 & 0.027 & 0.037 & 0.042 & 0.044 & 0.074 & 0.074 & 0.090 \\
		\vspace{-0.3cm}\\
		& {$p$-value} & 0.328 & 0.227 & 0.370 & 0.550 & 0.695 & 0.441 & 0.582 & 0.381 \\
		& {$g$ weak} &  &  &  &  &  &  &  &  \\
		\\
		\hline
		\hline
		\vspace{-2.5mm}
	\end{tabular}
	\begin{minipage}{15.8cm}
		Notes: This table reports the results of the three-pass regression procedure of \cite{Giglio2019} to estimate the risk premium of the S\&P 500 option-implied equity tail risk measure $\text{TR}^{(eq)}$ in the Treasury bond market of United Kingdom (UK), Germany (DE), Switzerland (CH), France (FR), Italy (IT), and Spain (ES). $p$ denotes the number of latent factors used in the three-pass estimator. For each number of latent factors, we report the estimate of the market price of risk $\gamma_g$ of the observable factor $g = \text{TR}^{(eq)}$ with standard errors in parentheses, the $R$-squared of the time series regression of the observable factor $g$ onto the $p$ latent factors, and the $p$-value of the Wald test of testing the null hypothesis that the observable factor is weak. * (resp. **, and ***) denote statistical significance at the 10\% (resp. 5\%, and 1\%) level. 
	\end{minipage}
\end{table}

\clearpage \newpage

\begin{table}[!ht]
	\caption{In-sample forecasts of international bond returns with country-specific equity tail risk}
	\label{tab:IS_inter_vs_cs}
	\fontsize{9}{11} \selectfont
	\renewcommand{\tabcolsep}{7.4pt}  
	\sisetup{ input-symbols = {()},
		table-format = -0.2,
		table-space-text-post = ***,
		table-align-text-post = false,
		group-digits = false,
		explicit-sign}
	\centering 
	\begin{tabular}{llSSSSSSS} \hline \hline
		\vspace{-0.15cm}\\
		& & {$n=12$} & {$n=24$} & {$n=36$}  & {$n=48$} & {$n=60$} & {$n=84$} & {$n=120$} \vspace{+0.12cm}\\
		\hline
		\multicolumn{9}{l}{\rule[-2.2mm]{0mm}{6.5mm}{\textbf{Panel A: No control for bond return forecasting factors}}}\\  
		\hline
		\vspace{-0.15cm}\\
		{UK} & {$\beta$}  & 0.864 & 1.612 & 2.290 & 2.955 & 3.623 & 4.810 & 6.137 \\
		& {$p$-value}  & 0.004 & 0.000 & 0.000 & 0.000 & 0.000 & 0.002 & 0.016 \\
		& {$p$-value \tiny{(b)}} & 0.000 & 0.000 & 0.000 & 0.000 & 0.000 & 0.001 & 0.001 \\
		\\[-0.2cm]
		{DE} & {$\beta$}  & 0.581 & 1.128 & 1.563 & 1.927 & 2.231 & 2.667 & 2.909 \\
		& {$p$-value}  & 0.000 & 0.000 & 0.001 & 0.002 & 0.006 & 0.018 & 0.065 \\
		& {$p$-value \tiny{(b)}} & 0.001 & 0.002 & 0.004 & 0.006 & 0.010 & 0.022 & 0.048 \\
		\\[-0.2cm]
		{CH} & {$\beta$}  & 0.600 & 0.741 & 0.826 & 1.008 & 1.270 & 1.921 & 3.002 \\
		& {$p$-value}  & 0.117 & 0.012 & 0.009 & 0.016 & 0.015 & 0.010 & 0.005 \\
		& {$p$-value \tiny{(b)}} & 0.000 & 0.003 & 0.035 & 0.059 & 0.049 & 0.023 & 0.012 \\
		\\[-0.2cm]
		{FR} & {$\beta$}  & 0.432 & 0.561 & 0.558 & 0.557 & 0.614 & 0.966 & 2.124 \\
		& {$p$-value}  & 0.017 & 0.107 & 0.221 & 0.296 & 0.308 & 0.217 & 0.099 \\
		& {$p$-value \tiny{(b)}} & 0.008 & 0.115 & 0.310 & 0.458 & 0.517 & 0.479 & 0.327 \\
		\\[-0.2cm]
		{IT} & {$\beta$}  & 1.205 & 1.837 & 2.076 & 2.105 & 2.048 & 1.920 & 1.931 \\
		& {$p$-value}  & 0.000 & 0.007 & 0.046 & 0.123 & 0.218 & 0.379 & 0.494 \\
		& {$p$-value \tiny{(b)}} & 0.000 & 0.028 & 0.123 & 0.219 & 0.306 & 0.436 & 0.524 \\
		\\[-0.2cm]
		{ES} & {$\beta$}  & 0.726 & 0.752 & 0.560 & 0.515 & 0.531 & 0.487 & 0.187 \\
		& {$p$-value}  & 0.295 & 0.439 & 0.672 & 0.765 & 0.803 & 0.868 & 0.963 \\
		& {$p$-value \tiny{(b)}} & 0.127 & 0.492 & 0.749 & 0.804 & 0.831 & 0.892 & 0.972 \\
		\\[-0.15cm]
		\vspace{-0.15cm}\\
		\hline
		\multicolumn{9}{l}{\rule[-2.2mm]{0mm}{6.5mm}{\textbf{Panel B: Control for yield curve factors with 3 PCs (country-specific)}}}\\
		\hline
		\vspace{-0.15cm}\\
		{UK} & {$\beta$}  & 0.758 & 1.348 & 1.862 & 2.386 & 2.924 & 3.851 & 4.847 \\
		& {$p$-value}  & 0.015 & 0.001 & 0.001 & 0.002 & 0.006 & 0.034 & 0.094 \\
		& {$p$-value \tiny{(b)}} & 0.000 & 0.000 & 0.000 & 0.001 & 0.001 & 0.004 & 0.012 \\
		\\[-0.2cm]
		{DE} & {$\beta$}  & 0.634 & 1.160 & 1.512 & 1.754 & 1.915 & 2.036 & 1.798 \\
		& {$p$-value}  & 0.000 & 0.001 & 0.007 & 0.025 & 0.055 & 0.148 & 0.366 \\
		& {$p$-value \tiny{(b)}} & 0.000 & 0.001 & 0.005 & 0.012 & 0.030 & 0.097 & 0.293 \\
		\\[-0.2cm]
		{CH} & {$\beta$}  & 0.588 & 0.783 & 0.860 & 0.948 & 1.076 & 1.469 & 2.308 \\
		& {$p$-value}  & 0.139 & 0.019 & 0.004 & 0.009 & 0.022 & 0.045 & 0.039 \\
		& {$p$-value \tiny{(b)}} & 0.000 & 0.002 & 0.034 & 0.074 & 0.094 & 0.096 & 0.065 \\
		\\[-0.2cm]
		{FR} & {$\beta$}  & 0.320 & 0.280 & 0.076 & -0.138 & -0.293 & -0.346 & 0.283 \\
		& {$p$-value}  & 0.105 & 0.483 & 0.885 & 0.826 & 0.688 & 0.735 & 0.868 \\
		& {$p$-value \tiny{(b)}} & 0.062 & 0.468 & 0.892 & 0.861 & 0.767 & 0.816 & 0.907 \\
		\\[-0.2cm]
		{IT} & {$\beta$}  & 0.603 & 0.458 & 0.099 & -0.298 & -0.646 & -1.088 & -1.191 \\
		& {$p$-value}  & 0.061 & 0.621 & 0.941 & 0.852 & 0.717 & 0.607 & 0.652 \\
		& {$p$-value \tiny{(b)}} & 0.049 & 0.605 & 0.944 & 0.868 & 0.768 & 0.689 & 0.713 \\
		\\[-0.2cm]
		{ES} & {$\beta$}  & 0.419 & 0.197 & -0.188 & -0.366 & -0.444 & -0.628 & -1.127 \\
		& {$p$-value}  & 0.555 & 0.845 & 0.892 & 0.841 & 0.846 & 0.843 & 0.798 \\
		& {$p$-value \tiny{(b)}} & 0.371 & 0.854 & 0.911 & 0.858 & 0.869 & 0.858 & 0.832 \\
		\\
		\hline
		\vspace{-2.5mm}
	\end{tabular}
\end{table}

\clearpage \newpage

\begin{table}[!ht]
	\ContinuedFloat
	\caption{In-sample forecasts of international bond returns with country-specific equity tail risk (cont.)}
	\fontsize{9}{11} \selectfont
	\renewcommand{\tabcolsep}{7.4pt}  
	\sisetup{ input-symbols = {()},
		table-format = -0.2,
		table-space-text-post = ***,
		table-align-text-post = false,
		group-digits = false,
		explicit-sign}
	\centering 
	\begin{tabular}{llSSSSSSS} \hline \hline
		\vspace{-0.15cm}\\
		& & {$n=12$} & {$n=24$} & {$n=36$}  & {$n=48$} & {$n=60$} & {$n=84$} & {$n=120$} \vspace{+0.12cm}\\
		\hline
		\multicolumn{9}{l}{\rule[-2.2mm]{0mm}{6.5mm}{\textbf{Panel C: Control for yield curve factors with 5 PCs (country-specific)}}}\\
		\hline
		\vspace{-0.15cm}\\
		{UK} & {$\beta$}  & 0.747 & 1.346 & 1.847 & 2.360 & 2.894 & 3.849 & 4.944 \\
		& {$p$-value}  & 0.018 & 0.002 & 0.001 & 0.001 & 0.005 & 0.032 & 0.093 \\
		& {$p$-value \tiny{(b)}} & 0.000 & 0.000 & 0.001 & 0.001 & 0.001 & 0.003 & 0.011 \\
		\\[-0.2cm]
		{DE} & {$\beta$}  & 0.604 & 1.151 & 1.514 & 1.751 & 1.903 & 2.018 & 1.819 \\
		& {$p$-value}  & 0.000 & 0.001 & 0.009 & 0.028 & 0.059 & 0.156 & 0.370 \\
		& {$p$-value \tiny{(b)}} & 0.000 & 0.001 & 0.007 & 0.013 & 0.036 & 0.107 & 0.297 \\
		\\[-0.2cm]
		{CH} & {$\beta$}  & 0.515 & 0.729 & 0.815 & 0.886 & 0.987 & 1.342 & 2.175 \\
		& {$p$-value}  & 0.015 & 0.000 & 0.001 & 0.014 & 0.036 & 0.046 & 0.017 \\
		& {$p$-value \tiny{(b)}} & 0.000 & 0.002 & 0.042 & 0.100 & 0.122 & 0.132 & 0.080 \\
		\\[-0.2cm]
		{FR} & {$\beta$}  & 0.254 & 0.086 & -0.291 & -0.672 & -0.970 & -1.247 & -0.977 \\
		& {$p$-value}  & 0.174 & 0.810 & 0.511 & 0.201 & 0.132 & 0.209 & 0.574 \\
		& {$p$-value \tiny{(b)}} & 0.138 & 0.825 & 0.636 & 0.413 & 0.355 & 0.411 & 0.680 \\
		\\[-0.2cm]
		{IT} & {$\beta$}  & 0.639 & 0.669 & 0.545 & 0.353 & 0.149 & -0.196 & -0.476 \\
		& {$p$-value}  & 0.066 & 0.467 & 0.677 & 0.821 & 0.932 & 0.924 & 0.854 \\
		& {$p$-value \tiny{(b)}} & 0.037 & 0.443 & 0.691 & 0.841 & 0.945 & 0.948 & 0.884 \\
		\\[-0.2cm]
		{ES} & {$\beta$}  & 0.430 & 0.287 & -0.055 & -0.177 & -0.149 & 0.004 & 0.145 \\
		& {$p$-value}  & 0.509 & 0.787 & 0.970 & 0.924 & 0.947 & 0.999 & 0.972 \\
		& {$p$-value \tiny{(b)}} & 0.354 & 0.793 & 0.973 & 0.933 & 0.955 & 0.999 & 0.983 \\
		\\[-0.1cm]
		\hline
		\hline
		\vspace{-2.5mm}
	\end{tabular}
	\begin{minipage}{15.8cm}
		\footnotesize
		Notes: This table reports the slope estimates and $p$-values associated with country-specific equity tail risk measures used in return predictive regressions of Treasury bonds in United Kingdom (UK), Germany (DE), Switzerland (CH), France (FR), Italy (IT), and Spain (ES). $n$ denotes the maturity of the bonds in months. The country-specific equity tail risk measures are calculated using options on the FTSE 100 (UK), DAX (DE), SMI (CH), CAC 40 (FR), FTSE MIB (IT) and IBEX 35 (ES) equity index. Panel A reports the results of a regression that only uses the country-specific equity tail risk measure as predictor. Panel B (resp.~C) reports the results of a predictive regression that controls for country-specific yield curve factors represented by the first three (resp.~five) principal components of Treasury bond yields. All predictors have been normalized to have mean zero and unit variance. We report the Newey-West $p$-values computed with a 12-lag standard error correction, and the $p$-value \tiny{(b)} \footnotesize computed with the bootstrap procedure of \cite{Bauer2018}. The in-sample period is 2002:01--2018:12 in UK, DE and CH, 2007:01--2018:12 in IT and FR, 2007:05--2018:12 in ES.
	\end{minipage}
\end{table}

\clearpage \newpage

\begin{table}[!ht]
	\caption{Market price of country-specific equity tail risk in international bond markets}
	\label{tab:GX_RiskPremia_inter_vs_cs}
	\fontsize{9}{11} \selectfont
	\renewcommand{\tabcolsep}{5.4pt}  
	\sisetup{ input-symbols = {()},
		table-format = -0.2,
		table-space-text-post = ***,
		table-align-text-post = false,
		group-digits = false,
		explicit-sign}
	\centering 
	\begin{tabular}{llSSSSSSSS} \hline \hline
		\vspace{-0.15cm}\\
		& & {$p=1$} & {$p=2$} & {$p=3$}  & {$p=4$} & {$p=5$} & {$p=6$} & {$p=7$} & {$p=8$} \vspace{+0.12cm}\\
		\hline
		\vspace{-0.15cm}\\
		{UK} & {$\gamma_g$} & 0.054{*} & 0.091{*} & 0.086 & 0.085 & 0.083 & 0.084 & 0.087 & 0.087 \\
		& & (0.028) & (0.053) & (0.056) & (0.057) & (0.059) & (0.059) & (0.059) & (0.060) \\
		\vspace{-0.3cm}\\
		& {$R^{2}_g$} & 0.087 & 0.180 & 0.231 & 0.239 & 0.269 & 0.275 & 0.278 & 0.278 \\
		\vspace{-0.3cm}\\
		& {$p$-value} & 0.000 & 0.000 & 0.000 & 0.000 & 0.000 & 0.000 & 0.000 & 0.000 \\
		& {$g$ weak} &  &  &  &  &  &  &  &  \\
		\\[-0.2cm]
		{DE} & {$\gamma_g$} & 0.075{**} & 0.093 & 0.099 & 0.120{*} & 0.119{*} & 0.114{*} & 0.120{*} & 0.123{*} \\
		& & (0.038) & (0.057) & (0.062) & (0.065) & (0.066) & (0.068) & (0.071) & (0.072) \\
		\vspace{-0.3cm}\\
		& {$R^{2}_g$} & 0.081 & 0.189 & 0.246 & 0.263 & 0.279 & 0.324 & 0.327 & 0.329 \\
		\vspace{-0.3cm}\\
		& {$p$-value} & 0.003 & 0.001 & 0.000 & 0.000 & 0.000 & 0.000 & 0.000 & 0.000 \\
		& {$g$ weak} &  &  &  &  &  &  &  &  \\
		\\[-0.2cm]
		{CH} & {$\gamma_g$} & 0.077{**} & 0.088{**} & 0.084{**} & 0.111{**} & 0.137{**} & 0.135{**} & 0.197{**} & 0.208{**} \\
		& & (0.034) & (0.038) & (0.041) & (0.056) & (0.056) & (0.058) & (0.091) & (0.094) \\
		\vspace{-0.3cm}\\
		& {$R^{2}_g$} & 0.083 & 0.099 & 0.149 & 0.204 & 0.216 & 0.216 & 0.259 & 0.268 \\
		\vspace{-0.3cm}\\
		& {$p$-value} & 0.000 & 0.000 & 0.000 & 0.000 & 0.000 & 0.000 & 0.000 & 0.000 \\
		& {$g$ weak} &  &  &  &  &  &  &  &  \\
		\\[-0.2cm]
		{FR} & {$\gamma_g$} & 0.053{*} & 0.151{*} & 0.150{*} & 0.184{*} & 0.184{*} & 0.183{*} & 0.195{*} & 0.195{*} \\
		& & (0.032) & (0.087) & (0.087) & (0.101) & (0.101) & (0.097) & (0.101) & (0.102) \\
		\vspace{-0.3cm}\\
		& {$R^{2}_g$} & 0.042 & 0.213 & 0.214 & 0.283 & 0.283 & 0.299 & 0.322 & 0.322 \\
		\vspace{-0.3cm}\\
		& {$p$-value} & 0.027 & 0.008 & 0.016 & 0.007 & 0.014 & 0.000 & 0.013 & 0.023 \\
		& {$g$ weak} &  &  &  &  &  &  &  &  \\
		\\[-0.2cm]
		{IT} & {$\gamma_g$} & -0.011 & -0.006 & 0.010 & 0.044 & 0.046 & 0.047 & 0.048 & 0.048 \\
		& & (0.010) & (0.015) & (0.023) & (0.055) & (0.057) & (0.061) & (0.061) & (0.060) \\
		\vspace{-0.3cm}\\
		& {$R^{2}_g$} & 0.005 & 0.011 & 0.062 & 0.146 & 0.149 & 0.155 & 0.156 & 0.156 \\
		\vspace{-0.3cm}\\
		& {$p$-value} & 0.369 & 0.506 & 0.047 & 0.017 & 0.013 & 0.004 & 0.000 & 0.000 \\
		& {$g$ weak} &  &  &  &  &  &  &  &  \\
		\\[-0.2cm]
		{ES} & {$\gamma_g$} & 0.009 & 0.020 & 0.016 & 0.004 & 0.020 & 0.065 & 0.073 & 0.092 \\
		& & (0.009) & (0.025) & (0.025) & (0.025) & (0.034) & (0.068) & (0.070) & (0.070) \\
		\vspace{-0.3cm}\\
		& {$R^{2}_g$} & 0.004 & 0.009 & 0.018 & 0.020 & 0.022 & 0.029 & 0.035 & 0.058 \\
		\vspace{-0.3cm}\\
		& {$p$-value} & 0.286 & 0.564 & 0.660 & 0.762 & 0.788 & 0.851 & 0.840 & 0.000 \\
		& {$g$ weak} &  &  &  &  &  &  &  &  \\
		\\
		\hline
		\hline
		\vspace{-2.5mm}
	\end{tabular}
	\begin{minipage}{15.8cm}
		Notes: This table reports the results of the three-pass regression procedure of \cite{Giglio2019} to estimate the risk premium of country-specific equity tail risk measures in the Treasury bond market of United Kingdom (UK), Germany (DE), Switzerland (CH), France (FR), Italy (IT), and Spain (ES). The country-specific equity tail risk measures are calculated using options on the FTSE 100 (UK), DAX (DE), SMI (CH), CAC 40 (FR), FTSE MIB (IT) and IBEX 35 (ES) equity index. $p$ denotes the number of latent factors used in the three-pass estimator. For each number of latent factors, we report the estimate of the market price of risk $\gamma_g$ of the observable factor $g = \text{TR}^{(eq)}$ with standard errors in parentheses, the $R$-squared of the time series regression of the observable factor $g$ onto the $p$ latent factors, and the $p$-value of the Wald test of testing the null hypothesis that the observable factor is weak. * (resp. **, and ***) denote statistical significance at the 10\% (resp. 5\%, and 1\%) level. 
	\end{minipage}
\end{table}



\clearpage \newpage

\clearpage\newpage

\clearpage\newpage

\clearpage\newpage

\clearpage\newpage

\clearpage\newpage

\clearpage\newpage

\clearpage\newpage

\clearpage\newpage

\end{document}